\documentclass{aa}

\usepackage{txfonts}
\usepackage{color}
\usepackage[dvipsnames]{xcolor}
\usepackage{float}
\usepackage{graphicx}
\usepackage[breaklinks=true,colorlinks=true,linkcolor=blue,filecolor=blue,citecolor=blue,urlcolor=blue]{hyperref}
\usepackage{multirow}

\usepackage{siunitx}
\usepackage{caption}
\DeclareCaptionFormat{cont}{#1#2#3\par}

\newcommand{\citel}[1]{\citeauthor{#1}\,\citeyear{#1}}

\begin{document}

   \title{Constraints to neutron-star kicks in High-Mass X-ray Binaries with {\bf {\emph Gaia}} EDR3}

   \author{Francis Fortin \inst{1}
          \and
          Federico Garc\'ia \inst{2}
          \and
          Sylvain Chaty \inst{1}
          \and
          Eric Chassande-Mottin \inst{1}
          \and
          Adolfo Simaz Bunzel \inst{2}
          }

   \institute{Universit\'e Paris Cit\'e, CNRS, Astroparticule et Cosmologie, F-75013 Paris, France
     \and
     Instituto Argentino de Radioastronom\'ia (CCT La Plata, CONICET; CICPBA; UNLP), C.C.5, (1894) Villa Elisa, Buenos Aires, Argentina
   }

   \date{received 22/03/2021 ; accepted 07/06/2022}

  \abstract
   {All neutron star progenitors in neutron-star High-Mass X-ray Binaries (NS HMXBs) undergo a supernova event that may lead to a significant natal kick impacting the motion of the whole binary system. The space observatory {\it Gaia} performs a deep optical survey with exquisite astrometric accuracy, for both position and proper motions, that can be used to study natal kicks in NS HMXBs.}
   {We aim to survey the observed Galactic NS HMXB population and to quantify the magnitude of the kick imparted onto their NSs, and to highlight any possible differences arising in between the various HMXB types.}
    {We perform a census of Galactic NS HMXBs and cross-match existing detections in X-rays, optical and infrared with the {\it Gaia} Early Data Release 3 database.
    We retrieve their parallaxes, proper motions, and radial velocities (when available), and perform a selection based on the quality of the parallax measurement.
    We then compute their peculiar velocities with respect to the rotating reference frame of the Milky Way, and including their respective masses and periods, we estimate their kick velocities through Markov Chain Monte Carlo simulations of the orbit undergoing a supernova event.}
   {We infer the posterior kick distributions of 35 NS HMXBs. After an inconclusive attempt at characterising the kick distributions with Maxwellian statistics, we find that the observed NS kicks are best reproduced by a Gamma distribution of mean $116^{+18}_{-15}$\,km\,s$^{-1}$. We note that supergiant systems tend to have higher kick velocities than Be High-Mass X-ray Binaries. The peculiar velocity versus non-degenerate companion mass plane hints at a similar trend, supergiant systems having a higher peculiar velocity independently of their companion mass.
   }
   {}


   \maketitle

\section{Introduction}

High-mass X-ray binaries (HMXBs) are gravitationally-bound systems composed of a primary compact object --either a neutron star (NS) or a black hole (BH)-- accreting material from a massive O-B secondary star ($M \gtrsim 8$~M$_{\odot}$), hereafter referred to as the companion. In this work, we focus on HMXBs hosting a NS. Several types of HMXBs are defined depending on the nature of the donor and the mode of accretion. Here we distinguish 3 categories based on the nature of the companion star. Firstly, Be High-Mass X-ray Binaries (BeHMXBs) host a Be star (see review by \citeauthor{rivinius_classical_2013}\,\citeyear{rivinius_classical_2013}) that feeds a compact object through its decretion disk. Secondly, Oe High-Mass X-ray Binaries (OeHMXBs) are the hotter, O-type counterparts to BeHMXBs, and share the same accretion mechanisms. Finally, companion stars that evolve up to the supergiant stage deplete part of their intense stellar wind into their compact object, forming supergiant high-mass X-ray binaries (sgHMXBs, see review by \citeauthor{chaty_optical/infrared_2013}\,\citeyear{chaty_optical/infrared_2013}).

HMXBs are intrinsically young objects, and as such their distribution along the Milky Way is mainly concentrated in the Galactic Plane, towards the tangential directions of the spiral arms \citep{grimm_milky_2002}. More recently, \cite{2013ApJ...764..185C} show that HMXBs tend to closely follow the spiral arm structures, and that their positions can be correlated with close-by stellar-forming complexes, to estimate their migration distances. Such distances can be covered thanks to the velocity imprinted on the binary at formation, and/or by the kick experienced during the supernova (SN) event that precedes the HMXB phase, i.e. the \textit{natal kick} (see the review by \citel{lai_pulsar_2001} and the works of \citel{2004ApJ...612.1044P}, \citeyear{2005ASPC..328..327P}).

The natal kick is a key
phenomenon
in the life of a binary, as it affects both its orbital parameters and its systemic velocity \citep{1995MNRAS.274..461B}. Quantifying the exact impact of natal kicks can thus help answering several questions tied to the formation and evolution of the HMXB population: where do these systems form ? What are the characteristics of the progenitors of HMXBs ?
How prone are they to disruption in the event of a supernova kick ?

Naturally, natal kicks are tied to the mechanisms of SN explosions; a better understanding of kicks may, for instance, bring constraints to whether SN are preferentially ejecta-driven or neutrino-driven \citep{fryer_effects_2006}.
Providing observations of NS kick distributions in HMXBs could also impact closely-related research fields such as the estimation and understanding of compact merger rates as seen by the current gravitational wave detectors LIGO/Virgo (see \citel{abbott_population_2021} for both binary BH and binary NS mergers), and could help refining the population synthesis models that predict those rates (see e.g. \citel{baibhav_gravitational-wave_2019}).

Observation and modelling of natal kicks is performed on a variety of different sources, ranging from isolated pulsars \citep{1994Natur.369..127L,2005MNRAS.360..974H} to BH X-ray binaries (e.g. \citel{2013MNRAS.434.1355J}). \cite{tauris_circinus_1999} combine Monte-Carlo simulations with observations of the binary Cir X-1 to determine that an important kick ($\sim$500\,km\,s$^{-1}$) is necessary to explain its current velocity and orbit. Hubble observations led by \cite{mirabel_runaway_2002} on the BH X-ray binary GRO\,J1655$-$40 show that a high runaway velocity may be explained by the natal kick imparted during the formation of the compact object. More recently, \cite{repetto_galactic_2017} use a population-synthesis model to compare the expected vertical distribution of both BH and NS HMXBs in the Galaxy. Another study by \cite{2018MNRAS.479.4849M} make use of extremely precise interferometric data from the Australian Long Baseline Array to determine the position, velocity and orbit of PSR B1259$-$63. The authors conclude that the binary has travelled about 8\,pc from its birthplace due to a moderate kick from the first SN event. Lately, \cite{atri_potential_2019} combine radio interferometry with {\it Gaia} DR2 observables on a sample of 16 BH X-ray binaries, and infer their distribution of potential kick velocity. This population of sources has an average kick greater than 100\,km\,s$^{-1}$ with no correlation to the BH mass, potentially ruling out any link between the formation mechanism of stellar mass BHs and their mass\footnote{For BH masses ranging from 3 to 15\,M$_{\odot}$.}.

In this paper, we propose to infer the magnitude of the natal kicks of the known NS HMXBs in the Galaxy using observations of their kinematics. Our aim is to characterise the observed NS kick distributions in various types of HMXBs and provide the community constraints to those NS kicks for modelling the current population of HMXBs in the scope of binary evolution.

To perform this study, we require high-precision determination of the intrinsic binary parameters (such as companion masses, orbital period, eccentricity and systemic radial velocity) and astrometry (parallax and proper motion), which can be used to infer the distance and peculiar velocity of HMXBs. The {\it Gaia} satellite \citep{gaia_collaboration_mission_2016} provides estimations of the latter observables over a large number of sources in its latest data release \citep{brown_gaia_2021}.

As such, the Early Data Release 3 (EDR3) of {\it Gaia} offers the opportunity to robustly study the galactic distribution of NS HMXBs in five dimensions (position + proper motion). We aim to push it to a 6-dimension population study by recovering the missing systemic radial velocity of the binaries from the literature. From those observables we infer the magnitude of the natal kick imparted by the first SN explosion and the masses of the progenitors. In Section 2, we start by building a census of the known Galactic NS HMXBs and identifying their {\it Gaia} counterparts. Then, we use the observables from {\it Gaia} to compute the peculiar velocity of HMXBs in Section 3. In Section 4, we use the peculiar velocity to estimate the magnitude of the NS kicks through a Monte Carlo Markov Chain (MCMC) inversion method. We finish by discussing our results in Section 5 and conclude in Section 6.

\section{HMXB compilation and {\bf {\emph Gaia}} counterparts}

In this section we focus on building an up-to-date list of known HMXBs in the Milky Way, and look for their {\it Gaia} EDR3 counterparts. When doing this we only consider HMXBs having a referenced high-energy detection (in gamma- and/or X-rays), a single optical/near-infrared (nIR) counterpart other than {\it Gaia}, and a spectroscopic determination of the companion's spectral type,
hereafter referred to as "confirmed" HMXBs.
.

This is motivated by the fact that the latest catalogue dedicated to HMXBs
in the Milky Way
was published 15 years ago \citep{liu_catalogue_2006}; since then, many efforts were made to identify new binaries from hard and/or soft X-ray detections (eg. \citel{2013A&A...560A.108C}, \citel{masetti_unveiling_2013}, \citel{landi_investigating_2017}, \citel{fortin_spectroscopic_2018}, \citel{schwope_identification_2020}) and, when dealing with accreting binaries, to retrieve important parameters such as the spectral type of the companion, masses, periods and eccentricities. As a result, many of those parameters are missing from the catalogue
compiled by \cite{liu_catalogue_2006}
.

First we perform a cross-match of current catalogues of X-ray binaries and X-ray sources to get the starting set. Then, we query the SIMBAD database to update in bulk the information about each binary, followed by a manual, minute search for each missing values or measurement references. After that we confirm the HMXB nature of our binaries by looking for their soft X-ray counterpart as well as their associated optical/infrared counterpart. We finally look for a single, unambiguous counterpart within the {\it Gaia} EDR3.

\subsection{Cross-matching existing catalogues of X-ray binaries}

The 114 HMXBs catalogued by \cite{liu_catalogue_2006} constitute a base set of sources, that we supplement with the hard X-ray sources from the latest \textit{INTEGRAL} catalogue \citep{bird_ibis_2016}. The 939 sources in the \textit{INTEGRAL} catalogue can be of various nature, hence we only consider a subset that contains sources identified as HMXBs, Low-Mass X-ray Binaries (LMXB) or Cataclysmic Variables (CV), as well as unidentified sources. LMXBs and CVs are also accreting binaries; misidentification between HMXB, LMXB or CV is not uncommon, which is why we keep them at this stage.

To account for sources that appear in both \cite{liu_catalogue_2006} and \cite{bird_ibis_2016}, we perform a cross-match between these two lists with TOPCAT \citep{taylor_topcat_2005}. We use a positional match with the sky coordinates and the positional uncertainty of the sources (90\% confidence radius). We make sure to check the consistency of the match by looking at the identifiers of the sources, which are often common to both catalogues. This also allows us to spot any duplicates that make it through the sky match.

\subsection{Update using SIMBAD}

The SIMBAD Astronomical Database \citep{wenger_simbad_2000} from the Strasbourg astronomical Data Center (CDS) is a service that collects a wide array of information on astronomical sources. We thus chose this service in order to retrieve any update in coordinate,
companion
spectral type and
systemic
radial velocity on the list of sources we build in the previous subsection.

At this point, the coordinates we retrieved for the sources come from various instruments with significantly different astrometric precision from source to source. As such, a coordinate cross-match between this list and the CDS would not be efficient at consistently finding the correct counterparts in the CDS. Because we use different catalogues as inputs for our list, we have at our disposal up to three different identifiers for each source. Since SIMBAD offers a complete record of the different source identifiers of its sources, we query using identifiers only, testing for each and every source all existing identifiers to check the consistency of the queries.

This allows us to update the coordinates for some of the sources, and retrieve the many additional information available in SIMBAD such as the spectral type of the
companion
, optical to infrared magnitudes, proper motions and radial velocity. Yet, the localisation and identification of X-ray binaries and their optical/nIR counterparts require multi-wavelength observations, which are often led by independent teams. As a result, there is a significant amount of X-ray binaries in SIMBAD for which the information and references are not quite up to date, and need to be manually checked for more recent results.

\subsection{Optical/nIR counterparts to high-energy detections}

It is crucial to confirm for each X-ray binary whether the counterpart observed in the optical or infrared wavelengths is indeed associated to the X-ray or gamma detections. Indeed, during the listing of the HMXBs, we came across several cases in which the initial high energy detection was too uncertain (either in positional accuracy or in flux) to be classified as a confirmed HMXB according to our criteria.
Thus, we start by tracing back to the first high-energy detections and retrieve the position and uncertainty of each source. Then, we query the field in the 3XMM 8$^{\rm th}$ data release\footnote{\url{http://xmm-catalog.irap.omp.eu}}, in the 1SXPS Swift XRT point source catalogue \citep{evans_1SXPS_2014} and in the current Chandra Source Catalogue \citep[CSC 2.0,][]{evans_chandra_2019} using the dedicated CSC View software, aiming to find at least one soft X-ray counterpart. We retrieve the position with the best spatial precision (which goes from 0.6\arcsec with Chandra, 1.6\arcsec with XMM and up to $\sim$3\arcsec with Swift).

Then, we look for an IR counterpart using the 2MASS Point Source Catalogue \citep{skrutskie_two_2006}, which has a full sky coverage and the limiting magnitudes are deep enough (down to J$_{mag}$=15.8 at SNR=10) to find faint or absorbed sources. We confirm the counterpart if the optical/infrared source is located strictly within the 90\% positional error circle in X-rays, and if no other source is compatible with it.

We are only interested in HMXBs for this study; as such, it is necessary to confirm that the secondaries in our list of HMXBs have a well-constrained spectral type compatible with a massive star. Moreover, the type and luminosity class of the companion star in an HMXB can hint at how accretion happens (Roche Lobe overflow, wind accretion or decretion disk). Therefore, retrieving the spectral type of the companion also allow us to divide our sample into relevant categories and to see how they may be affected by natal kicks. This is why we only keep HMXBs with secondaries that have been characterised with spectroscopic data (see references in Table \ref{TableSources}), and not just photometry, since they provide better constraints on the spectral type.

\subsection{Gaia counterparts and binary parameters}\label{section:Gaia_counterparts}

At this point, we have at our disposal a set of confirmed HMXBs with homogeneous positional data coming from the 2MASS catalogue. We now aim to find out if a single, unambiguous {\it Gaia} counterpart exists for each of the HMXBs we collected.

Because {\it Gaia} provides extremely accurate astrometric data compared to 2MASS, we allow a certain amount of discrepancy between the positions of the 2MASS counterparts and the {\it Gaia} candidates. Firstly, we consider the presence of potential systematic error between the astrometric solutions of both surveys of 0.5\arcsec. This is backed by a study of correct matches between the {\it Gaia} DR2 versus the 2MASS catalogue by \cite{marrese_gaia_2019}, where 90\% of the correct matches are found below 0.5\arcsec from one another.
Secondly, we also control the possibility of an ambiguous association by counting the number of neighbouring {\it Gaia} sources within a radius of 0.5\arcsec around the 2MASS counterparts. All the sources we present have a single {\it Gaia} counterpart, with no other {\it Gaia} source closer than 0.5\arcsec. This ensures that there is no ambiguity in the association of our sources to their {\it Gaia} counterpart.

All the counterparts we found in {\it Gaia} EDR3 have outstanding positional constraints due to {\it Gaia}'s performance in astrometry. Their proper motion is usually given with a Signal-to-Noise Ratio
(SNR)
of 10 or more. Most have a well-constrained parallax, although its quality tends to worsen for faint sources.

Because of the orbital motion in HMXBs, single or median radial velocity measurements available in {\it Gaia} EDR3 or elsewhere in the literature are not representative of the actual systemic radial velocity. For this reason, we carefully looked for studies in which a proper radial velocity spectroscopic followup was performed, and retrieved the systemic radial velocity from the best orbital solution given. Parameters such as orbital period, eccentricity and mass ratio are also often available in the aforementioned studies. We added those to the data set as they allow us to lift degeneracies in the calculation of the natal kick.

Stellar masses, crucial in the MCMC schemes developed in Section \ref{section:kicks} for kick determination, can be constrained by orbital solutions or stellar spectra fitting. In the case none of those measurements are available, when possible, we refer to spectral type tables to get an estimation of the mass of B stars \citep{1996MNRAS.280L..31P} and O stars \citep{2005A&A...436.1049M}; this concerns 20 sources among the ones presented in Table\,\ref{TableSources}.

\subsection{The final HMXB sample}\label{sect:HMXBsample}

The sample of HMXBs we built follows two main criteria: the presence of an optical/infrared spectrum in the literature, and an unambiguous association between the high-energy and optical/infrared detections of the binaries. The latter, while seeming obvious, is far from being met for the entire population of HMXBs candidates we retrieved. The former criterion ensures that the spectral type is based on an accurate analysis of lines, and allows us to split HMXBs according to their companion being cool (Be), hot (Oe) or evolved (sg) stars.

We note that five sources (1H 0739-529, 1H 0749-600, 1H 1249-637, 1H 1253-761, 1H 1255-567) previously identified as HMXBs or candidate HMXBs ended up having a very high parallax, making them very close (closer than 700\,pc, and as close as 112\,pc for 1H 1255-567). Consequently, their resulting X-ray luminosity in the 2--10\,keV band from the 43$^{th}$ INTEGRAL General Reference Catalogue\footnote{\url{https://www.isdc.unige.ch/integral/catalog/latest/catalog.html}} is between $0.1$--$3\times10^{31}$\,erg\,s$^{-1}$, i.e. 7 to 8 orders of magnitude lower than the Galactic average for HMXBs \citep{grimm_milky_2002}. As such, we discarded them from our list as such close-by and dim HMXBs would be very unlikely.

Also, we removed from our list four binaries (Cyg X-3, Cyg X-1, SS 433 and MWC 656) as they are either confirmed or candidate BH HMXBs in BlackCAT \citep{corral-santana_blackcat_2016}.

In total, we retrieve 58 confirmed systems in the Milky Way that have an unambiguous {\it Gaia} EDR3 counterpart. The uncertainty of the parallax can vary from excellent (SNR$\sim$50) to quite poor (SNR$<$1) depending on the source, which is why we settle to work on a subsample of HMXBs that meet two additional parallax quality criteria. First, they must have a parallax estimation with an SNR above 2. Second, we also put a condition on the astrometric excess noise, which measures the disagreement between the observations and the best-fitting astrometric model. We chose to keep only sources with a ratio of astrometric excess noise over parallax below 0.5. This results in the subset of 44 HMXBs listed in Table \ref{TableSources}. We note that this procedure may give rise to selection effects, as we tend to discard far away HMXBs with poor parallax measurement. These effects are not explicitly taken into account in this study. However, we still probe HMXBs at distance greater than 8\,kpc, and reckon they should be representative of the whole population of HMXBs in the Milky Way.
According to the data we compiled from the literature, 28 of the HMXBs listed in Table\,\ref{TableSources} have a measured spin period, strongly hinting at the presence of a NS primary. For the remaining 16 HMXBs, we assume they host a NS primary after carefully checking they are not referenced in BlackCAT. Although BH BeHMXBs were recently proven to exist (e.g. MWC 656, \citel{casares_be-type_2014}), it is reasonable to assume that BeHMXBs bear a NS; the existence of BH sgHMXBs is however already well-established, which might pose a caveat for this subtype of HMXBs. In our case, 9 out of the 11 sgHMXBs we present have a confirmed spin period strongly hinting at the presence of a neutron star.

\begin{table*}
\small
\centering 
\caption{List of the 44 selected NS-HMXB sources and their {\it Gaia} EDR3 counterparts.}\label{TableSources}
\begin{tabular}{lcccccccccc}
\hline\hline
Name & Type & Sp. Type & Ref. & $M$ [M$_\odot$] & Ref. & $P_{\rm orb}$ [d]& Ref. & ecc. & Ref. & {\it Gaia}.eDR3 source id. \ \\
\hline
IGR J00370+6122$^a$ & Be & B0.5II-III & [1] & 22.0 & [2] & 15.7 & [3]  & 0.5 & [2]  & 427234969757165952 \\
1A 0114+650$^b$ & sg & B1Iae & [4] & 16.0$\pm$2.0 & [5] & 11.6 & [6]  & 0.2 & [7]  & 524924310153249920 \\
4U 0115+634$^c$ & Be & B0.2Ve & [8] & 17.5 & [9] & 24.3 & [10]  & 0.3 & [10]  & 524677469790488960 \\
IGR J01363+6610 & Be & B1Ve & [1] & 12.5 & [9] & 159.0 & [11]  &  -- & --  & 519352324516039680 \\
IGR J01583+6713$^d$ & Be & B2IVe+ & [12] & 12.5 & [9] &  -- & --  &  -- & --  & 518990967445248256 \\
LS I+61 303$^e$ & Be & B0Ve & [13] & 12.5 & [14] & 26.5 & [15]  & 0.5 & [16]  & 465645515129855872 \\
EXO 0331+530$^c$ & Oe & O8.5Ve & [17] & 18.8 & [18] & 36.5 & [10]  & 0.4 & [10]  & 444752973131169664 \\
X Per$^c$ & Be & O9.5III & [19] & 15.5 & [20] & 250.3 & [21]  & 0.1 & [21]  & 168450545792009600 \\
XTE J0421+560 & sg & B0/2I[e]:sgB[e] & [22] &  -- & --  &  -- & --  &  -- & --  & 276644757710014976 \\
RX J0440.9+4431$^d$ & Be & B0.2Ve & [23] & 17.5 & [9] & 150.0 & [24]  &  -- & --  & 252878401557369088 \\
EXO 051910+3737.7 & Be & B0III-IVe & [25] & 17.5 & [9] &  -- & --  &  -- & --  & 184497471323752064 \\
1A 0535+262$^c$ & Oe & O9.5III-Ve & [26] & 20.0 & [27] & 110.3 & [27]  & 0.5 & [28]  & 3441207615229815040 \\
HD 259440 & Be & B0pe & [29] & 15.7$\pm$2.5 & [30] & 308.0 & [31]  & 0.6 & [31]  & 3131822364779745536 \\
SAX J0635.2+0533$^f$ & Be & B2V-B1IIIe & [32] & 9.6 & [9] & 11.2 & [33]  & 0.3 & [33]  & 3131755947406031104 \\
IGR J08262-3736 & Be & OBV & [34] &  -- & --  &  -- & --  &  -- & --  & 5541793213959987968 \\
IGR J08408-4503 & sg & O8.5Ib-II(f)p & [35] & 33.0 & [36] & 9.5 & [36]  & 0.6 & [36]  & 5522306019626566528 \\
Vela X-1$^c$ & sg & B0.5Iae-1b & [37] & 26.0$\pm$1.0 & [38] & 9.0 & [39]  & 0.1 & [39]  & 5620657678322625920 \\
GRO J1008-57$^d$ & Be & B0e & [40] & 17.5 & [9] & 249.5 & [41]  & 0.7 & [41]  & 5258414192353423360 \\
2FGL J1019.0-5856 & Oe & O6V & [42] & 23.0$\pm$3.0 & [43] & 16.5 & [44]  & 0.3 & [44]  & 5255509901121774976 \\
1A 1118-615$^d$ & Oe & O9.5III-Ve & [45] & 18.0 & [18] & 24.0 & [46]  &  -- & --  & 5336957010898124160 \\
1E 1145.1-6141$^d$ & sg & B2Iae & [47] & 14.0$\pm$4.0 & [48] & 14.4 & [49]  & 0.2 & [49]  & 5334851450481641088 \\
2E 1145.5-6155 & Be & B1III-Ve & [50] & 12.5 & [9] & 187.5 & [41]  & 0.5 & [41]  & 5334823859608495104 \\
EXMS B1210-645 & Be & B2V & [51] & 9.6 & [9] & 6.7 & [46]  &  -- & --  & 6053076566300433920 \\
GX 301-2$^c$ & sg & B1.5Iaeq & [52] & 43.0$\pm$10.0 & [53] & 41.5 & [53]  & 0.5 & [53]  & 6054569565614460800 \\
GX 304-1$^c$ & Be & B2Vne & [54] & 9.6 & [9] & 132.5 & [46]  &  -- & --  & 5863533199843070208 \\
PSR B1259-63$^d$ & Oe & O9.5Ve & [55] & 22.5$\pm$7.5 & [56] & 1236.7 & [56]  & 0.9 & [56]  & 5862299960127967488 \\
IGR J14331-6112 & Be & BV-IIIe & [57] &  -- & --  &  -- & --  &  -- & --  & 5878377736381364608 \\
4U 1538-522$^g$ & sg & B0.2Ia & [58] & 20.0 & [59] & 3.7 & [59]  & 0.1 & [59]  & 5886085557746480000 \\
1H 1555-552 & Be & B2IIIn & [60] & 9.6 & [9] &  -- & --  &  -- & --  & 5884544931471259136 \\
IGR J16195-4945 & Oe & ON9.7Iab & [61] & 27.8 & [18] & 16.0 & [46]  &  -- & --  & 5935509395659726592 \\
IGR J16465-4507$^d$ & sg & O9.5Ia & [62] & 27.8 & [18] & 30.2 & [41]  &  -- & --  & 5943246345430928512 \\
4U 1700-377$^c$ & sg & O6Iafcp & [35] & 46.0$\pm$5.0 & [38] & 3.4 & [63]  & 0.0 & [63]  & 5976382915813535232 \\
RX J1744.7-2713 & Be & B0.5V-IIIe & [64] & 14.6 & [9] &  -- & --  &  -- & --  & 4060784345959549184 \\
IGR J17544-2619$^c$ & sg & O9Ib & [65] & 23.0$\pm$2.0 & [66] & 12.2 & [67]  & 0.4 & [67]  & 4063908810076415872 \\
LS5039 & Oe & ON6.5V(f) & [68] & 23.0 & [69] & 3.9 & [70]  & 0.3 & [70]  & 4104196427943626624 \\
XTE J1855-026$^d$ & sg & B0Iaep & [71] &  -- & --  & 6.0 & [71]  &  -- & --  & 4255891924062617088 \\
IGR J20006+3210$^h$ & Be & BV-III & [57] &  -- & --  &  -- & --  &  -- & --  & 2033989790047905024 \\
RX J2030.5+4751 & Be & BVe & [72] &  -- & --  & 46.0 & [41]  & 0.4 & [41]  & 2083644392294059520 \\
GRO J2058+42$^i$ & Be & O9.5-B0IV-Ve & [1] & 18.0 & [73] & 55.0 & [74]  &  -- & --  & 2065653598916388352 \\
SAX J2103.5+4545$^c$ & Be & B0Ve & [75] & 17.5 & [9] & 12.7 & [76]  & 0.4 & [76]  & 2162805896614571904 \\
IGR J21347+4737 & Be & B3V & [51] & 7.7 & [9] &  -- & --  &  -- & --  & 1978365123143522176 \\
Cep X-4$^c$ & Be & B1.5Ve:B1-B2Ve & [77] & 10.8 & [9] & 20.9 & [78]  &  -- & --  & 2178178409188167296 \\
4U 2206+543$^d$ & Oe & O9.5Vep & [79] & 18.0 & [73] & 9.6 & [80]  & 0.3 & [80]  & 2005653524280214400 \\
SAX J2239.3+6116$^g$ & Be & B0Ve & [81] & 17.5 & [9] & 262.0 & [82]  &  -- & --  & 2201091578667140352 \\
\hline
\end{tabular}

\tablefoot{Systems with a measured pulse-period according to: $^a$\cite{uchida_study_2021}, $^b$\cite{sanjurjo-ferrrin_xmm-newton_2017}, $^c$\cite{staubert_cyclotron_2019}, $^d$\cite{walter_high-mass_2015},$^e$\cite{weng_radio_2022},$^f$\cite{la_palombara_swift_2017}, $^g$\cite{van_den_eijnden_new_2021}, $^h$\cite{pradhan_revisiting_2013}, $^i$\cite{kabiraj_broad-band_2020}. Companion masses with no uncertainty provided come from photometric masses taken from [9] or [73]; a 20\% systematic error is assumed in the rest of the paper. This table is available online\footnote{\url{https://apc.u-paris.fr/~fortin/HMXB_catalogue.html}} alongside other HMXBs detected by {\it Gaia} that didn't meet the quality criterion for this study.}
\tablebib{
[1] \cite{2005A&A...440..637R}; [2] \cite{2014A&A...563A...1G}; [3] \cite{2007A&A...469.1063I}; [4] \cite{2015A&A...579A.111K}; [5] \cite{2017ApJ...844...16H}; [6] \cite{1985ApJ...299..839C}; [7] \cite{2006A&A...458..513K}; [8] \cite{2001A&A...369..108N}; [9] \cite{1996MNRAS.280L..31P}; [10] \cite{2010MNRAS.406.2663R}; [11] \cite{2010ATel.3079....1C}; [12] \cite{2008MNRAS.386.2253K}; [13] \cite{1981PASP...93..741H}; [14] \cite{2005MNRAS.360.1105C}; [15] \cite{2002ApJ...575..427G}; [16] \cite{2009ApJ...698..514A}; [17] \cite{1999MNRAS.307..695N}; [18] \cite{2005A&A...436.1049M}; [19] \cite{1982ApJS...50...55S}; [20] \cite{1997MNRAS.286..549L}; [21] \cite{2001ApJ...546..455D}; [22] \cite{2002A&A...392..991H}; [23] \cite{2005A&A...440.1079R}; [24] \cite{2013A&A...553A.103F}; [25] \cite{1990A&A...231..354P}; [26] \cite{1998PASP..110.1310W}; [27] \cite{2001A&A...377..161O}; [28] \cite{1996ApJ...459..288F}; [29] \cite{1982IAUS...98..261J}; [30] \cite{2010ApJ...724..306A}; [31] \cite{2018PASJ...70...61M}; [32] \cite{1999ApJ...523..197K}; [33] \cite{2000ApJ...542L..41K}; [34] \cite{2010A&A...519A..96M}; [35] \cite{2014ApJS..211...10S}; [36] \cite{2015A&A...583L...4G}; [37] \cite{1978mcts.book.....H}; [38] \cite{2015A&A...577A.130F}; [39] \cite{1997MNRAS.286L..21S}; [40] \cite{1994MNRAS.270L..57C}; [41] \cite{2018MNRAS.481.2779S}; [42] \cite{2015ApJ...813L..26S}; [43] \cite{2015ApJ...805...18W}  \& \cite{2015ApJ...813L..26S}; [44] \cite{2015ApJ...806..166A}; [45] \cite{1981A&A....99..274J}; [46] \cite{2015A&ARv..23....2W}; [47] \cite{1982MNRAS.201..171D}; [48] \cite{1987PASP...99..420H}; [49] \cite{2002ApJ...581.1293R}; [50] \cite{1982IAUS...98..151P}; [51] \cite{2009A&A...495..121M}; [52] \cite{1982PASP...94..541H}; [53] \cite{2006A&A...457..595K}; [54] \cite{1980MNRAS.190..537P}; [55] \cite{2011ApJ...732L..11N}; [56] \cite{2018MNRAS.479.4849M}; [57] \cite{2008A&A...482..113M}; [58] \cite{1978MNRAS.184P..73P}; [59] \cite{2004ARep...48...89A}; [60] \cite{2003AJ....126.2971V}; [61] \cite{2013A&A...560A.108C}; [62] \cite{2008A&A...486..911N}; [63] \cite{2016MNRAS.461..816I}; [64] \cite{2006A&A...454..265L}; [65] \cite{2006A&A...455..653P}; [66] \cite{2017AstL...43..664B}; [67] \cite{2013BCrAO.109...27N}; [68] \cite{2011MNRAS.416.1556T}; [69] \cite{2005MNRAS.364..899C}; [70] \cite{2011ASSP...21..559C}; [71] \cite{2008ATel.1876....1N}; [72] \cite{1988MNRAS.232..865C}; [73] \cite{2005A&A...441..235Z}; [74] \cite{1998ApJ...499..820W}; [75] \cite{2004A&A...421..673R}; [76] \cite{2007MNRAS.374.1108B}; [77] \cite{1998A&A...332L...9B}; [78] \cite{2007A&A...470.1065M}; [79] \cite{2006A&A...446.1095B}; [80] \cite{2014AN....335.1060S}; [81] \cite{2017A&A...598A..16R}; [82] \cite{2000A&A...361...85I}; } \end{table*}

\section{Galactic view of NS HMXBs}\label{section:vpec}

We retrieved the data from {\it Gaia} archive\footnote{\url{http://gea.esac.esa.int/archive/}} including the re-normalized unit weight errors (RUWE) and the corresponding updated parameter uncertainties\footnote{see {\it Gaia} technical note: Gaia-C3-TN-LU-LL-124-01}. We estimated distances following \cite{bailer-jones_estimating_2021}, using the new formula given for the distance prior.
The authors use MCMC to construct posterior probability distributions of the geometric distance to the {\it Gaia} sources, using the measured parallax combined with a direction-dependent distance prior built from a Galactic model.
We also took into account the variations of the parallax zeropoint of {\it Gaia} as described in \cite{Lindegren_gaia_2021}. The size of our sample allows us to compute more MCMCs per source than \cite{bailer-jones_estimating_2021}; while they use a burn in period of 50 samples and restrict their analysis to 500 post burn-in samples, we use a burn in period of $10^3$ samples and keep $10^6$ samples to obtain posterior distributions of distances, for each individual source in our list. In Table~\ref{TableKinematics} we present the {\it Gaia} EDR3 data retrieved and our distance estimations $r_{\rm est}$ to each of the selected HMXB sources. Lower and upper error bars are respectively taken at the $16^{th}$ and $84^{th}$ percentiles of the posteriors, corresponding to a 68\% confidence interval.

\begin{figure}
    \centering
    \includegraphics[width=0.95\columnwidth,trim={0 4cm 0 0},clip]{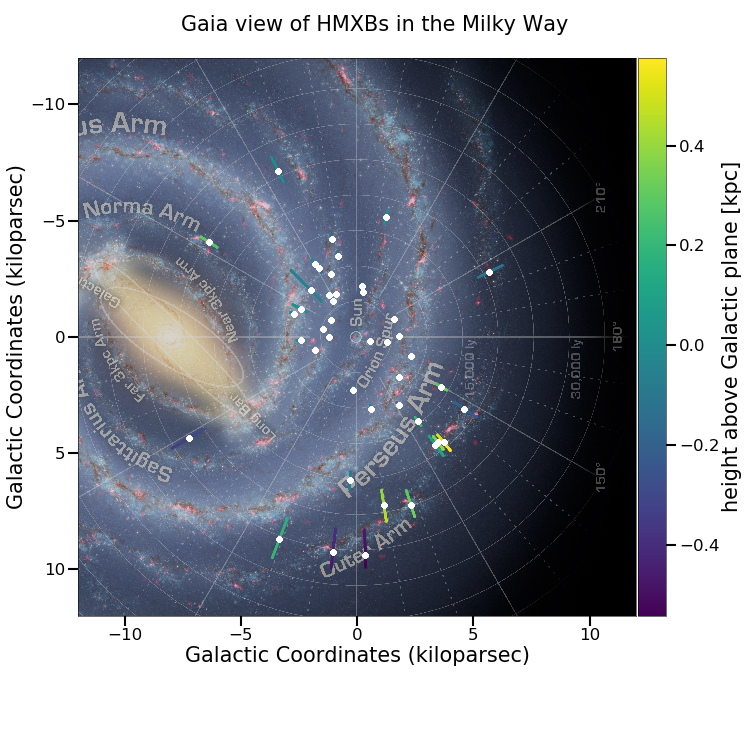}
    \caption{{\it Gaia} EDR3 view of the 44 NS-HMXBs with good {\it Gaia} astrometry. Distance error-bars correspond to 68\% confidence intervals for the posterior of the re-weighted parallax errors, obtained following \citet{bailer-jones_estimating_2021} and \citet{Lindegren_gaia_2021}. The colour scheme of the error bars is tied to the distance to the Galactic plane.}
    \label{fig:MWplane}
\end{figure}

Using this information, on Fig.~\ref{fig:MWplane} we present the Galactic distribution of the 44 NS HMXBs in our list. As previously shown by \cite{2013ApJ...764..185C} the HMXBs trace the spiral arms of the Galaxy. The unprecedented accuracy of {\it Gaia} parallaxes allows to extend this result from Perseus to the outer arm.
Since the HMXBs are part of a young stellar population, they are good tracers of recent stellar-formation regions in the Galactic plane. In our
Milky Way
map, the trend found for galactic height is consistent with the Galactic warp \citep[see][]{2019A&A...627A.150R}.

In order to compute galactocentric velocities from heliocentric equatorial coordinates, we use the \textsc{galactocentric} module from \textsc{astropy.coordinates}. For this, we assume the position of the Galactic Centre in ICRS coordinates given by \cite[][$\alpha=17^h45^m37^s.224$,  $\delta=-28^\circ56'10''.23$]{2004ApJ...616..872R}, and the height of the Sun above the mid Galactic plane of 27~pc \citep{2001ApJ...553..184C}, with a null roll angle. Furthermore, following \cite{2019MNRAS.482...40K}, we consider the solar distance to the Galactic centre to be $R_\odot = 8.2$~kpc, with a Galactic rotation speed at the local standard of rest of $v_{\rm LSR} = 236.0$~km~s$^{-1}$, and a solar peculiar velocity of $v_\odot = (8.0, 12.4, 7.7)$~km~s$^{-1}$ in Galactic coordinates. Once we obtain the galactocentric position and velocity of each source, in order to obtain their peculiar velocities $v_{\rm pec}$, we subtract the corresponding Galactic co-rotation velocity at each galactocentric position using the corresponding \textsc{MWPotential2014} solution from \cite{2015ApJS..216...29B}. 

\begin{figure}
    \centering
    \includegraphics[width=0.95\columnwidth]{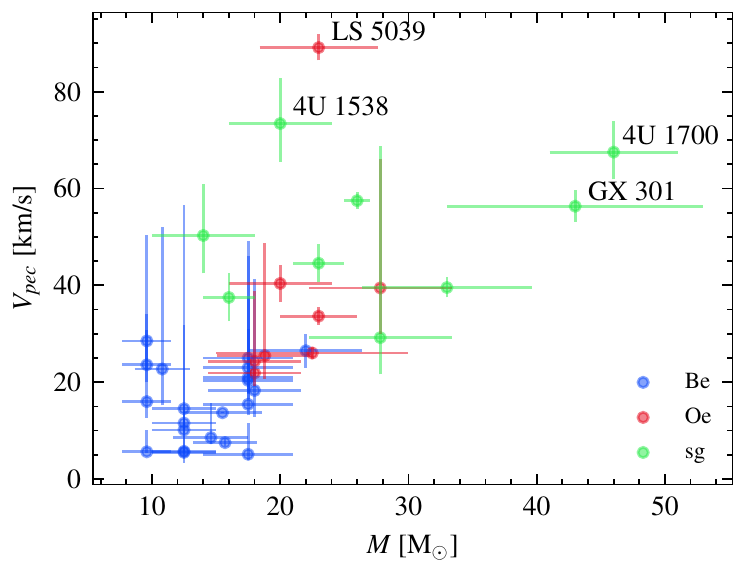}
    \caption{Inferred peculiar velocities versus companion masses in the selected HMXBs. The uncertainty in mass either come from the literature (see Table \ref{TableSources}) or from the 20\% systematic described in Section \ref{section:kicks}; the uncertainty in peculiar velocity is the 68\% confidence interval from the posteriors inferred in Section \ref{section:vpec}.}
    \label{fig:MassVsVpec}
\end{figure}

For this purpose
, during the MCMC calculations to infer the distance to the {\it Gaia} sources, 
we convert the proper motions to Galactic coordinates and, combining those values with the distance estimation, we obtain both the latitudinal and longitudinal peculiar velocities. Added to the tangential peculiar velocity $v_{\rm pec}^{\rm tan}$ make up the full peculiar velocity vector. For sources which have a systemic radial velocity (RV) measurement, we can directly estimate the full peculiar velocity $v_{\rm pec}$
posterior
. When no RV estimation is available (27 out of 44 sources), we fill in the peculiar velocity by assuming a random inclination for $v_{\rm pec}$, based on the estimated $v_{\rm pec}^{\rm tan}$ on the plane of the sky, and adopting a uniform prior in the 0--500\,km\,s$^{-1}$ range for the full peculiar velocity. On Table~\ref{TableKinematics} we present mean values and 68\% uncertainties for each of these
posterior probability distributions
for each source. Figure \ref{fig:MassVsVpec} illustrates the relation between companion masses and the inferred peculiar velocity for each HMXB that have a companion mass indicated in Table\,\ref{TableSources}.

We note that for the kick determination in the following section, we add an extra source of error to our inferences of peculiar velocities akin to a least count error. The peculiar velocity posteriors are broadened by a Gaussian with $\sigma$=15\,km\,s$^{-1}$. This is to account for the fact that when we infer v$_{\rm pec} \leq$ 15\,km\,s$^{-1}$, the statistical error returned by our method becomes much lower than the typical velocity dispersion of the massive binary population.
This dispersion was estimated for the population of young stars (0--150\,Myr) in the thin disk of the Galaxy by \cite{robin_synthetic_2003} to be between 10-20\,km\,s$^{-1}$, which should be applicable to our sample of HMXBs which are intrinsically young objects.
Hence our choice to apply a broadening of 15\,km\,s$^{-1}$ for the peculiar velocities in the following calculations.

\begin{table*}
\renewcommand{\arraystretch}{1.12}
\centering 
\caption{{\it Gaia} information, inferred distances and peculiar velocities of the 44 NS-HMXB sources with good astrometric data quality}\label{TableKinematics}
\begin{tabular}{lccccccccc}
\hline\hline
Name & pmra & pmdec & parallax & a.e.n.$^{\dagger}$ & $\sigma_{\rm a.e.n.}^{\dagger\dagger}$& $r_{\rm est}$ & RV & Ref. & $v_{\rm pec}$ \\
     & [mas yr$^{-1}$] & [mas yr$^{-1}$] & [mas] & [mas] & & [kpc] & [km~s$^{-1}$] &  & [km s$^{-1}$] \\
\hline
 IGR J00370+6122 &  -1.80$\pm$0.01 &  -0.53$\pm$0.01 &  0.272$\pm$0.012 & 0.07 & 7.91 &  $3.4^{+0.2}_{-0.1}$ & -80.0$\pm$3.0 & [1] &  $26.5^{+3.5}_{-3.6}$  \\
 1A 0114+650 &  -1.24$\pm$0.01 &  0.76$\pm$0.01 &  0.196$\pm$0.011 & 0.06 & 6.20 &  $4.5^{+0.2}_{-0.2}$ & -31.0$\pm$5.0 & [2] &  $37.5^{+5.0}_{-4.9}$  \\
 4U 0115+634 &  -1.68$\pm$0.01 &  0.50$\pm$0.01 &  0.136$\pm$0.016 & -- & -- &  $5.7^{+0.6}_{-0.5}$ &  -- & -- &  $23.0^{+26.1}_{-5.3}$  \\
 IGR J01363+6610 &  -1.63$\pm$0.01 &  -0.03$\pm$0.01 &  0.167$\pm$0.011 & 0.06 & 4.70 &  $5.7^{+0.4}_{-0.3}$ &  -- & -- &  $14.5^{+42.0}_{-4.2}$  \\
 IGR J01583+6713 &  -1.20$\pm$0.01 &  0.30$\pm$0.01 &  0.133$\pm$0.013 & -- & -- &  $5.9^{+0.5}_{-0.4}$ &  -- & -- &  $5.7^{+9.6}_{-2.4}$  \\
 LS I+61 303 &  -0.42$\pm$0.01 &  -0.26$\pm$0.01 &  0.378$\pm$0.013 & 0.07 & 8.11 &  $2.5^{+0.1}_{-0.1}$ & -41.4$\pm$0.6 & [3] &  $5.5^{+0.7}_{-0.5}$  \\
 EXO 0331+530 &  -0.27$\pm$0.02 &  0.44$\pm$0.01 &  0.134$\pm$0.020 & -- & -- &  $5.6^{+0.7}_{-0.5}$ &  -- & -- &  $25.5^{+23.3}_{-4.8}$  \\
 X Per &  -1.28$\pm$0.05 &  -1.87$\pm$0.02 &  1.627$\pm$0.037 & 0.29 & 167.97 &  $0.6^{+0.0}_{-0.0}$ & 1.0$\pm$0.9 & [4] &  $13.7^{+0.1}_{-0.1}$  \\
 XTE J0421+560 &  -0.47$\pm$0.02 &  -0.51$\pm$0.01 &  0.210$\pm$0.015 & 0.09 & 15.08 &  $4.2^{+0.3}_{-0.3}$ &  -- & -- &  $13.8^{+1.5}_{-1.2}$  \\
 RX J0440.9+4431 &  0.10$\pm$0.02 &  -1.19$\pm$0.01 &  0.379$\pm$0.015 & 0.07 & 7.05 &  $2.5^{+0.1}_{-0.1}$ &  -- & -- &  $5.1^{+6.4}_{-1.3}$  \\
 EXO 051910+3737.7 &  1.30$\pm$0.04 &  -4.00$\pm$0.02 &  0.721$\pm$0.030 & 0.18 & 23.83 &  $1.3^{+0.1}_{-0.1}$ &  -- & -- &  $20.4^{+4.1}_{-3.9}$  \\
 1A 0535+262 &  -0.59$\pm$0.03 &  -2.88$\pm$0.02 &  0.525$\pm$0.023 & 0.12 & 25.19 &  $1.8^{+0.1}_{-0.1}$ & -30.0$\pm$4.0 & [5] &  $40.4^{+3.9}_{-3.9}$  \\
 HD 259440 &  -0.03$\pm$0.02 &  -0.43$\pm$0.02 &  0.540$\pm$0.023 & 0.10 & 10.64 &  $1.8^{+0.1}_{-0.1}$ & 36.9$\pm$0.8 & [6] &  $7.5^{+0.8}_{-0.8}$  \\
 SAX J0635.2+0533 &  -0.42$\pm$0.01 &  0.40$\pm$0.01 &  0.142$\pm$0.015 & 0.05 & 2.84 &  $6.3^{+0.6}_{-0.5}$ &  -- & -- &  $28.5^{+22.0}_{-5.9}$  \\
 IGR J08262-3736 &  -2.37$\pm$0.01 &  3.18$\pm$0.01 &  0.178$\pm$0.010 & 0.04 & 2.81 &  $5.3^{+0.3}_{-0.3}$ &  -- & -- &  $9.2^{+14.7}_{-2.6}$  \\
 IGR J08408-4503 &  -7.46$\pm$0.02 &  6.10$\pm$0.01 &  0.443$\pm$0.017 & 0.14 & 16.19 &  $2.2^{+0.1}_{-0.1}$ & 15.3$\pm$0.5 & [7] &  $39.6^{+2.2}_{-2.0}$  \\
 Vela X-1 &  -4.82$\pm$0.01 &  9.28$\pm$0.01 &  0.496$\pm$0.015 & 0.10 & 12.15 &  $2.0^{+0.1}_{-0.1}$ & -3.2$\pm$0.9 & [8] &  $57.5^{+1.7}_{-1.6}$  \\
 GRO J1008-57 &  -4.70$\pm$0.02 &  3.56$\pm$0.01 &  0.243$\pm$0.013 & 0.03 & 0.47 &  $3.6^{+0.2}_{-0.2}$ &  -- & -- &  $21.0^{+16.5}_{-3.3}$  \\
 2FGL J1019.0-5856 &  -6.45$\pm$0.01 &  2.26$\pm$0.01 &  0.227$\pm$0.010 & 0.05 & 3.30 &  $4.3^{+0.2}_{-0.2}$ & 33.0$\pm$3.0 & [9] &  $33.6^{+1.9}_{-1.8}$  \\
 1A 1118-615 &  -5.42$\pm$0.01 &  1.37$\pm$0.01 &  0.329$\pm$0.011 & 0.07 & 6.38 &  $2.9^{+0.1}_{-0.1}$ &  -- & -- &  $24.3^{+14.4}_{-5.1}$  \\
 1E 1145.1-6141 &  -6.23$\pm$0.01 &  2.36$\pm$0.01 &  0.127$\pm$0.010 & 0.03 & 1.04 &  $7.9^{+0.7}_{-0.6}$ & -13.0$\pm$3.0 & [10] &  $50.3^{+10.6}_{-7.8}$  \\
 2E 1145.5-6155 &  -6.23$\pm$0.02 &  1.60$\pm$0.01 &  0.475$\pm$0.017 & 0.11 & 10.14 &  $2.1^{+0.1}_{-0.1}$ &  -- & -- &  $10.1^{+2.4}_{-0.6}$  \\
 EXMS B1210-645 &  -5.95$\pm$0.02 &  0.45$\pm$0.02 &  0.264$\pm$0.018 & 0.11 & 6.77 &  $3.4^{+0.2}_{-0.2}$ &  -- & -- &  $16.0^{+14.8}_{-3.3}$  \\
 GX 301-2 &  -5.23$\pm$0.02 &  -2.07$\pm$0.01 &  0.251$\pm$0.016 & 0.11 & 12.94 &  $3.6^{+0.2}_{-0.2}$ & 4.1$\pm$2.4 & [11] &  $56.3^{+3.4}_{-3.2}$  \\
 GX 304-1 &  -4.34$\pm$0.01 &  -0.24$\pm$0.01 &  0.540$\pm$0.014 & 0.09 & 10.30 &  $1.8^{+0.0}_{-0.0}$ &  -- & -- &  $23.6^{+10.5}_{-3.7}$  \\
 PSR B1259-63 &  -7.09$\pm$0.01 &  -0.34$\pm$0.01 &  0.443$\pm$0.013 & 0.07 & 4.90 &  $2.2^{+0.1}_{-0.1}$ & 0.0$\pm$1.0 & [12]$^*$ &  $26.0^{+1.2}_{-1.2}$  \\
 IGR J14331-6112 &  -7.25$\pm$0.11 &  -3.98$\pm$0.09 &  0.415$\pm$0.116 & -- & -- &  $2.8^{+1.2}_{-0.7}$ &  -- & -- &  $46.7^{+36.6}_{-12.2}$  \\
 4U 1538-522 &  -6.71$\pm$0.01 &  -4.11$\pm$0.01 &  0.128$\pm$0.015 & -- & -- &  $7.6^{+0.4}_{-0.4}$ & -158.0$\pm$11.0 & [13] &  $73.4^{+9.5}_{-8.0}$  \\
 1H 1555-552 &  -3.12$\pm$0.02 &  -3.22$\pm$0.01 &  0.732$\pm$0.018 & 0.09 & 9.58 &  $1.3^{+0.0}_{-0.0}$ &  -- & -- &  $5.6^{+4.4}_{-0.3}$  \\
 IGR J16195-4945 &  -0.18$\pm$0.06 &  -0.54$\pm$0.03 &  0.359$\pm$0.051 & 0.08 & 0.63 &  $2.7^{+0.4}_{-0.3}$ &  -- & -- &  $39.5^{+26.6}_{-9.5}$  \\
 IGR J16465-4507 &  -1.76$\pm$0.02 &  -3.06$\pm$0.01 &  0.296$\pm$0.017 & -- & -- &  $2.9^{+0.2}_{-0.1}$ &  -- & -- &  $29.2^{+39.6}_{-7.6}$  \\
 4U 1700-377 &  2.41$\pm$0.03 &  5.02$\pm$0.01 &  0.633$\pm$0.026 & 0.15 & 25.81 &  $1.5^{+0.1}_{-0.1}$ & -60.0$\pm$10.0 & [14] &  $67.5^{+6.5}_{-5.6}$  \\
 RX J1744.7-2713 &  -0.86$\pm$0.02 &  -2.30$\pm$0.02 &  0.787$\pm$0.024 & 0.09 & 4.67 &  $1.2^{+0.0}_{-0.0}$ &  -- & -- &  $8.6^{+7.1}_{-1.4}$  \\
 IGR J17544-2619 &  -0.51$\pm$0.03 &  -0.67$\pm$0.02 &  0.396$\pm$0.027 & 0.18 & 36.39 &  $2.4^{+0.2}_{-0.1}$ & -46.8$\pm$4.0 & [15] &  $44.6^{+3.9}_{-4.0}$  \\
 LS5039 &  7.43$\pm$0.01 &  -8.15$\pm$0.01 &  0.490$\pm$0.015 & -- & -- &  $1.9^{+0.1}_{-0.1}$ & 17.3$\pm$0.5 & [16] &  $89.1^{+2.8}_{-2.6}$  \\
 XTE J1855-026 &  -2.52$\pm$0.02 &  -6.77$\pm$0.01 &  0.078$\pm$0.016 & -- & -- &  $8.5^{+0.8}_{-0.8}$ &  -- & -- &  $34.6^{+16.6}_{-4.6}$  \\
 IGR J20006+3210 &  -2.85$\pm$0.02 &  -4.30$\pm$0.02 &  0.067$\pm$0.021 & -- & -- &  $9.3^{+0.9}_{-1.0}$ &  -- & -- &  $20.6^{+28.1}_{-6.1}$  \\
 RX J2030.5+4751 &  -2.71$\pm$0.02 &  -4.54$\pm$0.02 &  0.419$\pm$0.016 & 0.15 & 30.89 &  $2.3^{+0.1}_{-0.1}$ &  -- & -- &  $1.9^{+2.4}_{-0.4}$  \\
 GRO J2058+42 &  -2.21$\pm$0.02 &  -3.35$\pm$0.01 &  0.078$\pm$0.015 & -- & -- &  $9.3^{+0.7}_{-1.0}$ &  -- & -- &  $18.3^{+23.0}_{-5.4}$  \\
 SAX J2103.5+4545 &  -3.51$\pm$0.01 &  -3.16$\pm$0.01 &  0.131$\pm$0.013 & -- & -- &  $6.2^{+0.4}_{-0.4}$ &  -- & -- &  $25.0^{+21.0}_{-1.8}$  \\
 IGR J21347+4737 &  -2.21$\pm$0.01 &  -2.56$\pm$0.01 &  0.084$\pm$0.014 & 0.02 & 0.22 &  $9.4^{+0.6}_{-1.1}$ &  -- & -- &  $11.6^{+20.3}_{-6.0}$  \\
 Cep X-4 &  -2.96$\pm$0.01 &  -2.20$\pm$0.01 &  0.105$\pm$0.013 & -- & -- &  $7.3^{+0.8}_{-0.6}$ &  -- & -- &  $22.7^{+29.3}_{-7.5}$  \\
 4U 2206+543 &  -4.17$\pm$0.02 &  -3.32$\pm$0.01 &  0.305$\pm$0.014 & 0.11 & 14.00 &  $3.1^{+0.1}_{-0.1}$ & -54.5$\pm$1.0 & [17] &  $21.9^{+0.7}_{-0.6}$  \\
 SAX J2239.3+6116 &  -2.34$\pm$0.02 &  -1.02$\pm$0.01 &  0.104$\pm$0.014 & 0.04 & 1.04 &  $7.6^{+0.5}_{-0.7}$ &  -- & -- &  $15.4^{+15.5}_{-2.3}$  \\
\hline
\end{tabular}
\tablefoot{\scriptsize{pmra and pmdec are respectively the proper motion in RA and Dec available in {\it Gaia} EDR3. $^\dagger$\texttt{ astrometric\_excess\_noise}, $^{\dagger\dagger}$\texttt{astrometric\_excess\_noise\_sig}, see the online documentation\footnote{\url{https://gea.esac.esa.int/archive/documentation/GDR3/}}. $r_{\rm est}$ is the inferred distance, $RV$ the systemic radial velocity and $v_{\rm pec}$ the inferred peculiar velocity. $^*$Since the systemic radial velocity is not given in this reference \citep{1994MNRAS.268..430J}, we fit the radial velocity curve in figure 2 using WebPlotDigitizer \citep{Rohatgi2020} to retrieve the value in the table.}}
\tablebib{
[1] \cite{2014A&A...563A...1G}; [2] \cite{2003RMxAA..39...17K}; [3] \cite{2009ApJ...698..514A}; [4] \cite{2007ApJ...660.1398G}; [5] \cite{1984PASP...96..312H}; [6] \cite{2018PASJ...70...61M}; [7] \cite{2015A&A...583L...4G}; [8] \cite{1997MNRAS.286L..21S}; [9] \cite{2015ApJ...813L..26S}; [10] \cite{1987PASP...99..420H}; [11] \cite{2006A&A...457..595K}; [12] \cite{1994MNRAS.268..430J} 
[13] \cite{2004ARep...48...89A}; [14] \cite{1986ApJS...61..419G}; [15] \cite{2013BCrAO.109...27N}; [16] \cite{2011ASSP...21..559C}; [17] \cite{2014AN....335.1060S}; } \end{table*}

\section{NS kick inference}\label{section:kicks}

Peculiar velocities carry indirect information about the natal kick incurred by the binary system. In this section we construct a MCMC algorithm to invert the orbital equations for an asymmetric SN explosion following the classical works from \cite{1995MNRAS.274..461B,1996ApJ...471..352K}, to infer both the natal kick ($\vec{w}$, whose magnitude we hereafter denote $w$) experienced by the NSs in HMXBs, as well as the pre-supernova masses ($M_{\rm preSN}$) of the NS progenitors.

This method requires the orbital period of the NS HMXBs to be one of the inputs. From the 44 systems listed in Table\,\ref{TableSources}, 35 have an orbital period estimation. As such, in the following, we will be working with those 35 sources only.
Moreover, among those, there are two systems for which we were not able to recover a companion mass using the means described in Section \ref{section:Gaia_counterparts}.
Their companion's spectral type is not constrained enough to assign them a mass using the spectral type tables.
During the MCMC simulations, we chose to fill the missing values by a uniform random draw in an interval consistent with the spectral type of the companion: between 10--18\,M$_{\odot}$ for RX J2030.5+4751 (Be companion), and 10--53\,M$_{\odot}$ for XTE J1855-026 (supergiant companion). The mass ranges are based on the minimum and maximum companion masses we retrieved for Be and sg HMXBs. In all the
other 22
cases where no companion mass uncertainty is available, we use a 20\% systematic error during the MCMC calculations, that we include through the likelihood.
This value was chosen to reflect the relative error distribution on the companion mass in Table\,\ref{TableSources}, which averages at 17$\pm$9\%.

In our treatment, we assume that the orbit of each binary system is circular prior to SN core collapse. Most of binaries born with relatively short orbital period (up to a couple thousand days, depending on masses and metallicities, see e.g. \citel{2015ApJ...805...20S}) will experience mass transfer through Roche-Lobe-Overflow (RLO) or even a common-envelope (CE) phase. With the help of tidal forces \citep{2016ApJ...825...70D}, the timescale for circularisation in such events should be shorter than the main sequence lifetime \citep{1996A&A...309..179P}, so that just before the SN explosion the binary is likely to have already been circularised. Binaries that do not circularise their orbit because of their initial large separations instead will more likely be disrupted by the SN event \citep{1996ApJ...471..352K}. It is therefore unlikely that systems in our NS HMXB list had significant eccentricity before the SN, even though binaries can start their evolution in non-circular orbits.

In the absence of an asymmetry in the supernova explosion ($w=0$), and considering no eccentricity prior to SN explosion, binary systems survive if less than half of the total system mass is lost during the SN \citep{1961BAN....15..265B}. In such a case, the post-SN eccentricity is directly proportional to the systemic velocity, $v_{\rm sys} = e_{\rm postSN} v_{\rm orb}$ \citep{1996ApJ...471..352K}. Thus, it becomes evident that, under those assumptions, binary systems showing large systemic velocities and low eccentricities, as well as those systems with low systemic velocities and high eccentricities are hardly explained without invoking an asymmetric SN kick at birth ($w>>0$). 

The direction of the asymmetric kick could be affected by the orbital or rotational properties of the collapsing core. No clear connection between them has been found or modelled in \cite{1996ApJ...471..352K}. However, \cite{ng_birth_2007} find that isolated pulsars may be preferentially kicked along their rotation axis if they are born with short (P$_{\rm spin} < 20$\,ms) or long (P$_{\rm spin} > 100$\,ms) spin periods. Intermediate spin pulsars would on the contrary suffer from kicks at almost perpendicular directions from their rotation axis.

Since we are dealing with binary systems, and since we do not have sufficient information on the spin axis and period of the NSs we study, we prefer to remain agnostic in the modelling of the kick. We thus assume that the direction of the asymmetric kick is random, i.e. uniformly distributed in the sphere defined by the polar ($\theta$) and azimutal ($\phi$) angles with respect to the orbital velocity. In this scheme, we ignore any possible interaction between the expanding supernova shell and its binary companion.

The equations from \cite{1996ApJ...471..352K} relate the pre-SN system ($M_{1, i}$, $M_{2, i}$, $P_i$, $e_i$) with those of the post-SN ($M_{1, f}$, $M_{2, f}$, $P_{f}$, $e_{f}$, $\cos(i)$, $v_{\rm sys}$), after an arbitrary natal kick is applied ($\vec{w}$, given by its magnitude, $w$, and polar angles, $\theta$ and $\phi$), and the system loses a certain amount of mass, $\Delta M = M_{1, f} - M_{1, i}$.
We note that "pre-SN" denotes the stage right before the SN event, hence the pre-SN parameters take into account the potential mass transfer episodes preceding the SN event.
In this work, we assume that the companion mass is not affected by the SN explosion ($M_{2, i} = M_{2, f} = M_2$). This is backed by \cite{liu_interaction_2015} in which the effect of the SN onto a companion star is shown get smaller as the companion mass gets greater (simulations were performed using a 0.9 then a 3.5\,M$_{\odot}$ companion). Extrapolating to companions $>8$\,M$_{\odot}$ would make the SN have an insignificant impact on the surviving companion stars in our sample. We also assume that the binary system circularised before the SN explosion ($e_i = 0$), as explained earlier; and that the NS mass is fixed to the canonical NS mass, $M_{1, f} = 1.4$~M$_\odot$. The latter choice is backed by the observed NS mass distribution presented in \cite{kiziltan_neutron_2013} for double NS systems and NS-white dwarf binaries, which average respectively at 1.33$\pm$0.1\,M$_{\odot}$ and 1.55$\pm$0.25\,M$_{\odot}$. We thus used the in-between value of 1.4\,M$_\odot$ which happens to be the one commonly used to assign a mass to a NS when no other information is available.

This leaves us with 10 ``free'' parameters, namely: $M_{1, i}$, $M_2$, $P_i$, $P_f$, $e_f$, $\cos(i)$, $v_{\rm sys}$, $w$, $\theta$, and $\phi$. On the one hand, part of these parameters can be directly compared to observed data: the companion mass ($M_2$), the orbital period and eccentricity ($P_f$, $e_f$), and the systemic velocity ($v_{\rm sys}$) inferred from the proper motion and radial velocity, when available. On the other hand, the pre-SN mass, $M_{1, i} = M_{\rm pre-SN}$, the pre-SN period, $P_i$, the inclination angle $\cos(i)$ and the natal kick $\vec{w}$ remain unknown, but can be inferred from those equations, in a Bayesian scheme. We then construct a likelihood comparing $M_2$, $P_f$, $e_f$, and $v_{\rm sys}$ with the observables, and infer the posterior-probability distributions of $M_{1, i}$, $P_i$, $\cos(i)$, $w$, $\theta$ and $\phi$, assuming certain priors: isotropicity (by sampling $\theta$ and $\phi$ from a sphere), uniform natal kick in the $0-500$~km~s$^{-1}$ range, and broad uniform priors on the pre-SN mass ($M_{1, i}$ uniform in $1.4-25$~M$_\odot$), and the pre-SN orbital period ($P_i$ uniform in $1-10^3$~d). We choose these simple priors in such a way to introduce the least amount of assumptions, trying to bias the results as little as possible.

We assume that the likelihoods on period, eccentricity and companion mass follow normal distributions. The width of the distributions are tied to the uncertainty of those parameters. For the companion masses, we used the documented uncertainties in Table \ref{TableSources} when available; otherwise we set it to 20\% of the companion mass. The statistical errors on orbital period and eccentricity are very small; we chose to apply a 20\% systematic as the width of the Gaussian likelihood. We note that for almost-circularised systems ($e<0.2$), the likelihood for eccentricity is uniform in 0--0.25 instead of Gaussian. Choosing to apply large systematics to very constrained parameters is motivated by the potential evolution of those parameters during the post-SN epoch, even more so that we are dealing with accreting systems in which the masses and the orbits are affected by the mass and angular momentum transfer. This is backed by the observations of orbital period evolution in HMXBs that can go from a relative change of
$\dot{P}_{orb}/P_{orb}$=
10$^{-7}$\,yr$^{-1}$ \citep{1993ApJ...410..328L,1996ApJ...456L..37S,2000ApJ...541..194L} up to 10$^{-5}$\,yr$^{-1}$ \citep{1986ASIC..167...75K}. On an expected timescale since SN in HMXBs of a few 10$^{5}$ to 10$^{6}$\,yr, this warrants the consideration of a difference between the orbital parameters in the current and post-SN epochs. The value we chose (20\%) is empiric and could be addressed in the scope of more accurate binary evolution simulations.

\begin{figure}
    \centering
    \includegraphics[width=0.95\columnwidth]{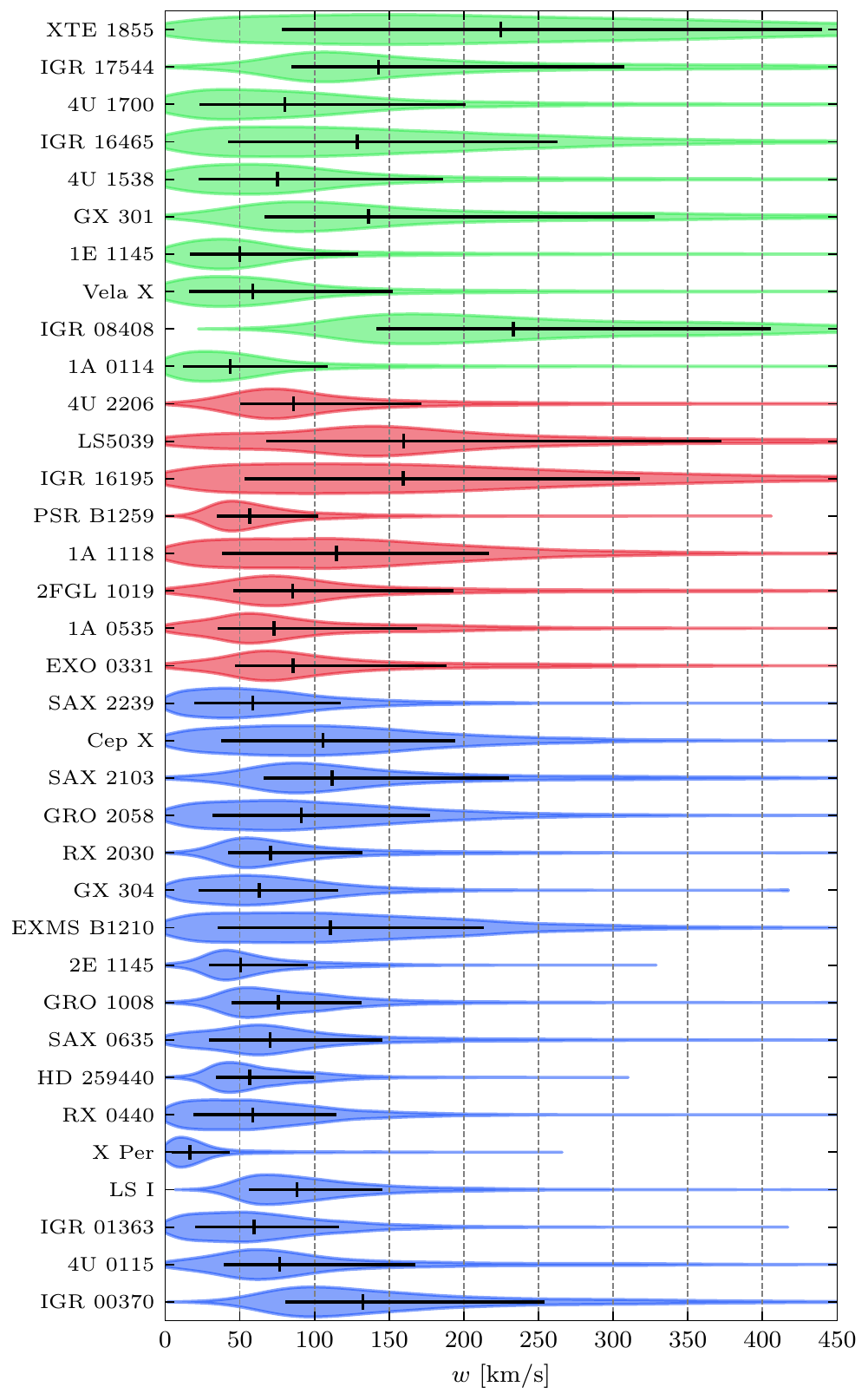}
    \caption{Mirrored density plots of kick posteriors for individual binaries. Green, red and blue colours are used for {\em sg}, {\em Oe} and {\em Be} systems, respectively. Posteriors are normalized to their maximum values.
    Source identifiers are cut for simplicity.
    }
    \label{fig:mcmc_kick_violin}
\end{figure}

\begin{figure}
    \centering
    \includegraphics[width=0.95\columnwidth]{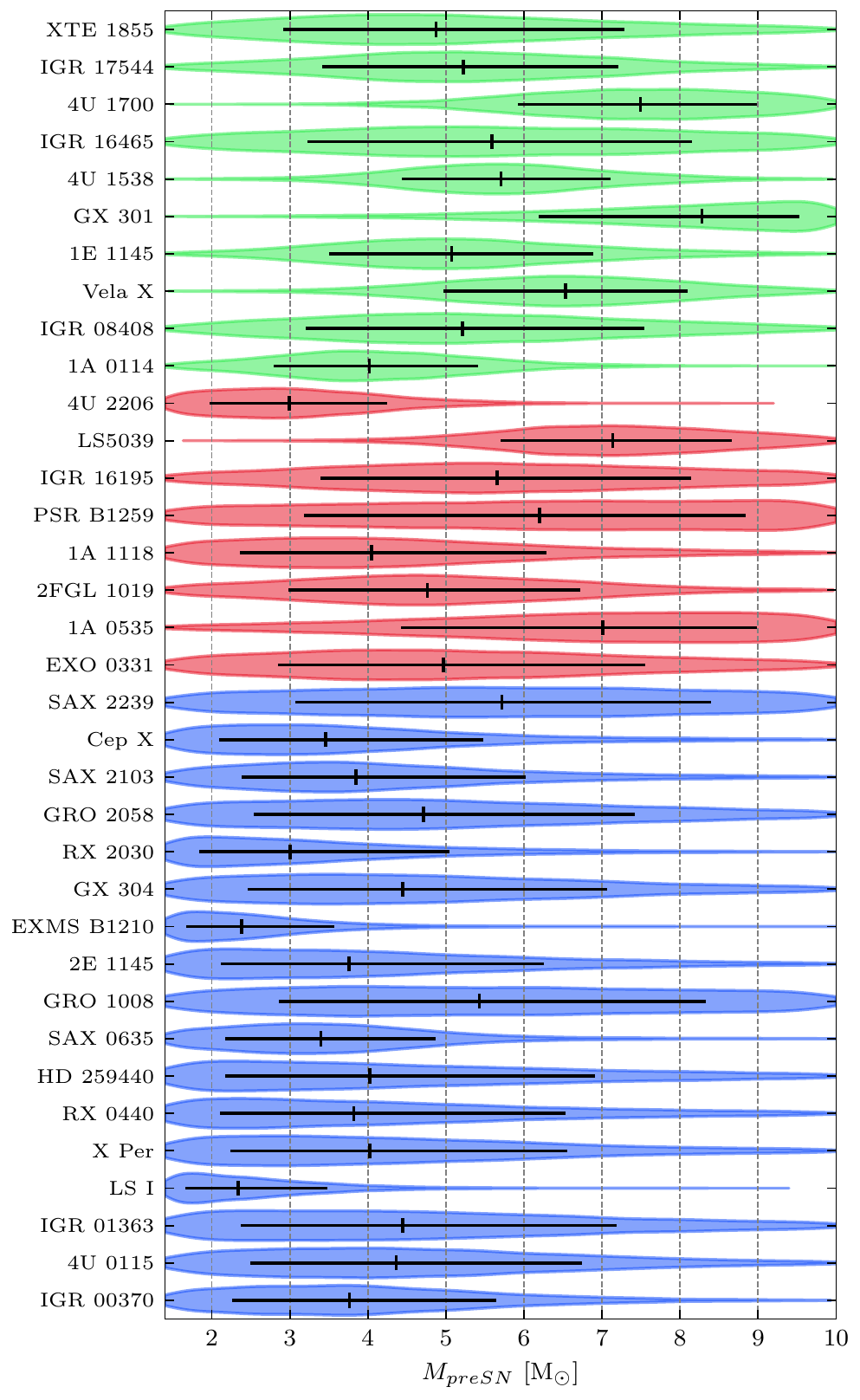}
    \caption{Mirrored density plots of pre-SN mass posteriors. Green, red and blue colours are used for {\em sg}, {\em Oe} and {\em Be} systems, respectively. Posteriors are normalized to their maximum values.
    Source identifiers are cut for simplicity.
    }
    \label{fig:mcmc_mpre_violin}
\end{figure}

Using the \texttt{emcee} sampler, we ran 200 walkers for $50 \times 10^3$ steps each. We burn-in the first $20 \times 10^3$ steps of each individual chain, and combine and thin them into a unique MCMC chain by randomly choosing $10^5$ samples, that we use to obtain median and 68\% credible intervals reported in Table~\ref{TableKicks}. On Figure~\ref{fig:mcmc_kick_violin} and~\ref{fig:mcmc_mpre_violin} we present the posterior probability distributions of the kick magnitude ($w$) and pre-SN mass ($M_{\rm preSN}$) of each HMXBs considered in this work. We use three different colours to separate the three different types of HMXBs according to the spectral type of the stellar companion (Be, Oe, sg).

From Figure \ref{fig:mcmc_kick_violin} we can note the importance of having a complete set of observables to make a reliable inference of kick velocity. For instance, IGR J08408$-$4503 and XTE J1855$-$026 are two noteworthy sgHMXBs that we infer to have high kicks ($>200$\,km\,s$^{-1}$). Yet, the latter is quite poorly constrained since we only have the proper motion, parallax and orbital period to infer the kick. The extra companion mass, radial systemic velocity and eccentricity make a much stronger case for the NS in IGR J08408$-$4503 to have experienced a very high natal kick.

\begin{table*}
\renewcommand{\arraystretch}{1.20}
\centering 
\caption{Inferred pre-SN properties and NS-kick strength for the 35 sources with determined orbital period and good astrometry}\label{TableKicks}
\begin{tabular}{lcccccc}
\hline\hline
Name & Type & $p_{\rm retrograde}$ & $f_{\rm runaway}$ & $P_{\rm preSN}$ [d] & $M_{\rm preSN}$ [M$_\odot$] & $w$ [km s$^{-1}$] \\
\hline
 IGR J00370+6122 & Be & 0.09 & 0.29 &  $18.9^{+8.8}_{-11.0}$ &  $3.8^{+1.9}_{-1.5}$ &  $132^{+121}_{-51}$  \\
 4U 0115+634 & Be & 0.06 & 0.20 &  $21.6^{+12.9}_{-9.2}$ &  $4.4^{+2.4}_{-1.8}$ &  $77^{+91}_{-36}$  \\
 IGR J01363+6610 & Be & 0.15 & 0.36 &  $165.6^{+125.6}_{-86.3}$ &  $4.4^{+2.8}_{-2.0}$ &  $60^{+55}_{-39}$  \\
 LS I+61 303 & Be & 0.04 & 0.24 &  $34.9^{+15.9}_{-22.5}$ &  $2.3^{+1.1}_{-0.7}$ &  $88^{+57}_{-31}$  \\
 X Per & Be & 0.03 & 0.10 &  $213.3^{+70.1}_{-54.8}$ &  $4.1^{+2.5}_{-1.8}$ &  $17^{+25}_{-11}$  \\
 RX J0440.9+4431 & Be & 0.12 & 0.30 &  $168.5^{+121.3}_{-80.4}$ &  $3.8^{+2.7}_{-1.7}$ &  $59^{+56}_{-39}$  \\
 HD 259440 & Be & 0.14 & 0.41 &  $405.7^{+197.8}_{-272.1}$ &  $4.0^{+2.9}_{-1.8}$ &  $57^{+42}_{-22}$  \\
 SAX J0635.2+0533 & Be & 0.03 & 0.15 &  $9.3^{+5.6}_{-3.3}$ &  $3.4^{+1.4}_{-1.2}$ &  $70^{+72}_{-40}$  \\
 GRO J1008-57 & Be & 0.21 & 0.49 &  $302.4^{+186.3}_{-220.8}$ &  $5.5^{+2.8}_{-2.6}$ &  $76^{+54}_{-31}$  \\
 2E 1145.5-6155 & Be & 0.09 & 0.32 &  $206.5^{+114.9}_{-126.9}$ &  $3.7^{+2.5}_{-1.6}$ &  $51^{+44}_{-21}$  \\
 EXMS B1210-645 & Be & 0.06 & 0.22 &  $7.5^{+5.6}_{-3.4}$ &  $2.4^{+1.1}_{-0.7}$ &  $111^{+102}_{-74}$  \\
 GX 304-1 & Be & 0.15 & 0.39 &  $136.3^{+105.1}_{-72.6}$ &  $4.4^{+2.6}_{-2.0}$ &  $63^{+53}_{-39}$  \\
 RX J2030.5+4751 & Be & 0.05 & 0.19 &  $50.5^{+25.1}_{-26.7}$ &  $3.0^{+2.0}_{-1.1}$ &  $70^{+61}_{-27}$  \\
 GRO J2058+42 & Be & 0.14 & 0.35 &  $59.5^{+44.2}_{-29.8}$ &  $4.7^{+2.7}_{-2.1}$ &  $91^{+86}_{-59}$  \\
 SAX J2103.5+4545 & Be & 0.07 & 0.25 &  $13.0^{+6.9}_{-6.8}$ &  $3.9^{+2.1}_{-1.5}$ &  $114^{+121}_{-47}$  \\
 Cep X-4 & Be & 0.12 & 0.33 &  $23.0^{+17.1}_{-11.5}$ &  $3.5^{+2.1}_{-1.3}$ &  $105^{+90}_{-67}$  \\
 SAX J2239.3+6116 & Be & 0.17 & 0.38 &  $254.2^{+210.7}_{-136.7}$ &  $5.7^{+2.7}_{-2.7}$ &  $59^{+58}_{-39}$  \\
 EXO 0331+530 & Oe & 0.09 & 0.28 &  $34.3^{+21.5}_{-17.7}$ &  $5.0^{+2.5}_{-2.1}$ &  $85^{+101}_{-38}$  \\
 1A 0535+262 & Oe & 0.14 & 0.36 &  $84.5^{+77.6}_{-44.3}$ &  $7.0^{+2.0}_{-2.6}$ &  $73^{+94}_{-37}$  \\
 2FGL J1019.0-5856 & Oe & 0.06 & 0.18 &  $15.0^{+8.1}_{-5.8}$ &  $4.8^{+1.9}_{-1.8}$ &  $85^{+109}_{-39}$  \\
 1A 1118-615 & Oe & 0.13 & 0.32 &  $27.4^{+19.7}_{-13.1}$ &  $4.0^{+2.3}_{-1.7}$ &  $115^{+104}_{-75}$  \\
 PSR B1259-63 & Oe & 0.29 & 0.56 &  $1592.0^{+1081.4}_{-1264.2}$ &  $6.2^{+2.6}_{-3.0}$ &  $57^{+44}_{-21}$  \\
 IGR J16195-4945 & Oe & 0.16 & 0.33 &  $17.3^{+13.0}_{-8.2}$ &  $5.6^{+2.5}_{-2.3}$ &  $158^{+154}_{-105}$  \\
 LS5039 & Oe & 0.07 & 0.24 &  $3.2^{+2.1}_{-1.2}$ &  $7.1^{+1.5}_{-1.4}$ &  $162^{+202}_{-94}$  \\
 4U 2206+543 & Oe & 0.02 & 0.11 &  $9.2^{+4.5}_{-3.6}$ &  $3.0^{+1.3}_{-1.0}$ &  $87^{+85}_{-35}$  \\
 1A 0114+650 & sg & 0.02 & 0.08 &  $9.9^{+3.3}_{-2.5}$ &  $4.0^{+1.3}_{-1.2}$ &  $44^{+66}_{-30}$  \\
 IGR J08408-4503 & sg & 0.17 & 0.41 &  $13.5^{+6.0}_{-8.3}$ &  $5.2^{+2.3}_{-2.0}$ &  $235^{+170}_{-91}$  \\
 Vela X-1 & sg & 0.03 & 0.09 &  $7.4^{+2.5}_{-1.8}$ &  $6.5^{+1.6}_{-1.5}$ &  $59^{+92}_{-41}$  \\
 1E 1145.1-6141 & sg & 0.04 & 0.12 &  $11.8^{+4.0}_{-3.0}$ &  $5.1^{+1.8}_{-1.5}$ &  $50^{+75}_{-32}$  \\
 GX 301-2 & sg & 0.18 & 0.36 &  $26.4^{+29.8}_{-12.1}$ &  $8.3^{+1.2}_{-2.1}$ &  $136^{+194}_{-68}$  \\
 4U 1538-522 & sg & 0.03 & 0.10 &  $3.1^{+1.0}_{-0.8}$ &  $5.7^{+1.4}_{-1.3}$ &  $76^{+112}_{-53}$  \\
 IGR J16465-4507 & sg & 0.16 & 0.34 &  $32.3^{+24.5}_{-15.8}$ &  $5.6^{+2.6}_{-2.4}$ &  $129^{+128}_{-86}$  \\
 4U 1700-377 & sg & 0.00 & 0.05 &  $2.9^{+0.9}_{-0.7}$ &  $7.5^{+1.4}_{-1.6}$ &  $79^{+113}_{-56}$  \\
 IGR J17544-2619 & sg & 0.11 & 0.31 &  $13.0^{+6.6}_{-7.1}$ &  $5.2^{+1.9}_{-1.8}$ &  $142^{+161}_{-57}$  \\
 XTE J1855-026 & sg & 0.13 & 0.30 &  $7.2^{+4.9}_{-3.3}$ &  $4.9^{+2.3}_{-1.9}$ &  $223^{+206}_{-147}$  \\
\hline
\end{tabular}
\tablefoot{$p_{\rm retrograde}$ is the fraction of MCMC simulations that end up with opposite sense of rotation with respect to the pre-SN orbit. $f_{\rm runaway}$ is the expected fraction of unbound systems assuming a random direction of the same inferred NS kick for each MCMC simulation. The higher this fraction, the less likely it is that such a system is found as a survival binary, in an HMXB phase. $P_{\rm pre SN}$ and $M_{\rm pre SN}$ are respectively the inferred pre-SN orbital period and the pre-SN progenitor mass, assuming a NS mass of 1.4~M$_\odot$ and a uniform prior in the 1.4--10~M$_\odot$ mass range. $w$ is the magnitude of the kick experienced by the NS after the SN event.}
\end{table*}

\section{Discussion}

\subsection{Peculiar velocities}
On Figure~\ref{fig:MassVsVpec} we present the relation between the mass of the companion star and the inferred peculiar velocity. Taken as a whole, the sample shows a slight positive correlation between companion mass and peculiar velocity. The HMXBs containing Be donor stars have the lowest masses and lowest peculiar velocities in the sample, while the slightly more massive HMXBs hosting Oe donors reach the highest peculiar velocities. The sg HMXBs have a systematically high peculiar velocity, with little to no dependence on mass. Based on the obtained $M_{\rm preSN}$ distributions for each HMXB type (see Figure \ref{fig:mcmc_mpre_violin} and Table~\ref{TableKicks}) and under the assumption that the newly formed compact object is a NS, there is a wide range of ejecta masses associated to each HMXB type. BeHMXBs are associated with lower $M_{\rm preSN}$ compared to Oe and sgHMXBs.

A study on BeHMXBs and sgHMXBs \citep{van_den_heuvel_origin_2000} using Hipparcos data already anticipated a dichotomy between the two populations concerning their peculiar velocities. The authors provide two reasons, the first being linked to the higher fractional helium core mass of sgHMXB progenitors compared to BeHMXBs. The authors show that the higher primary mass in sgHMXBs progenitors leads to a higher helium core mass; in turn, the initial mass transfer from the primary to the companion leads to a lower increase in orbital period compared to BeHMXB progenitors. As such, sgHMXBs tend to have tighter pre-SN orbits and higher orbital velocities compared to BeHMXBs. The second reason comes from the proportionally lower mass ejection in Be systems compared to supergiants during the SN event. Figure\,\ref{fig:MassVsVpec} shows this dichotomy between lower mass and supergiant systems, although we reckon the number of sources is too low to define a clear region in the mass-velocity plane. Roughly, the Be systems do not go beyond 40\,km\,s$^{-1}$ while only 2 out of 9 supergiants are found below that limit.

The highest peculiar velocity we inferred is attributed to the OeHMXB LS 5039 (89\,km\,s$^{-1}$), and it is rather well-constrained thanks to the accuracy of the distance and radial velocity measurements for this source. It is noteworthy since all the 7 other OeHMXBs are inferred to have peculiar velocities between 20 and 40\,km\,s$^{-1}$.
However it is difficult to tell if the very high peculiar velocity of LS 5039 compared to the other OeHMXBs is an outlier, or actually probe a specific mechanism that leads some systems to reach high peculiar velocities. One way to verify that would be to combine both observations such as the ones used here and a detailed binary evolution study of the population of HMXBs with NS, which can set better constraints on the natal-kicks for this population.

\subsection{Kick velocity and populations}\label{sect:discuss:subsect:kick}

In this section we aim to characterise the observed kick distributions of the three HMXB types (Be, Oe and sg) by using a bootstrapping method. With each iteration, we randomly pick one kick velocity according to each individual posterior inferred in Section \ref{section:kicks}. One iteration thus results in a collection of $N$ kick measurements for a population of $N$ sources. For each type of binary, we perform $10^3$ iterations to build a set of possible kick posteriors.

We fit the probability density (PDF) of each iteration with a Gamma distribution (see Equation \eqref{eq:gamma}), which we chose to parametrize with its mean $\mu$ and its skewness $s$. The best-fitting parameters are found by minimising a $\chi^2$ function with a Levenberg-Marquardt algorithm using \textsc{Scipy} \citep{scipy}.

\begin{equation}
\label{eq:gamma}
\begin{split}
&PDF(x) = \frac{1}{\Gamma(k)\theta^k}x^{k-1}e^{-\frac{x}{\theta}}\\
&\mu = k\theta, s = \frac{2}{\sqrt{k}}
\end{split}
\end{equation}

\begin{figure}
\includegraphics[width=0.95\columnwidth]{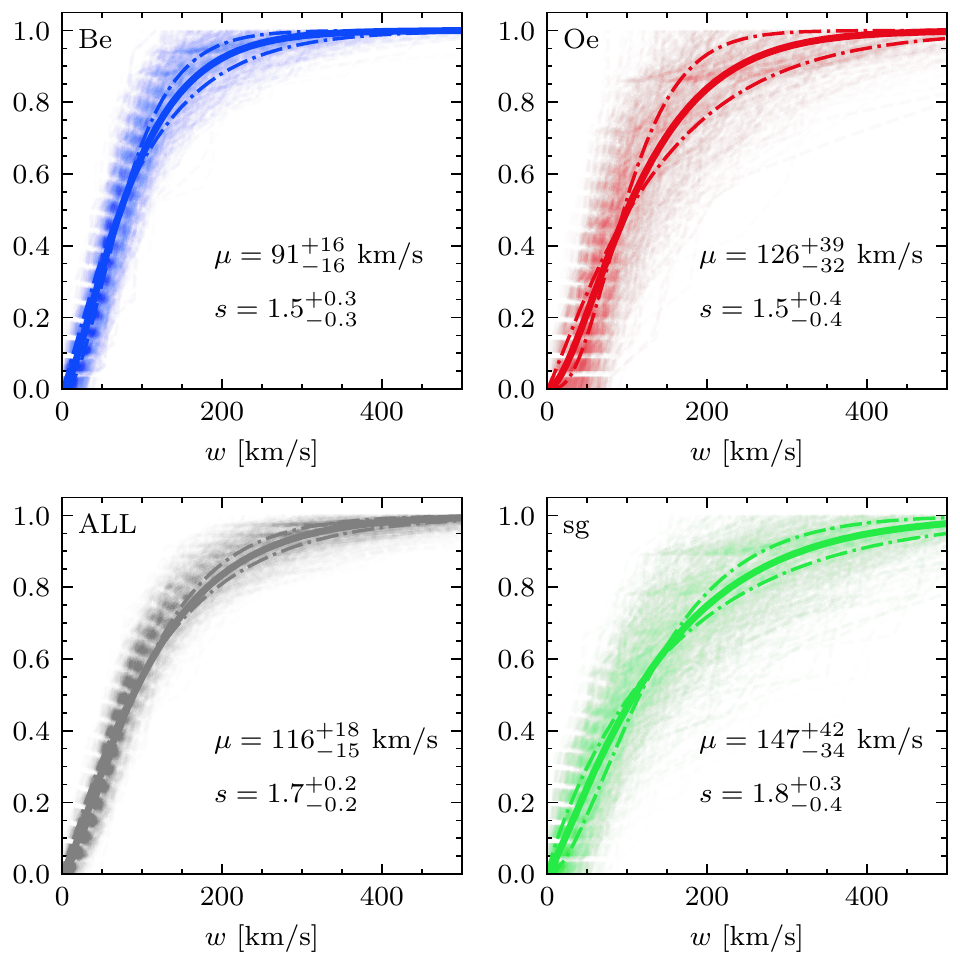}
\caption{Cumulative distributions of kicks obtained with the bootstrapping over each population (transparent) against the median results of the fit by a Gamma function (opaque lines, 1-$\sigma$ errors are dotted).}\label{fig:KickCDF}
\end{figure}

As such, for each type of HMXB, we obtain a distribution of $\mu$ and $s$ that best represent their kick velocities. We extract the median and the 68\% confidence interval of these two parameters using the same percentiles described in Section \ref{section:vpec}. The bootstraps over each kick probability distributions are plotted against a Gamma function with the above-mentioned median parameters in Figure \ref{fig:KickCDF}, in the form of cumulative distributions (CDF). For completeness, we applied the same method to the whole set of 35 HMXBs ($ALL$).

To quantify the goodness of fit over each population, we performed KS-tests of the median Gamma distributions against each iteration of the bootstrap. The resulting $p$-value distributions are all peaking over 0.9, and less than 10\% of the tests score lower than $p=0.1$, indicating that the median Gamma distributions accurately represent the data. We note that we first performed this bootstrapping method using Maxwellian distributions instead of Gamma, and we found that the fit are poor across all the HMXB types. This attempt at reproducing the NS HMXB kicks with Maxwellians is presented in Appendix \ref{sect:appendix1}.

We note a difference in the mean velocity of the Be population versus the supergiant ($\mu_{Be}=91\pm16$\,km\,s$^{-1}$ and $\mu_{sg}=147^{+42}_{-34}$\,km\,s$^{-1}$) while the Oe lay in-between at $\mu_{Oe}=126^{+39}_{-32}$\,km\,s$^{-1}$, indistinguishable from the other two. This is reflected in the 2-sample KS tests we performed between the three HMXB types, in which p-values for Be-Oe and Oe-sg both peak above 0.9, while the p-values of Be-sg are both flatter between 0.3-1 and extend further below 0.1. Possible dichotomic kick distributions according to different binary evolution channels for HMXBs have been explored in \cite{1995MNRAS.274..461B}, notably regarding the stability of mass transfer events prior to the first SN explosion. Our results tend to agree with the presence of a dichotomy, although we reckon that a bigger sample of binaries treated with our method could help to better constrain it.

The distributions of NS kicks in our list of HMXBs have an extra contribution at low velocities that may be explained by two reasons: i) we cannot account for unbound systems that suffered very high kicks and ii)
while this is still under debate,
there is a possibility that the binaries we study have interacted before the SN event, leading to a stripped progenitor and resulting in overall smaller kicks
\citep{willcox_constraints_2021}
.

Lastly, this extra contribution at low velocities could challenge our hypothesis of the presence of a NS in each of our HMXBs. As stated in Section\,\ref{sect:HMXBsample}, 28 of them are confirmed to have a pulse period. We have inferred NS kick magnitudes for 24 of them. If the low velocity contribution arises from the presence of a BH instead of a NS in some of the remaining 11 HMXBs of Table\,\ref{TableKicks}, we can roughly estimate how many BH HMXBs by re-fitting the Gamma distributions to this sub-sample of 11 sources. After performing this, we obtain seemingly identical results as the ones presented in Figure\,\ref{fig:KickCDF}, albeit with higher error bars on the fitted parameters in accordance with the lower sample size.

We also checked the consistency between the inferred kicks and the orbital period and eccentricity of the binaries, as we would expect BH HMXBs to have low kicks, low orbital periods and low eccentricities. None of the binaries in our sample meet the three criteria. The only binary with a low kick ($<$ 50\,km\,s$^{-1}$) and low eccentricity (e$\sim$0.1) is X Per, which is confirmed to bear a NS since the discovery of its pulse period \citep{staubert_cyclotron_2019}. The three binaries with low orbital periods and eccentricities (P$_{orb}<$10\,d, e$<$0.1) are Vela X-1, 4U 1538-522 and 4U 1700-377; they all have higher kicks and are also confirmed to have a pulse period \citep{staubert_cyclotron_2019,van_den_eijnden_new_2021}, thus harbouring a NS.

Based on these arguments and our search for catalogued BH HMXBs in Section\,\ref{sect:HMXBsample}, we conclude that it is unlikely that our sample of HMXBs contain any significant number of BH HMXBs that may have impacted the kick magnitude distributions.

\subsection{Impact of missing systemic radial velocity}\label{sect:discuss:subsect:rv}

Our full sample of NS HMXBs (17 Be, 8 Oe, 10 sg) is quite heterogeneous concerning the available measurements on systemic radial velocity (4 Be, 5 Oe, 8 sg), which is necessary to properly infer the full peculiar velocity and NS kick vectors. We discuss in Section \ref{section:vpec} the method we used to palliate this issue; here we present an overview of the impact of such method.

We only focus on systems with a measure of the systemic radial velocity, and perform the same analysis as described in Subsection \ref{sect:discuss:subsect:kick}. Again, the Gamma distribution represents best the distributions of kick posteriors for individual types of NS HMXBs (Be, Oe, sg) as well as for the full sample ($ALL$). The fitted parameters of the Gamma distributions are compatible within 1\,$\sigma$ of the ones we derive in Subsection \ref{sect:discuss:subsect:kick} on the entire set of binaries. We fit mean velocities of $\mu_{\rm Be}$ = 79$^{+29}_{-19}$~km~s$^{-1}$, $\mu_{\rm Oe} = 109^{+53}_{-34}$~km~s$^{-1}$ and $\mu_{\rm sg} = 135^{+48}_{-37}$~km~s$^{-1}$. The mean kick velocity of all the HMXB types confounded is fitted at $\mu = 117^{+25}_{-21}$~km~s$^{-1}$, which is seemingly the same as the one derived in Subsection \ref{sect:discuss:subsect:kick} although with larger uncertainties (by a factor of $\sim 1.4$). At $N=35$ for the full sample and $N=17$ for binaries with systemic radial velocity only, the change in uncertainty is exactly equal to the value we expect if we consider the change in sample size ($\sqrt{35/17}\sim 1.4$). At this stage, considering only NS HMXBs with an accurate measure of the systemic radial velocity does not allow for a finer determination of their kick as a population.

As for why the number of measurements on systemic radial velocity are so heterogeneous, and in particular highly in favour of sgHMXBs, a first explanation could be simply tied to the low sample numbers and statistical fluctuation. A second explanation could come from a selection effect: supergiant companions may tend to be brighter, and thus easier to get time series of spectra of. However, according to the distribution of G magnitudes in our list of HMXBs, the two sgHMXBs that do not have any RV measurement fall within the magnitude limit in which we start to see other HMXBs with a determined RV (about G=13.5). So their brightness should not be an obstacle for RV determination. The rest of the sgHMXB with determined RV are not brighter in average than the Be and Oe with determined RV. Looking at the dates of publications of the studies we found RVs in, most happened during the time the INTEGRAL satellite was operational (2003 - today). Since this mission discovered many sgHMXBs, it may also have caused an incentive to focus more on observing those systems in optical/infrared, hence the high fraction of sgHMXBs with an RV determination (especially compared to BeHMXBs); however this is purely speculative.

\subsection{Impact of companion mass sampling}

As stated in Section\,\ref{section:kicks}, we do not have a mass determination for the BeHMXB RXJ2030.5+4751 and for the sgHMXB XTE J1855-026; to infer their natal kick magnitude, we adopt a uniform sampling of the companion mass during the MCMC calculations, according to the range of masses found in their respective HMXB type (Be and sg).
This method produces two main effects. First, the effective companion mass uncertainty is greater compared to the rest of our sources. Second, this arbitrarily sets the average companion mass at the centre of the mass range, which may heavily bias our kick magnitude estimation in the case the true companion mass lies at either of the range boundaries.
Compared to all the other HMXBs kick estimation, the first effect appears negligible, as the kick magnitude error bars are not significantly higher for the two aforementioned sources (see Table\,\ref{TableKicks}). Mass uncertainty is thus not a predominant source of statistical error in the final inference in this case.
As for the potential presence of a systematic error in the companion mass, this could heavily change the magnitude of the kick in either direction if the true mass is different from the mean of the mass range. While this effect should still be marginal for the BeHMXB since the mass range is rather small, the mass range for the sgHMXB is much greater. For XTE J1855-026, if we refer to the kick magnitude posteriors in Figure\,\ref{fig:mcmc_kick_violin} and compare the the other sgHMXB kick posteriors, we notice it is one of the two sgHMXBs with very high kicks ($>200$\,km\,s$^{-1}$). It is more likely that we overestimate the NS kick magnitude for this particular source, and that the true mass is in fact on the lower side of the 10--53\,M$_{\odot}$ range.

\subsection{Neutron star velocities from disrupted systems}

The work of \cite{tauris_runaway_1998} provides analytical equations for the resulting space velocity of a neutron star that is kicked out of a binary after undergoing a supernova event. We implemented those equations in our MCMC calculations, in the cases where the kick disrupts the HMXBs, i.e. when the post-SN eccentricity is $\geq$\,1 or negative. For each of our HMXBs, we obtain the distribution of NS velocity at infinity in the case of disruption (which can vary from 5 to 56\%\, of the outcomes depending on the system, see the runaway fraction in Table\,\ref{TableKicks}). As expected, we find that the average NS velocity in each case is proportional to the kick magnitude we derive.

We compared the full NS velocity distribution across our HMXB sample to the observed velocity distributions of isolated radio pulsars. \cite{2005MNRAS.360..974H} show that the 3-D velocity of 233 known pulsars can be reproduced by a Maxwellian distribution at $\sigma$=265\,km\,s$^{-1}$. More recently, \cite{igoshev_observed_2020} report on distributions fitted either by a single Maxwellian of width $\sigma$ = 229$^{+16}_{-14}$\,km\,s$^{-1}$, or a bimodal Maxwellian with $\sigma_1$ = 146\,km\,s$^{-1}$ ($\sim$60\%) and $\sigma_2$ = 317\,km\,s$^{-1}$ ($\sim$40\%). Our data cannot be reproduced by a bimodal Maxwellian; a unimodal Maxwellian fits rather poorly, with a width $\sigma \sim$\,110\,km\,s$^{-1}$. We note that a Gamma distribution does fit slightly better than a Maxwellian, and provides a mean velocity of 240\,km\,s$^{-1}$ with a skewness of 1.2, but the fit is still not satisfactory.

Since the systems we study here have by definition survived the supernova event, it is not surprising that their pre-SN orbital configuration is not prone to produce high velocity isolated NSs; hence the isolated NS velocities we derive are lower on average than the observed population of isolated radio pulsars. Although binary evolution can produce high velocity NSs, our data suggests that less than 3\% of the outcomes results in NSs with velocities greater than 500\,km\,s$^{-1}$, and these come from systems with pre-SN orbital periods of at least a few hundred days, if not upward of 1000 days.

\section{Conclusion}

Aiming to put constraints on natal kicks experienced in observed NS HMXBs using geometric and kinematic data, we gathered a sample of HMXBs from the literature that were observed by the {\it Gaia} satellite. We derived their peculiar velocities by using their position (3D) and proper motion (2D) from {\it Gaia} data, and by retrieving their systemic radial velocity (1D) from literature. This 6D data allowed us to infer posterior distributions of NS natal kicks by running MCMC simulations of the HMXB orbits when undergoing their first SN event. In our sample of binaries, 44 had sufficient accuracy in their astrometry to derive peculiar velocity, and 35 had additional orbital parameters needed to infer natal kick and pre-SN primary mass posteriors. Differences in kick velocities between types of HMXBs were anticipated \citep{1995MNRAS.274..461B}, and our results hint at a dichotomy between Be and sg HMXBs. Be systems tend to have low NS kick compared to systems hosting a supergiant companion. In agreement with \cite{van_den_heuvel_origin_2000}, we find that the peculiar velocity of sgHMXBs is systematically high, even for supergiant companions of mass comparable to Be systems (10--18 M$_{\odot}$).

The kick distributions we infer cannot be reproduced by Maxwellian statistics. We find that the inferred distributions are better represented by Gamma functions, which incorporate an extra degree of freedom, and helps to fit for a excess at low kick velocities ($\lesssim$50~km~s$^{-1}$) which coexists with a broad distribution with significant support at high kick velocities (>200~km~s$^{-1}$, see Appendix~\ref{sect:appendix1}). This low-velocity excess could be related to the fact that we study the sub-sample of NSs which are effectively observed as members of HMXBs, compared to the full NS population. Lower kick velocities are favoured in these systems firstly because high kick velocities should lead to a higher fraction of unbound systems, and secondly because most NSs present in HMXBs might evolve from stripped-envelope progenitors from mass transfer to their companion stars. Moreover, our inference is based on isotropicity in the kicks, an hypothesis that has been challenged for isolated pulsars in more recent studies (e.g. \citel{ng_birth_2007} and \citel{noutsos_pulsar_2013}), thus possibly pointing to more complex underlying kick distributions.

Our sample of HMXBs might suffer from unaccounted selection effects coming from cut we perform in the quality of the {\it Gaia} parallax; we however have HMXBs lying at distances greater than 8\,kpc, so we are likely probing a representative sample of the general population of HMXBs in the Milky Way. In terms of binary evolution, there are two limiting factors in this study. First and foremost, each system has a peculiar velocity imprinted at birth that we do not measure individually, but instead chose to model as a velocity dispersion following \cite{2017A&A...605A...1R}. Knowing the individual birth conditions of each binary could inform us on that, but would also require studies dedicated to finding how HMXBs are born (isolated, in clusters, etc). Secondly, the orbital parameters of HMXBs might change from post-SN to what is observed today, due to mass and angular momentum exchange, wind mass loss from the secondary, proportionally to the time spent since the SN event. A better modelling of this could be done on individual HMXBs using binary evolution simulations, which are usually performed on this type of sources in the scope of forming double compact merger candidates (see e.g. \citel{marchant_new_2016}).

With the upcoming release of gravitational wave (GW) sources catalogues from the LIGO/Virgo collaborations (e.g. GWTC-2, \citeauthor{abbott_gwtc-2_2021}\,\citeyear{abbott_gwtc-2_2021}), we will soon have access to a very wide range of observables on binaries, from the high energy radiation coming from the vicinity of compact objects during accreting phases to the gravitational waves emitted during the last moments of a compact binary. When combined, they will span most of the evolutionary stages of binaries. More precisely, GW observations will bring information on the orbital parameters of compact binaries, which in turn will help to better constrain the key phases that affect those parameters, such as the kicks imparted by SN events and the common envelope episodes (see the review by \citeauthor{ivanova_common_2013}\,\citeyear{ivanova_common_2013}). For instance, \cite{garcia_progenitors_2021} study the impact of both kicks and common envelope on the recreation of the merger events GW 151226 and GW 170608 using hydrodynamical simulations of stellar evolution.

Lastly, we reckon that we should keep pushing to perform classical observations on HMXBs, as we are still in great need of homogeneous data on these sources. Direct measurements done at evolutionary stages close to the SN event are especially valuable since they are less affected by binary evolution than they are at the endpoint of their life. Observables such as systemic radial velocities, companion masses, periods and eccentricities are yet to be determined on well-known HMXBs in the Milky Way: photometry and spectroscopy in X-rays, optical, IR and radio still have a lot to bring. Depicting the full picture of today's X-ray binaries is necessary for us to have a more informed view of their evolution.

\begin{acknowledgements}
The authors were supported by the LabEx UnivEarthS, Interface project I10, "From evolution of binaries to merging of compact objects" and Interface project "Binary rEvolution".

The authors would like to thank Thierry Foglizzo, J\'er\^ome Guillet and Thomas Tauris for fruitful discussions during the writing of this paper.

This research has made use of the SIMBAD database,
operated at CDS, Strasbourg, France.
This work has made use of data from the European Space Agency (ESA) mission
{\it Gaia} (\url{https://www.cosmos.esa.int/Gaia}), processed by the {\it Gaia}
Data Processing and Analysis Consortium (DPAC,
\url{https://www.cosmos.esa.int/web/Gaia/dpac/consortium}). Funding for the DPAC
has been provided by national institutions, in particular the institutions
participating in the {\it Gaia} Multilateral Agreement.

{\em Software:} {\sc ipython/jupyter} \citep{2007CSE.....9c..21P}, {\sc matplotlib} \citep{2007CSE.....9...90H}, {\sc NumPy} \citep{2011CSE....13b..22V}, {\sc scipy} \citep{scipy} and {\sc Python} from \url{python.org}. MCMC samples were obtained with the {\sc emcee} package \citep{2013PASP..125..306F}. This research made use of {\sc astropy}, a community-developed core {\sc Python} package for astronomy \citep{2013A&A...558A..33A,2018AJ....156..123A}.

\end{acknowledgements}

\bibliographystyle{aa}
\bibliography{aanda}

\begin{thebibliography}{169}
\expandafter\ifx\csname natexlab\endcsname\relax\def\natexlab#1{#1}\fi

\bibitem[{Abbott {et~al.}(2021{\natexlab{a}})Abbott, Abbott, Abraham, Acernese,
  Ackley, Adams, Adams, Adhikari, Adya, Affeldt, Agathos, Agatsuma, Aggarwal,
  Aguiar, Aiello, Ain, Ajith, Akcay, Allen, Allocca, Altin, Amato, Anand,
  Ananyeva, Anderson, Anderson, Angelova, Ansoldi, Antelis, Antier, Appert,
  Arai, Araya, Areeda, Arène, Arnaud, Aronson, Arun, Asali, Ascenzi, Ashton,
  Aston, Astone, Aubin, Aufmuth, AultONeal, Austin, Avendano, Babak, Badaracco,
  Bader, Bae, Baer, Bagnasco, Baird, Ball, Ballardin, Ballmer, Bals, Balsamo,
  Baltus, Banagiri, Bankar, Bankar, Barayoga, Barbieri, Barish, Barker, Barneo,
  Barnum, Barone, Barr, Barsotti, Barsuglia, Barta, Bartlett, Bartos, Bassiri,
  Basti, Bawaj, Bayley, Bazzan, Becher, Bécsy, Bedakihale, Bejger, Belahcene,
  Beniwal, Benjamin, Bennett, Bentley, Bergamin, Berger, Bergmann, Bernuzzi,
  Berry, Bersanetti, Bertolini, Betzwieser, Bhandare, Bhandari, Bhattacharjee,
  Bidler, Bilenko, Billingsley, Birney, Birnholtz, Biscans, Bischi, Biscoveanu,
  Bisht, Bitossi, Bizouard, Blackburn, Blackman, Blair, Blair, Blair, Blanch,
  Bobba, Bode, Boer, Boetzel, Bogaert, Boldrini, Bondu, Bonilla, Bonnand,
  Booker, Boom, Bork, Boschi, Bose, Bossilkov, Boudart, Bouffanais, Bozzi,
  Bradaschia, Brady, Bramley, Branchesi, Brau, Breschi, Briant, Briggs,
  Brighenti, Brillet, Brinkmann, Brockill, Brooks, Brooks, Brown, Brunett,
  Bruno, Bruntz, Buikema, Bulik, Bulten, Buonanno, Buscicchio, Buskulic, Byer,
  Cabero, Cadonati, Caesar, Cagnoli, Cahillane, Calderón~Bustillo, Callaghan,
  Callister, Calloni, Camp, Canepa, Cannon, Cao, Cao, Carapella, Carbognani,
  Carney, Carpinelli, Carullo, Carver, Casanueva~Diaz, Casentini, Caudill,
  Cavaglià, Cavalier, Cavalieri, Cella, Cerdá-Durán, Cesarini, Chaibi,
  Chakravarti, Chan, Chan, Chandra, Chanial, Chao, Charlton, Chase,
  Chassande-Mottin, Chatterjee, Chattopadhyay, Chaturvedi, Chatziioannou, Chen,
  Chen, Chen, Chen, Cheng, Cheong, Chia, Chiadini, Chierici, Chincarini,
  Chiummo, Cho, Cho, Cho, Choate, Christensen, Chu, Chua, Chung, Chung, Ciani,
  Ciecielag, Cieślar, Cifaldi, Ciobanu, Ciolfi, Cipriano, Cirone, Clara,
  Clark, Clark, Clarke, Clearwater, Clesse, Cleva, Coccia, Cohadon, Cohen,
  Colleoni, Collette, Collins, Colpi, Constancio, Conti, Cooper, Corban,
  Corbitt, Cordero-Carrión, Corezzi, Corley, Cornish, Corre, Corsi, Cortese,
  Costa, Cotesta, Coughlin, Coughlin, Coulon, Countryman, Cousins, Couvares,
  Covas, Coward, Cowart, Coyne, Coyne, Creighton, Creighton, Croquette,
  Crowder, Cudell, Cullen, Cumming, Cummings, Cunningham, Cuoco, Curyło,
  Canton, Dálya, Dana, DaneshgaranBajastani, D'Angelo, Danila, Danilishin,
  D'Antonio, Danzmann, Darsow-Fromm, Dasgupta, Datrier, Dattilo, Dave, Davier,
  Davies, Davis, Daw, Dean, DeBra, Deenadayalan, Degallaix, De~Laurentis,
  Deléglise, Del~Favero, De~Lillo, De~Lillo, Del~Pozzo, DeMarchi, De~Matteis,
  D'Emilio, Demos, Denker, Dent, Depasse, De~Pietri, De~Rosa, De~Rossi,
  DeSalvo, de~Varona, Dhurandhar, Díaz, Diaz-Ortiz, Didio, Dietrich, Di~Fiore,
  DiFronzo, Di~Giorgio, Di~Giovanni, Di~Giovanni, Di~Girolamo, Di~Lieto, Ding,
  Di~Pace, Di~Palma, Di~Renzo, Divakarla, Dmitriev, Doctor, D'Onofrio, Donovan,
  Dooley, Doravari, Dorrington, Downes, Drago, Driggers, Du, Ducoin, Dupej,
  Durante, D'Urso, Duverne, Dwyer, Easter, Eddolls, Edelman, Edo, Edy, Effler,
  Eichholz, Eikenberry, Eisenmann, Eisenstein, Ejlli, Errico, Essick,
  Estellés, Estevez, Etienne, Etzel, Evans, Evans, Ewing, Fafone, Fair,
  Fairhurst, Fan, Farah, Farinon, Farr, Farr, Fauchon-Jones, Favata, Fays,
  Fazio, Feicht, Fejer, Feng, Fenyvesi, Ferguson, Fernandez-Galiana, Ferrante,
  Ferreira, Fidecaro, Figura, Fiori, Fiorucci, Fishbach, Fisher, Fishner,
  Fittipaldi, Fitz-Axen, Fiumara, Flaminio, Floden, Flynn, Fong, Font, Forsyth,
  Fournier, Frasca, Frasconi, Frei, Freise, Frey, Frey, Fritschel, Frolov,
  Fronzé, Fulda, Fyffe, Gabbard, Gadre, Gaebel, Gair, Gais, Galaudage, Gamba,
  Ganapathy, Ganguly, Gaonkar, Garaventa, García-Quirós, Garufi, Gateley,
  Gaudio, Gayathri, Gemme, Gennai, George, George, George, Gergely, Ghonge,
  Ghosh, Ghosh, Ghosh, Giacomazzo, Giacoppo, Giaime, Giardina, Gibson, Gier,
  Gill, Giri, Glanzer, Gleckl, Godwin, Goetz, Goetz, Gohlke, Goncharov,
  González, Gopakumar, Gossan, Gosselin, Gouaty, Grace, Grado, Granata,
  Granata, Grant, Gras, Grassia, Gray, Gray, Greco, Green, Green, Gretarsson,
  Griggs, Grignani, Grimaldi, Grimes, Grimm, Grote, Grunewald, Gruning,
  Guerrero, Guidi, Guimaraes, Guixé, Gulati, Guo, Gupta, Gupta, Gupta,
  Gustafson, Gustafson, Guzman, Haegel, Halim, Hall, Hamilton, Hammond, Haney,
  Hanke, Hanks, Hanna, Hannam, Hannuksela, Hannuksela, Hansen, Hansen, Hanson,
  Harder, Hardwick, Haris, Harms, Harry, Harry, Hartwig, Hasskew, Haster,
  Haughian, Hayes, Healy, Heidmann, Heintze, Heinze, Heinzel, Heitmann,
  Hellman, Hello, Helmling-Cornell, Hemming, Hendry, Heng, Hennes, Hennig,
  Hennig, Hernandez~Vivanco, Heurs, Hild, Hill, Hines, Hochheim, Hofgard,
  Hofman, Hohmann, Holgado, Holland, Hollows, Holmes, Holt, Holz, Hopkins,
  Horst, Hough, Howell, Hoy, Hoyland, Huang, Hübner, Huddart, Huerta, Hughey,
  Hui, Husa, Huttner, Hutzler, Huxford, Huynh-Dinh, Idzkowski, Iess, Imperato,
  Inchauspe, Ingram, Intini, Isi, Iyer, JaberianHamedan, Jacqmin, Jadhav,
  Jadhav, James, Jani, Janssens, Janthalur, Jaranowski, Jariwala, Jaume,
  Jenkins, Jeunon, Jiang, Johns, Johnson-McDaniel, Jones, Jones, Jones, Jones,
  Jones, Jonker, Ju, Junker, Kalaghatgi, Kalogera, Kamai, Kandhasamy, Kang,
  Kanner, Kapadia, Kapasi, Karathanasis, Karki, Kashyap, Kasprzack, Kastaun,
  Katsanevas, Katsavounidis, Katzman, Kawabe, Kéfélian, Keitel, Key, Khadka,
  Khalili, Khan, Khan, Khazanov, Khetan, Khursheed, Kijbunchoo, Kim, Kim, Kim,
  Kim, Kim, Kim, Kimball, King, Kinley-Hanlon, Kirchhoff, Kissel, Kleybolte,
  Klimenko, Knowles, Knyazev, Koch, Koehlenbeck, Koekoek, Koley, Kolstein,
  Komori, Kondrashov, Kontos, Koper, Korobko, Korth, Kovalam, Kozak, Krämer,
  Kringel, Krishnendu, Królak, Kuehn, Kumar, Kumar, Kumar, Kumar, Kuns, Kwang,
  Lackey, Laghi, Lalande, Lam, Lamberts, Landry, Lane, Lang, Lange, Lantz,
  Lanza, La~Rosa, Lartaux-Vollard, Lasky, Laxen, Lazzarini, Lazzaro, Leaci,
  Leavey, Lecoeuche, Lee, Lee, Lee, Lee, Lehmann, Leon, Leroy, Letendre, Levin,
  Li, Li, Li, Li, Li, Linde, Linker, Linley, Littenberg, Liu, Liu,
  Llorens-Monteagudo, Lo, Lockwood, London, Longo, Lorenzini, Loriette,
  Lormand, Losurdo, Lough, Lousto, Lovelace, Lück, Lumaca, Lundgren, Ma,
  Macas, MacInnis, Macleod, MacMillan, Macquet, Magaña~Hernandez,
  Magaña-Sandoval, MagazzÃ¹, Magee, Majorana, Maksimovic, Maliakal, Malik,
  Man, Mandic, Mangano, Mansell, Manske, Mantovani, Mapelli, Marchesoni,
  Marion, Márka, Márka, Markakis, Markosyan, Markowitz, Maros, Marquina,
  Marsat, Martelli, Martin, Martin, Martinez, Martinez, Martynov, Masalehdan,
  Mason, Massera, Masserot, Massinger, Masso-Reid, Mastrogiovanni, Matas,
  Mateu-Lucena, Matichard, Matiushechkina, Mavalvala, Maynard, McCann,
  McCarthy, McClelland, McCormick, McCuller, McGuire, McIsaac, McIver, McManus,
  McRae, McWilliams, Meacher, Meadors, Mehmet, Mehta, Melatos, Melchor,
  Mendell, Menendez-Vazquez, Mercer, Mereni, Merfeld, Merilh, Merritt,
  Merzougui, Meshkov, Messenger, Messick, Metzdorff, Meyers, Meylahn, Mhaske,
  Miani, Miao, Michaloliakos, Michel, Middleton, Milano, Miller, Millhouse,
  Mills, Milotti, Milovich-Goff, Minazzoli, Minenkov, Mir, Mishkin, Mishra,
  Mistry, Mitra, Mitrofanov, Mitselmakher, Mittleman, Mo, Mogushi, Mohapatra,
  Mohite, Molina, Molina-Ruiz, Mondin, Montani, Moore, Moraru, Morawski,
  Moreno, Morisaki, Mours, Mow-Lowry, Mozzon, Muciaccia, Mukherjee, Mukherjee,
  Mukherjee, Mukherjee, Mukund, Mullavey, Munch, Muñiz, Murray, Nadji, Nagar,
  Nardecchia, Naticchioni, Nayak, Neil, Neilson, Nelemans, Nelson, Nery,
  Neunzert, Nitz, Ng, Ng, Nguyen, Nguyen, Nguyen, Nichols, Nissanke, Nocera,
  Noh, North, Nothard, Nuttall, Oberling, O'Brien, O'Dell, Oganesyan, Ogin, Oh,
  Oh, Ohme, Ohta, Okada, Olivetto, Oppermann, Oram, O'Reilly, Ormiston, Ortega,
  O'Shaughnessy, Ossokine, Osthelder, Ottaway, Overmier, Owen, Pace, Pagano,
  Page, Pagliaroli, Pai, Pai, Palamos, Palashov, Palomba, Pan, Panda, Pang,
  Pankow, Pannarale, Pant, Paoletti, Paoli, Paolone, Parker, Pascucci,
  Pasqualetti, Passaquieti, Passuello, Patel, Patricelli, Payne, Pechsiri,
  Pedraza, Pegoraro, Pele, Penn, Perego, Perez, Périgois, Perreca, Perriès,
  Petermann, Petterson, Pfeiffer, Pham, Phukon, Piccinni, Pichot, Piendibene,
  Piergiovanni, Pierini, Pierro, Pillant, Pilo, Pinard, Pinto, Piotrzkowski,
  Pirello, Pitkin, Placidi, Plastino, Pluchar, Poggiani, Polini, Pong,
  Ponrathnam, Popolizio, Porter, Poverman, Powell, Pracchia, Prajapati, Prasai,
  Prasanna, Pratten, Prestegard, Principe, Prodi, Prokhorov, Prosposito,
  Prudenzi, Puecher, Punturo, Puosi, Puppo, Pürrer, Qi, Quetschke, Quinonez,
  Quitzow-James, Raab, Raaijmakers, Radkins, Radulesco, Raffai, Rafferty, Rail,
  Raja, Rajan, Rajbhandari, Rakhmanov, Ramirez, Ramirez, Ramos-Buades, Rana,
  Rao, Rapagnani, Rapol, Ratto, Raymond, Razzano, Read, Regimbau, Rei, Reid,
  Reitze, Rettegno, Ricci, Richardson, Richardson, Richardson, Ricker,
  Riemenschneider, Riles, Rizzo, Robertson, Robinet, Rocchi, Rocha, Rodriguez,
  Rodriguez-Soto, Rolland, Rollins, Roma, Romanelli, Romano, Romel, Romero,
  Romero-Shaw, Romie, Ronchini, Rose, Rose, Rose, Rosell, Rosińska, Rosofsky,
  Ross, Rowan, Rowlinson, Roy, Roy, Ruggi, Ryan, Sachdev, Sadecki, Sadiq,
  Sakellariadou, Salafia, Salconi, Saleem, Samajdar, Sanchez, Sanchez, Sanchez,
  Sanchis-Gual, Sanders, Sandles, Santiago, Santos, Saravanan, Sarin, Sassolas,
  Sathyaprakash, Sauter, Savage, Savant, Sawant, Sayah, Schaetzl, Schale,
  Scheel, Scheuer, Schindler-Tyka, Schmidt, Schnabel, Schofield, Schönbeck,
  Schreiber, Schulte, Schutz, Schwarm, Schwartz, Scott, Scott, Seglar-Arroyo,
  Seidel, Sellers, Sengupta, Sennett, Sentenac, Sequino, Sergeev, Setyawati,
  Shaffer, Shahriar, Sharifi, Sharma, Sharma, Shawhan, Shen, Shikauchi, Shink,
  Shoemaker, Shoemaker, Shukla, ShyamSundar, Sieniawska, Sigg, Singer, Singh,
  Singh, Singha, Singhal, Sintes, Sipala, Skliris, Slagmolen, Slaven-Blair,
  Smetana, Smith, Smith, Somala, Son, Soni, Soni, Sorazu, Sordini, Sorrentino,
  Sorrentino, Soulard, Souradeep, Sowell, Spencer, Spera, Srivastava,
  Srivastava, Staats, Stachie, Steer, Steinhoff, Steinke, Steinlechner,
  Steinlechner, Steinmeyer, Stevenson, Stolle-McAllister, Stops, Stover,
  Strain, Stratta, Strunk, Sturani, Stuver, Südbeck, Sudhagar, Sudhir, Suh,
  Summerscales, Sun, Sun, Sunil, Sur, Suresh, Sutton, Swinkels, Szczepańczyk,
  Tacca, Tait, Talbot, Tanasijczuk, Tanner, Tao, Tapia, Tapia San~Martin,
  Tasson, Taylor, Tenorio, Terkowski, Thirugnanasambandam, Thomas, Thomas,
  Thomas, Thompson, Thondapu, Thorne, Thrane, Tiwari, Tiwari, Tiwari, Toland,
  Tolley, Tonelli, Tornasi, Torres-Forné, Torrie, e~Melo, Töyrä, Tran,
  Trapananti, Travasso, Traylor, Tringali, Tripathee, Trovato, Trudeau, Tsai,
  Tsang, Tse, Tso, Tsukada, Tsuna, Tsutsui, Turconi, Ubhi, Udall, Ueno,
  Ugolini, Unnikrishnan, Urban, Usman, Utina, Vahlbruch, Vajente, Vajpeyi,
  Valdes, Valentini, Valsan, van Bakel, van Beuzekom, van~den Brand, Van
  Den~Broeck, Vander-Hyde, van~der Schaaf, van Heijningen, Vardaro, Vargas,
  Varma, Vass, Vasúth, Vecchio, Vedovato, Veitch, Veitch, Venkateswara,
  Venneberg, Venugopalan, Verkindt, Verma, Veske, Vetrano, Viceré, Viets,
  Vijaykumar, Villa-Ortega, Vinet, Vitale, Vo, Vocca, Vorvick, Vyatchanin,
  Wade, Wade, Wade, Walet, Walker, Wallace, Wallace, Walsh, Wang, Wang, Wang,
  Wang, Ward, Warner, Was, Washington, Watchi, Weaver, Wei, Weinert, Weinstein,
  Weiss, Wellmann, Wen, Weßels, Westhouse, Wette, Whelan, White, White,
  Whiting, Whittle, Wilken, Williams, Williams, Williamson, Willis, Willke,
  Wilson, Wimmer, Winkler, Wipf, Woan, Woehler, Wofford, Wong, Wrangel, Wright,
  Wu, Wysocki, Xiao, Yamamoto, Yang, Yang, Yang, Yap, Yeeles, Yoon, Yu, Yu,
  Yuen, ZadroŻny, Zanolin, Zelenova, Zendri, Zevin, Zhang, Zhang, Zhang,
  Zhang, Zhao, Zhao, Zheng, Zhou, Zhou, Zhu, Zimmerman, Zlochower, Zucker,
  Zweizig, {LIGO Scientific Collaboration}, \& {Virgo
  Collaboration}}]{abbott_gwtc-2_2021}
Abbott, R., Abbott, T.~D., Abraham, S., {et~al.} 2021{\natexlab{a}}, Physical
  Review X, 11, 021053

\bibitem[{Abbott {et~al.}(2021{\natexlab{b}})Abbott, Abbott, Abraham, Acernese,
  Ackley, Adams, Adams, Adhikari, Adya, Affeldt, Agathos, Agatsuma, Aggarwal,
  Aguiar, Aiello, Ain, Ajith, Allen, Allocca, Altin, Amato, Anand, Ananyeva,
  Anderson, Anderson, Angelova, Ansoldi, Antelis, Antier, Appert, Arai, Araya,
  Areeda, Arène, Arnaud, Aronson, Arun, Asali, Ascenzi, Ashton, Aston, Astone,
  Aubin, Aufmuth, AultONeal, Austin, Avendano, Babak, Badaracco, Bader, Bae,
  Baer, Bagnasco, Baird, Ball, Ballardin, Ballmer, Bals, Balsamo, Baltus,
  Banagiri, Bankar, Bankar, Barayoga, Barbieri, Barish, Barker, Barneo, Barnum,
  Barone, Barr, Barsotti, Barsuglia, Barta, Bartlett, Bartos, Bassiri, Basti,
  Bawaj, Bayley, Bazzan, Becher, Bécsy, Bedakihale, Bejger, Belahcene,
  Beniwal, Benjamin, Bennett, Bentley, Bergamin, Berger, Bergmann, Bernuzzi,
  Berry, Bersanetti, Bertolini, Betzwieser, Bhandare, Bhandari, Bhattacharjee,
  Bidler, Bilenko, Billingsley, Birney, Birnholtz, Biscans, Bischi, Biscoveanu,
  Bisht, Bitossi, Bizouard, Blackburn, Blackman, Blair, Blair, Blair, Blanch,
  Bobba, Bode, Boer, Boetzel, Bogaert, Boldrini, Bondu, Bonilla, Bonnand,
  Booker, Boom, Bork, Boschi, Bose, Bossilkov, Boudart, Bouffanais, Bozzi,
  Bradaschia, Brady, Bramley, Branchesi, Brau, Breschi, Briant, Briggs,
  Brighenti, Brillet, Brinkmann, Brockill, Brooks, Brooks, Brown, Brunett,
  Bruno, Bruntz, Buikema, Bulik, Bulten, Buonanno, Buscicchio, Buskulic, Byer,
  Cabero, Cadonati, Caesar, Cagnoli, Cahillane, Calderón~Bustillo, Callaghan,
  Callister, Calloni, Camp, Canepa, Cannon, Cao, Cao, Carapella, Carbognani,
  Carney, Carpinelli, Carullo, Carver, Casanueva~Diaz, Casentini, Caudill,
  Cavaglià, Cavalier, Cavalieri, Cella, Cerdá-Durán, Cesarini, Chaibi,
  Chakravarti, Chan, Chan, Chandra, Chanial, Chao, Charlton, Chase,
  Chassande-Mottin, Chatterjee, Chattopadhyay, Chaturvedi, Chatziioannou, Chen,
  Chen, Chen, Chen, Cheng, Cheong, Chia, Chiadini, Chierici, Chincarini,
  Chiummo, Cho, Cho, Cho, Choate, Christensen, Chu, Chua, Chung, Chung, Ciani,
  Ciecielag, Cieślar, Cifaldi, Ciobanu, Ciolfi, Cipriano, Cirone, Clara,
  Clark, Clark, Clarke, Clearwater, Clesse, Cleva, Coccia, Cohadon, Cohen,
  Colleoni, Collette, Collins, Colpi, Constancio, Conti, Cooper, Corban,
  Corbitt, Cordero-Carrión, Corezzi, Corley, Cornish, Corre, Corsi, Cortese,
  Costa, Cotesta, Coughlin, Coughlin, Coulon, Countryman, Couvares, Covas,
  Coward, Cowart, Coyne, Coyne, Creighton, Creighton, Croquette, Crowder,
  Cudell, Cullen, Cumming, Cummings, Cunningham, Cuoco, Curylo, Dal~Canton,
  Dálya, Dana, DaneshgaranBajastani, D'Angelo, Danilishin, D'Antonio,
  Danzmann, Darsow-Fromm, Dasgupta, Datrier, Dattilo, Dave, Davier, Davies,
  Davis, Daw, Dean, DeBra, Deenadayalan, Degallaix, De~Laurentis, Deléglise,
  Del~Favero, De~Lillo, De~Lillo, Del~Pozzo, DeMarchi, De~Matteis, D'Emilio,
  Demos, Denker, Dent, Depasse, De~Pietri, De~Rosa, De~Rossi, DeSalvo,
  de~Varona, Dhurandhar, Díaz, Diaz-Ortiz, Didio, Dietrich, Di~Fiore,
  DiFronzo, Di~Giorgio, Di~Giovanni, Di~Giovanni, Di~Girolamo, Di~Lieto, Ding,
  Di~Pace, Di~Palma, Di~Renzo, Divakarla, Dmitriev, Doctor, D'Onofrio, Donovan,
  Dooley, Doravari, Dorrington, Downes, Drago, Driggers, Du, Ducoin, Dupej,
  Durante, D'Urso, Duverne, Dwyer, Easter, Eddolls, Edelman, Edo, Edy, Effler,
  Eichholz, Eikenberry, Eisenmann, Eisenstein, Ejlli, Errico, Essick,
  Estellés, Estevez, Etienne, Etzel, Evans, Evans, Ewing, Fafone, Fair,
  Fairhurst, Fan, Farah, Farinon, Farr, Farr, Fauchon-Jones, Favata, Fays,
  Fazio, Feicht, Fejer, Feng, Fenyvesi, Ferguson, Fernandez-Galiana, Ferrante,
  Ferreira, Fidecaro, Figura, Fiori, Fiorucci, Fishbach, Fisher, Fishner,
  Fittipaldi, Fitz-Axen, Fiumara, Flaminio, Floden, Flynn, Fong, Font, Forsyth,
  Fournier, Frasca, Frasconi, Frei, Freise, Frey, Frey, Fritschel, Frolov,
  Fronzé, Fulda, Fyffe, Gabbard, Gadre, Gaebel, Gair, Gais, Galaudage, Gamba,
  Ganapathy, Ganguly, Gaonkar, Garaventa, García-Quirós, Garufi, Gateley,
  Gaudio, Gayathri, Gemme, Gennai, George, George, Gergely, Ghonge, Ghosh,
  Ghosh, Ghosh, Giacomazzo, Giacoppo, Giaime, Giardina, Gibson, Gier, Gill,
  Giri, Glanzer, Gleckl, Godwin, Goetz, Goetz, Gohlke, Goncharov, González,
  Gopakumar, Gossan, Gosselin, Gouaty, Grace, Grado, Granata, Granata, Grant,
  Gras, Grassia, Gray, Gray, Greco, Green, Green, Gretarsson, Griggs, Grignani,
  Grimaldi, Grimes, Grimm, Grote, Grunewald, Gruning, Guerrero, Guidi,
  Guimaraes, Guixé, Gulati, Guo, Gupta, Gupta, Gupta, Gustafson, Gustafson,
  Guzman, Haegel, Halim, Hall, Hamilton, Hammond, Haney, Hanke, Hanks, Hanna,
  Hannuksela, Hannuksela, Hansen, Hansen, Hanson, Harder, Hardwick, Haris,
  Harms, Harry, Harry, Hartwig, Hasskew, Haster, Haughian, Hayes, Healy,
  Heidmann, Heintze, Heinze, Heinzel, Heitmann, Hellman, Hello,
  Helmling-Cornell, Hemming, Hendry, Heng, Hennes, Hennig, Hennig,
  Hernandez~Vivanco, Heurs, Hild, Hill, Hines, Hochheim, Hofgard, Hofman,
  Hohmann, Holgado, Holland, Hollows, Holmes, Holt, Holz, Hopkins, Horst,
  Hough, Howell, Hoy, Hoyland, Huang, Hübner, Huddart, Huerta, Hughey, Hui,
  Husa, Huttner, Hutzler, Huxford, Huynh-Dinh, Idzkowski, Iess, Imperato,
  Inchauspe, Ingram, Intini, Isi, Iyer, JaberianHamedan, Jacqmin, Jadhav,
  Jadhav, James, Jani, Janssens, Janthalur, Jaranowski, Jariwala, Jaume,
  Jenkins, Jeunon, Jiang, Johns, Jones, Jones, Jones, Jones, Jones, Jonker, Ju,
  Junker, Kalaghatgi, Kalogera, Kamai, Kandhasamy, Kang, Kanner, Kapadia,
  Kapasi, Karathanasis, Karki, Kashyap, Kasprzack, Kastaun, Katsanevas,
  Katsavounidis, Katzman, Kawabe, Kéfélian, Keitel, Key, Khadka, Khalili,
  Khan, Khan, Khazanov, Khetan, Khursheed, Kijbunchoo, Kim, Kim, Kim, Kim, Kim,
  Kim, Kimball, King, Kinley-Hanlon, Kirchhoff, Kissel, Kleybolte, Klimenko,
  Knowles, Knyazev, Koch, Koehlenbeck, Koekoek, Koley, Kolstein, Komori,
  Kondrashov, Kontos, Koper, Korobko, Korth, Kovalam, Kozak, Krämer, Kringel,
  Krishnendu, Królak, Kuehn, Kumar, Kumar, Kumar, Kumar, Kuns, Kwang, Lackey,
  Laghi, Lalande, Lam, Lamberts, Landry, Lane, Lang, Lange, Lantz, Lanza,
  La~Rosa, Lartaux-Vollard, Lasky, Laxen, Lazzarini, Lazzaro, Leaci, Leavey,
  Lecoeuche, Lee, Lee, Lee, Lee, Lehmann, Leon, Leroy, Letendre, Levin, Li, Li,
  Li, Li, Li, Linde, Linker, Linley, Littenberg, Liu, Liu, Llorens-Monteagudo,
  Lo, Lockwood, London, Longo, Lorenzini, Loriette, Lormand, Losurdo, Lough,
  Lousto, Lovelace, Lück, Lumaca, Lundgren, Ma, Macas, MacInnis, Macleod,
  MacMillan, Macquet, Magaña~Hernandez, Magaña-Sandoval, Magazzù, Magee,
  Majorana, Maksimovic, Maliakal, Malik, Man, Mandic, Mangano, Mansell, Manske,
  Mantovani, Mapelli, Marchesoni, Marion, Márka, Márka, Markakis, Markosyan,
  Markowitz, Maros, Marquina, Marsat, Martelli, Martin, Martin, Martinez,
  Martinez, Martynov, Masalehdan, Mason, Massera, Masserot, Massinger,
  Masso-Reid, Mastrogiovanni, Matas, Mateu-Lucena, Matichard, Matiushechkina,
  Mavalvala, Maynard, McCann, McCarthy, McClelland, McCormick, McCuller,
  McGuire, McIsaac, McIver, McManus, McRae, McWilliams, Meacher, Meadors,
  Mehmet, Mehta, Melatos, Melchor, Mendell, Menendez-Vazquez, Mercer, Mereni,
  Merfeld, Merilh, Merritt, Merzougui, Meshkov, Messenger, Messick, Metzdorff,
  Meyers, Meylahn, Mhaske, Miani, Miao, Michaloliakos, Michel, Middleton,
  Milano, Miller, Miller, Millhouse, Mills, Milotti, Milovich-Goff, Minazzoli,
  Minenkov, Mir, Mishkin, Mishra, Mistry, Mitra, Mitrofanov, Mitselmakher,
  Mittleman, Mo, Mogushi, Mohapatra, Mohite, Molina, Molina-Ruiz, Mondin,
  Montani, Moore, Moraru, Morawski, Moreno, Morisaki, Mours, Mow-Lowry, Mozzon,
  Muciaccia, Mukherjee, Mukherjee, Mukherjee, Mukherjee, Mukund, Mullavey,
  Munch, Muñiz, Murray, Nadji, Nagar, Nardecchia, Naticchioni, Nayak, Neil,
  Neilson, Nelemans, Nelson, Nery, Neunzert, Ng, Ng, Nguyen, Nguyen, Nguyen,
  Nichols, Nissanke, Nocera, Noh, North, Nothard, Nuttall, Oberling, O'Brien,
  O'Dell, Oganesyan, Ogin, Oh, Oh, Ohme, Ohta, Okada, Olivetto, Oppermann,
  Oram, O'Reilly, Ormiston, Ormsby, Ortega, O'Shaughnessy, Ossokine, Osthelder,
  Ottaway, Overmier, Owen, Pace, Pagano, Page, Pagliaroli, Pai, Pai, Palamos,
  Palashov, Palomba, Pan, Panda, Pang, Pankow, Pannarale, Pant, Paoletti,
  Paoli, Paolone, Parker, Pascucci, Pasqualetti, Passaquieti, Passuello, Patel,
  Patricelli, Payne, Pechsiri, Pedraza, Pegoraro, Pele, Penn, Perego, Perez,
  Périgois, Perreca, Perriès, Petermann, Petterson, Pfeiffer, Pham, Phukon,
  Piccinni, Pichot, Piendibene, Piergiovanni, Pierini, Pierro, Pillant, Pilo,
  Pinard, Pinto, Piotrzkowski, Pirello, Pitkin, Placidi, Plastino, Pluchar,
  Poggiani, Polini, Pong, Ponrathnam, Popolizio, Porter, Poverman, Powell,
  Pracchia, Prajapati, Prasai, Prasanna, Pratten, Prestegard, Principe, Prodi,
  Prokhorov, Prosposito, Puecher, Punturo, Puosi, Puppo, Pürrer, Qi,
  Quetschke, Quinonez, Quitzow-James, Raab, Raaijmakers, Radkins, Radulesco,
  Raffai, Rafferty, Rail, Raja, Rajan, Rajbhandari, Rakhmanov, Ramirez,
  Ramirez, Ramos-Buades, Rana, Rao, Rapagnani, Rapol, Ratto, Raymond, Razzano,
  Read, Regimbau, Rei, Reid, Reitze, Rettegno, Ricci, Richardson, Richardson,
  Richardson, Ricker, Riemenschneider, Riles, Rizzo, Robertson, Robinet,
  Rocchi, Rocha, Rodriguez, Rodriguez-Soto, Rolland, Rollins, Roma, Romanelli,
  Romano, Romel, Romero, Romero-Shaw, Romie, Ronchini, Rose, Rose, Rose,
  Rosell, Rosińska, Rosofsky, Ross, Rowan, Rowlinson, Roy, Roy, Ruggi, Ryan,
  Sachdev, Sadecki, Sakellariadou, Salafia, Salconi, Saleem, Samajdar, Sanchez,
  Sanchez, Sanchez, Sanchis-Gual, Sanders, Santiago, Santos, Saravanan, Sarin,
  Sassolas, Sathyaprakash, Sauter, Savage, Savant, Sawant, Sayah, Schaetzl,
  Schale, Scheel, Scheuer, Schindler-Tyka, Schmidt, Schnabel, Schofield,
  Schönbeck, Schreiber, Schulte, Schutz, Schwarm, Schwartz, Scott, Scott,
  Seglar-Arroyo, Seidel, Sellers, Sengupta, Sennett, Sentenac, Sequino,
  Sergeev, Setyawati, Shaffer, Shahriar, Sharifi, Sharma, Sharma, Shawhan,
  Shen, Shikauchi, Shink, Shoemaker, Shoemaker, Shukla, ShyamSundar,
  Sieniawska, Sigg, Singer, Singh, Singh, Singha, Singhal, Sintes, Sipala,
  Skliris, Slagmolen, Slaven-Blair, Smetana, Smith, Smith, Somala, Son, Soni,
  Sorazu, Sordini, Sorrentino, Sorrentino, Soulard, Souradeep, Sowell, Spencer,
  Spera, Srivastava, Srivastava, Staats, Stachie, Steer, Steinke, Steinlechner,
  Steinlechner, Steinmeyer, Stevenson, Stolle-McAllister, Stops, Stover,
  Strain, Stratta, Strunk, Sturani, Stuver, Südbeck, Sudhagar, Sudhir, Suh,
  Summerscales, Sun, Sun, Sunil, Sur, Suresh, Sutton, Swinkels, Szczepańczyk,
  Tacca, Tait, Talbot, Tanasijczuk, Tanner, Tao, Tapia, Tapia San~Martin,
  Tasson, Taylor, Tenorio, Terkowski, Thirugnanasambandam, Thomas, Thomas,
  Thomas, Thompson, Thondapu, Thorne, Thrane, Tiwari, Tiwari, Tiwari, Toland,
  Tolley, Tonelli, Tornasi, Torres-Forné, Torrie, Tosta~e Melo, Töyrä, Tran,
  Trapananti, Travasso, Traylor, Tringali, Tripathee, Trovato, Trudeau, Tsai,
  Tsang, Tse, Tso, Tsukada, Tsuna, Tsutsui, Turconi, Ubhi, Udall, Ueno,
  Ugolini, Unnikrishnan, Urban, Usman, Utina, Vahlbruch, Vajente, Vajpeyi,
  Valdes, Valentini, Valsan, van Bakel, Beuzekom, van~den Brand, Van
  Den~Broeck, Vander-Hyde, van~der Schaaf, van Heijningen, Vardaro, Vargas,
  Varma, Vass, Vasúth, Vecchio, Vedovato, Veitch, Veitch, Venkateswara,
  Venneberg, Venugopalan, Verkindt, Verma, Veske, Vetrano, Viceré, Viets,
  Villa-Ortega, Vinet, Vitale, Vo, Vocca, Vorvick, Vyatchanin, Wade, Wade,
  Wade, Walet, Walker, Wallace, Wallace, Walsh, Wang, Wang, Wang, Wang, Ward,
  Warner, Was, Washington, Watchi, Weaver, Wei, Weinert, Weinstein, Weiss,
  Wellmann, Wen, Weßels, Westhouse, Wette, Whelan, White, White, Whiting,
  Whittle, Wilken, Williams, Williams, Williamson, Willis, Willke, Wilson,
  Wimmer, Winkler, Wipf, Woan, Woehler, Wofford, Wong, Wrangel, Wright, Wu,
  Wysocki, Xiao, Yamamoto, Yang, Yang, Yang, Yap, Yeeles, Yoon, Yu, Yu, Yuen,
  Zadrożny, Zanolin, Zelenova, Zendri, Zevin, Zhang, Zhang, Zhang, Zhang,
  Zhao, Zhao, Zhou, Zhou, Zhu, Zimmerman, Zucker, Zweizig, {LIGO Scientific
  Collaboration}, \& {Virgo Collaboration}}]{abbott_population_2021}
Abbott, R., Abbott, T.~D., Abraham, S., {et~al.} 2021{\natexlab{b}}, ApJ, 913,
  L7

\bibitem[{{Abubekerov} {et~al.}(2004){Abubekerov}, {Antokhina}, \&
  {Cherepashchuk}}]{2004ARep...48...89A}
{Abubekerov}, M.~K., {Antokhina}, {\'E}.~A., \& {Cherepashchuk}, A.~M. 2004,
  Astronomy Reports, 48, 89

\bibitem[{{An} {et~al.}(2015){An}, {Bellm}, {Bhalerao}, {Boggs}, {Christensen},
  {Craig}, {Fuerst}, {Hailey}, {Harrison}, {Kaspi}, {Natalucci}, {Stern},
  {Tomsick}, \& {Zhang}}]{2015ApJ...806..166A}
{An}, H., {Bellm}, E., {Bhalerao}, V., {et~al.} 2015, \apj, 806, 166

\bibitem[{{Aragona} {et~al.}(2010){Aragona}, {McSwain}, \& {De
  Becker}}]{2010ApJ...724..306A}
{Aragona}, C., {McSwain}, M.~V., \& {De Becker}, M. 2010, \apj, 724, 306

\bibitem[{{Aragona} {et~al.}(2009){Aragona}, {McSwain}, {Grundstrom}, {Marsh},
  {Roettenbacher}, {Hessler}, {Boyajian}, \& {Ray}}]{2009ApJ...698..514A}
{Aragona}, C., {McSwain}, M.~V., {Grundstrom}, E.~D., {et~al.} 2009, \apj, 698,
  514

\bibitem[{{Astropy Collaboration} {et~al.}(2018){Astropy Collaboration},
  {Price-Whelan}, {Sip{\H{o}}cz}, {G{\"u}nther}, {Lim}, {Crawford}, {Conseil},
  {Shupe}, {Craig}, {Dencheva}, {Ginsburg}, {Vand erPlas}, {Bradley},
  {P{\'e}rez-Su{\'a}rez}, {de Val-Borro}, {Aldcroft}, {Cruz}, {Robitaille},
  {Tollerud}, {Ardelean}, {Babej}, {Bach}, {Bachetti}, {Bakanov}, {Bamford},
  {Barentsen}, {Barmby}, {Baumbach}, {Berry}, {Biscani}, {Boquien}, {Bostroem},
  {Bouma}, {Brammer}, {Bray}, {Breytenbach}, {Buddelmeijer}, {Burke},
  {Calderone}, {Cano Rodr{\'\i}guez}, {Cara}, {Cardoso}, {Cheedella}, {Copin},
  {Corrales}, {Crichton}, {D'Avella}, {Deil}, {Depagne}, {Dietrich}, {Donath},
  {Droettboom}, {Earl}, {Erben}, {Fabbro}, {Ferreira}, {Finethy}, {Fox},
  {Garrison}, {Gibbons}, {Goldstein}, {Gommers}, {Greco}, {Greenfield},
  {Groener}, {Grollier}, {Hagen}, {Hirst}, {Homeier}, {Horton}, {Hosseinzadeh},
  {Hu}, {Hunkeler}, {Ivezi{\'c}}, {Jain}, {Jenness}, {Kanarek}, {Kendrew},
  {Kern}, {Kerzendorf}, {Khvalko}, {King}, {Kirkby}, {Kulkarni}, {Kumar},
  {Lee}, {Lenz}, {Littlefair}, {Ma}, {Macleod}, {Mastropietro}, {McCully},
  {Montagnac}, {Morris}, {Mueller}, {Mumford}, {Muna}, {Murphy}, {Nelson},
  {Nguyen}, {Ninan}, {N{\"o}the}, {Ogaz}, {Oh}, {Parejko}, {Parley}, {Pascual},
  {Patil}, {Patil}, {Plunkett}, {Prochaska}, {Rastogi}, {Reddy Janga},
  {Sabater}, {Sakurikar}, {Seifert}, {Sherbert}, {Sherwood-Taylor}, {Shih},
  {Sick}, {Silbiger}, {Singanamalla}, {Singer}, {Sladen}, {Sooley},
  {Sornarajah}, {Streicher}, {Teuben}, {Thomas}, {Tremblay}, {Turner},
  {Terr{\'o}n}, {van Kerkwijk}, {de la Vega}, {Watkins}, {Weaver}, {Whitmore},
  {Woillez}, {Zabalza}, \& {Astropy Contributors}}]{2018AJ....156..123A}
{Astropy Collaboration}, {Price-Whelan}, A.~M., {Sip{\H{o}}cz}, B.~M., {et~al.}
  2018, \aj, 156, 123

\bibitem[{{Astropy Collaboration} {et~al.}(2013){Astropy Collaboration},
  {Robitaille}, {Tollerud}, {Greenfield}, {Droettboom}, {Bray}, {Aldcroft},
  {Davis}, {Ginsburg}, {Price-Whelan}, {Kerzendorf}, {Conley}, {Crighton},
  {Barbary}, {Muna}, {Ferguson}, {Grollier}, {Parikh}, {Nair}, {Unther},
  {Deil}, {Woillez}, {Conseil}, {Kramer}, {Turner}, {Singer}, {Fox}, {Weaver},
  {Zabalza}, {Edwards}, {Azalee Bostroem}, {Burke}, {Casey}, {Crawford},
  {Dencheva}, {Ely}, {Jenness}, {Labrie}, {Lim}, {Pierfederici}, {Pontzen},
  {Ptak}, {Refsdal}, {Servillat}, \& {Streicher}}]{2013A&A...558A..33A}
{Astropy Collaboration}, {Robitaille}, T.~P., {Tollerud}, E.~J., {et~al.} 2013,
  \aap, 558, A33

\bibitem[{Atri {et~al.}(2019)Atri, Miller-Jones, Bahramian, Plotkin, Jonker,
  Nelemans, Maccarone, Sivakoff, Deller, Chaty, Torres, Horiuchi, McCallum,
  Natusch, Phillips, Stevens, \& Weston}]{atri_potential_2019}
Atri, P., Miller-Jones, J. C.~A., Bahramian, A., {et~al.} 2019, MNRAS, 489,
  3116

\bibitem[{Baibhav {et~al.}(2019)Baibhav, Berti, Gerosa, Mapelli, Giacobbo,
  Bouffanais, \& Di~Carlo}]{baibhav_gravitational-wave_2019}
Baibhav, V., Berti, E., Gerosa, D., {et~al.} 2019, Physical Review D, 100,
  064060

\bibitem[{Bailer-Jones {et~al.}(2021)Bailer-Jones, Rybizki, Fouesneau,
  Demleitner, \& Andrae}]{bailer-jones_estimating_2021}
Bailer-Jones, C. A.~L., Rybizki, J., Fouesneau, M., Demleitner, M., \& Andrae,
  R. 2021, AS, 161, 147

\bibitem[{{Baykal} {et~al.}(2007){Baykal}, {Inam}, {Stark}, {Heffner},
  {Erkoca}, \& {Swank}}]{2007MNRAS.374.1108B}
{Baykal}, A., {Inam}, S.~{\c{C}}., {Stark}, M.~J., {et~al.} 2007, \mnras, 374,
  1108

\bibitem[{{Bikmaev} {et~al.}(2017){Bikmaev}, {Nikolaeva}, {Shimansky},
  {Galeev}, {Zhuchkov}, {Irtuganov}, {Melnikov}, {Sakhibullin}, {Grebenev}, \&
  {Sharipova}}]{2017AstL...43..664B}
{Bikmaev}, I.~F., {Nikolaeva}, E.~A., {Shimansky}, V.~V., {et~al.} 2017,
  Astronomy Letters, 43, 664

\bibitem[{Bird {et~al.}(2016)Bird, Bazzano, Malizia, Fiocchi, Sguera, Bassani,
  Hill, Ubertini, \& Winkler}]{bird_ibis_2016}
Bird, A.~J., Bazzano, A., Malizia, A., {et~al.} 2016, ApJS, 223, 15

\bibitem[{{Blaauw}(1961)}]{1961BAN....15..265B}
{Blaauw}, A. 1961, \bain, 15, 265

\bibitem[{{Blay} {et~al.}(2006){Blay}, {Negueruela}, {Reig}, {Coe}, {Corbet},
  {Fabregat}, \& {Tarasov}}]{2006A&A...446.1095B}
{Blay}, P., {Negueruela}, I., {Reig}, P., {et~al.} 2006, \aap, 446, 1095

\bibitem[{{Bonnet-Bidaud} \& {Mouchet}(1998)}]{1998A&A...332L...9B}
{Bonnet-Bidaud}, J.~M. \& {Mouchet}, M. 1998, \aap, 332, L9

\bibitem[{{Bovy}(2015)}]{2015ApJS..216...29B}
{Bovy}, J. 2015, \apjs, 216, 29

\bibitem[{{Brandt} \& {Podsiadlowski}(1995)}]{1995MNRAS.274..461B}
{Brandt}, N. \& {Podsiadlowski}, P. 1995, \mnras, 274, 461

\bibitem[{Brown {et~al.}(2021)Brown, Vallenari, Prusti, Bruijne, Babusiaux,
  Biermann, Creevey, Evans, Eyer, Hutton, Jansen, Jordi, Klioner, Lammers,
  Lindegren, Luri, Mignard, Panem, Pourbaix, Randich, Sartoretti, Soubiran,
  Walton, Arenou, Bailer-Jones, Bastian, Cropper, Drimmel, Katz, Lattanzi,
  Leeuwen, Bakker, Cacciari, Castañeda, Angeli, Ducourant, Fabricius,
  Fouesneau, Frémat, Guerra, Guerrier, Guiraud, Piccolo, Masana, Messineo,
  Mowlavi, Nicolas, Nienartowicz, Pailler, Panuzzo, Riclet, Roux, Seabroke,
  Sordo, Tanga, Thévenin, Gracia-Abril, Portell, Teyssier, Altmann, Andrae,
  Bellas-Velidis, Benson, Berthier, Blomme, Brugaletta, Burgess, Busso, Carry,
  Cellino, Cheek, Clementini, Damerdji, Davidson, Delchambre, Dell’Oro,
  Fernández-Hernández, Galluccio, García-Lario, Garcia-Reinaldos,
  González-Núñez, Gosset, Haigron, Halbwachs, Hambly, Harrison,
  Hatzidimitriou, Heiter, Hernández, Hestroffer, Hodgkin, Holl, Janßen,
  Fombelle, Jordan, Krone-Martins, Lanzafame, Löffler, Lorca, Manteiga,
  Marchal, Marrese, Moitinho, Mora, Muinonen, Osborne, Pancino, Pauwels, Petit,
  Recio-Blanco, Richards, Riello, Rimoldini, Robin, Roegiers, Rybizki, Sarro,
  Siopis, Smith, Sozzetti, Ulla, Utrilla, Leeuwen, Reeven, Abbas, Aramburu,
  Accart, Aerts, Aguado, Ajaj, Altavilla, Álvarez, Cid-Fuentes, Alves,
  Anderson, Varela, Antoja, Audard, Baines, Baker, Balaguer-Núñez, Balbinot,
  Balog, Barache, Barbato, Barros, Barstow, Bartolomé, Bassilana, Bauchet,
  Baudesson-Stella, Becciani, Bellazzini, Bernet, Bertone, Bianchi,
  Blanco-Cuaresma, Boch, Bombrun, Bossini, Bouquillon, Bragaglia, Bramante,
  Breedt, Bressan, Brouillet, Bucciarelli, Burlacu, Busonero, Butkevich, Buzzi,
  Caffau, Cancelliere, Cánovas, Cantat-Gaudin, Carballo, Carlucci, Carnerero,
  Carrasco, Casamiquela, Castellani, Castro-Ginard, Sampol, Chaoul, Charlot,
  Chemin, Chiavassa, Cioni, Comoretto, Cooper, Cornez, Cowell, Crifo, Crosta,
  Crowley, Dafonte, Dapergolas, David, David, Laverny, Luise, March, Ridder,
  Souza, Teodoro, Torres, Peloso, Pozo, Delbo, Delgado, Delgado, Delisle,
  Matteo, Diakite, Diener, Distefano, Dolding, Eappachen, Edvardsson, Enke,
  Esquej, Fabre, Fabrizio, Faigler, Fedorets, Fernique, Fienga, Figueras,
  Fouron, Fragkoudi, Fraile, Franke, Gai, Garabato, Garcia-Gutierrez,
  García-Torres, Garofalo, Gavras, Gerlach, Geyer, Giacobbe, Gilmore, Girona,
  Giuffrida, Gomel, Gomez, Gonzalez-Santamaria, González-Vidal, Granvik,
  Gutiérrez-Sánchez, Guy, Hauser, Haywood, Helmi, Hidalgo, Hilger, Hładczuk,
  Hobbs, Holland, Huckle, Jasniewicz, Jonker, Campillo, Julbe, Karbevska,
  Kervella, Khanna, Kochoska, Kontizas, Kordopatis, Korn, Kostrzewa-Rutkowska,
  Kruszyńska, Lambert, Lanza, Lasne, Campion, Fustec, Lebreton, Lebzelter,
  Leccia, Leclerc, Lecoeur-Taibi, Liao, Licata, Lindstrøm, Lister, Livanou,
  Lobel, Pardo, Managau, Mann, Marchant, Marconi, Santos, Marinoni, Marocco,
  Marshall, Polo, Martín-Fleitas, Masip, Massari, Mastrobuono-Battisti, Mazeh,
  McMillan, Messina, Michalik, Millar, Mints, Molina, Molinaro, Molnár,
  Montegriffo, Mor, Morbidelli, Morel, Morris, Mulone, Munoz, Muraveva, Murphy,
  Musella, Noval, Ordénovic, Orrù, Osinde, Pagani, Pagano, Palaversa,
  Palicio, Panahi, Pawlak, Esteller, Penttilä, Piersimoni, Pineau, Plachy,
  Plum, Poggio, Poretti, Poujoulet, Prša, Pulone, Racero, Ragaini, Rainer,
  Raiteri, Rambaux, Ramos, Ramos-Lerate, Fiorentin, Regibo, Reylé, Ripepi,
  Riva, Rixon, Robichon, Robin, Roelens, Rohrbasser, Romero-Gómez, Rowell,
  Royer, Rybicki, Sadowski, Sellés, Sahlmann, Salgado, Salguero, Samaras,
  Gimenez, Sanna, Santoveña, Sarasso, Schultheis, Sciacca, Segol, Segovia,
  Ségransan, Semeux, Shahaf, Siddiqui, Siebert, Siltala, Slezak, Smart,
  Solano, Solitro, Souami, Souchay, Spagna, Spoto, Steele, Steidelmüller,
  Stephenson, Süveges, Szabados, Szegedi-Elek, Taris, Tauran, Taylor,
  Teixeira, Thuillot, Tonello, Torra, Torra, Turon, Unger, Vaillant, Dillen,
  Vanel, Vecchiato, Viala, Vicente, Voutsinas, Weiler, Wevers, Wyrzykowski,
  Yoldas, Yvard, Zhao, Zorec, Zucker, Zurbach, \& Zwitter}]{brown_gaia_2021}
Brown, A. G.~A., Vallenari, A., Prusti, T., {et~al.} 2021, A\&A, 649, A1

\bibitem[{{Casares} {et~al.}(2011){Casares}, {Corral-Santana}, {Herrero},
  {Morales}, {Mu{\~n}oz-Darias}, {Negueruela}, {Paredes}, {Ribas}, {Rib{\'o}},
  {Steeghs}, {van Spaandonk}, \& {Vilardell}}]{2011ASSP...21..559C}
{Casares}, J., {Corral-Santana}, J.~M., {Herrero}, A., {et~al.} 2011, Ap\&SS
  Proceedings, 21, 599

\bibitem[{Casares {et~al.}(2014)Casares, Negueruela, Ribó, Ribas, Paredes,
  Herrero, \& Simón-Díaz}]{casares_be-type_2014}
Casares, J., Negueruela, I., Ribó, M., {et~al.} 2014, Nature, 505, 378

\bibitem[{{Casares} {et~al.}(2005{\natexlab{a}}){Casares}, {Ribas}, {Paredes},
  {Mart{\'\i}}, \& {Allende Prieto}}]{2005MNRAS.360.1105C}
{Casares}, J., {Ribas}, I., {Paredes}, J.~M., {Mart{\'\i}}, J., \& {Allende
  Prieto}, C. 2005{\natexlab{a}}, \mnras, 360, 1105

\bibitem[{{Casares} {et~al.}(2005{\natexlab{b}}){Casares}, {Rib{\'o}}, {Ribas},
  {Paredes}, {Mart{\'\i}}, \& {Herrero}}]{2005MNRAS.364..899C}
{Casares}, J., {Rib{\'o}}, M., {Ribas}, I., {et~al.} 2005{\natexlab{b}},
  \mnras, 364, 899

\bibitem[{Chaty(2013)}]{chaty_optical/infrared_2013}
Chaty, S. 2013, Advances in Space Research, 52, 2132

\bibitem[{{Chen} {et~al.}(2001){Chen}, {Stoughton}, {Smith}, {Uomoto}, {Pier},
  {Yanny}, {Ivezi{\'c}}, {York}, {Anderson}, {Annis}, {Brinkmann}, {Csabai},
  {Fukugita}, {Hindsley}, {Lupton}, {Munn}, \& {SDSS
  Collaboration}}]{2001ApJ...553..184C}
{Chen}, B., {Stoughton}, C., {Smith}, J.~A., {et~al.} 2001, \apj, 553, 184

\bibitem[{{Coe} {et~al.}(1988){Coe}, {Longmore}, {Payne}, \&
  {Hanson}}]{1988MNRAS.232..865C}
{Coe}, M.~J., {Longmore}, A., {Payne}, B.~J., \& {Hanson}, C.~G. 1988, \mnras,
  232, 865

\bibitem[{{Coe} {et~al.}(1994){Coe}, {Roche}, {Everall}, {Fabregat}, {Buckley},
  {Smith}, {Reynolds}, {Jupp}, \& {MacGillivray}}]{1994MNRAS.270L..57C}
{Coe}, M.~J., {Roche}, P., {Everall}, C., {et~al.} 1994, \mnras, 270, L57

\bibitem[{{Coleiro} \& {Chaty}(2013)}]{2013ApJ...764..185C}
{Coleiro}, A. \& {Chaty}, S. 2013, \apj, 764, 185

\bibitem[{{Coleiro} {et~al.}(2013){Coleiro}, {Chaty}, {Zurita Heras}, {Rahoui},
  \& {Tomsick}}]{2013A&A...560A.108C}
{Coleiro}, A., {Chaty}, S., {Zurita Heras}, J.~A., {Rahoui}, F., \& {Tomsick},
  J.~A. 2013, \aap, 560, A108

\bibitem[{{Corbet} \& {Krimm}(2010)}]{2010ATel.3079....1C}
{Corbet}, R.~H.~D. \& {Krimm}, H.~A. 2010, ATel, 3079, 1

\bibitem[{Corral-Santana {et~al.}(2016)Corral-Santana, Casares, Muñoz-Darias,
  Bauer, Martínez-Pais, \& Russell}]{corral-santana_blackcat_2016}
Corral-Santana, J.~M., Casares, J., Muñoz-Darias, T., {et~al.} 2016, \aap,
  587, A61

\bibitem[{{Crampton} {et~al.}(1985){Crampton}, {Hutchings}, \&
  {Cowley}}]{1985ApJ...299..839C}
{Crampton}, D., {Hutchings}, J.~B., \& {Cowley}, A.~P. 1985, \apj, 299, 839

\bibitem[{{Delgado-Mart{\'\i}} {et~al.}(2001){Delgado-Mart{\'\i}}, {Levine},
  {Pfahl}, \& {Rappaport}}]{2001ApJ...546..455D}
{Delgado-Mart{\'\i}}, H., {Levine}, A.~M., {Pfahl}, E., \& {Rappaport}, S.~A.
  2001, \apj, 546, 455

\bibitem[{{Densham} \& {Charles}(1982)}]{1982MNRAS.201..171D}
{Densham}, R.~H. \& {Charles}, P.~A. 1982, \mnras, 201, 171

\bibitem[{Dosopoulou \& Kalogera(2016)}]{2016ApJ...825...70D}
Dosopoulou, F. \& Kalogera, V. 2016, ApJ, 825, 70

\bibitem[{Evans {et~al.}(2019)Evans, Allen, Anderson, Budynkiewicz, Burke,
  Chen, Civano, D'Abrusco, Doe, Evans, Fabbiano, Gibbs, Glotfelty, Graessle,
  Grier, Hain, Hall, Harbo, Houck, Lauer, Laurino, Lee, Martínez-Galarza,
  McCollough, McDowell, Miller, McLaughlin, Morgan, Mossman, Nguyen, Nichols,
  Nowak, Paxson, Plummer, Primini, Rots, Siemiginowska, Sundheim, Tibbetts,
  Van~Stone, \& Zografou}]{evans_chandra_2019}
Evans, I.~N., Allen, C., Anderson, C.~S., {et~al.} 2019, American Astronomical
  Society Meeting Abstracts \#233, 233, 379.01

\bibitem[{Evans {et~al.}(2014)Evans, Osborne, Beardmore, Page, Willingale,
  Mountford, Pagani, Burrows, Kennea, Perri, Tagliaferri, \&
  Gehrels}]{evans_1SXPS_2014}
Evans, P.~A., Osborne, J.~P., Beardmore, A.~P., {et~al.} 2014, ApJS, 210, 8

\bibitem[{{Falanga} {et~al.}(2015){Falanga}, {Bozzo}, {Lutovinov},
  {Bonnet-Bidaud}, {Fetisova}, \& {Puls}}]{2015A&A...577A.130F}
{Falanga}, M., {Bozzo}, E., {Lutovinov}, A., {et~al.} 2015, \aap, 577, A130

\bibitem[{{Ferrigno} {et~al.}(2013){Ferrigno}, {Farinelli}, {Bozzo},
  {Pottschmidt}, {Klochkov}, \& {Kretschmar}}]{2013A&A...553A.103F}
{Ferrigno}, C., {Farinelli}, R., {Bozzo}, E., {et~al.} 2013, \aap, 553, A103

\bibitem[{{Finger} {et~al.}(1996){Finger}, {Wilson}, \&
  {Harmon}}]{1996ApJ...459..288F}
{Finger}, M.~H., {Wilson}, R.~B., \& {Harmon}, B.~A. 1996, \apj, 459, 288

\bibitem[{{Foreman-Mackey} {et~al.}(2013){Foreman-Mackey}, {Hogg}, {Lang}, \&
  {Goodman}}]{2013PASP..125..306F}
{Foreman-Mackey}, D., {Hogg}, D.~W., {Lang}, D., \& {Goodman}, J. 2013, \pasp,
  125, 306

\bibitem[{Fortin {et~al.}(2018)Fortin, Chaty, Coleiro, Tomsick, \&
  Nitschelm}]{fortin_spectroscopic_2018}
Fortin, F., Chaty, S., Coleiro, A., Tomsick, J.~A., \& Nitschelm, C. H.~R.
  2018, \aap, 618, A150

\bibitem[{Fryer \& Kusenko(2006)}]{fryer_effects_2006}
Fryer, C.~L. \& Kusenko, A. 2006, ApJS, 163, 335

\bibitem[{{Gaia Collaboration} {et~al.}(2016){Gaia Collaboration}, {Prusti},
  {de Bruijne}, {Brown}, {Vallenari}, {Babusiaux}, {Bailer-Jones}, {Bastian},
  {Biermann}, {Evans}, {Eyer}, {Jansen}, {Jordi}, {Klioner}, {Lammers},
  {Lindegren}, {Luri}, {Mignard}, {Milligan}, {Panem}, {Poinsignon},
  {Pourbaix}, {Randich}, {Sarri}, {Sartoretti}, {Siddiqui}, {Soubiran},
  {Valette}, {van Leeuwen}, {Walton}, {Aerts}, {Arenou}, {Cropper}, {Drimmel},
  {H{\o}g}, {Katz}, {Lattanzi}, {O'Mullane}, {Grebel}, {Holland}, {Huc},
  {Passot}, {Bramante}, {Cacciari}, {Casta{\~n}eda}, {Chaoul}, {Cheek}, {De
  Angeli}, {Fabricius}, {Guerra}, {Hern{\'a}ndez}, {Jean-Antoine-Piccolo},
  {Masana}, {Messineo}, {Mowlavi}, {Nienartowicz}, {Ord{\'o}{\~n}ez-Blanco},
  {Panuzzo}, {Portell}, {Richards}, {Riello}, {Seabroke}, {Tanga},
  {Th{\'e}venin}, {Torra}, {Els}, {Gracia-Abril}, {Comoretto},
  {Garcia-Reinaldos}, {Lock}, {Mercier}, {Altmann}, {Andrae}, {Astraatmadja},
  {Bellas-Velidis}, {Benson}, {Berthier}, {Blomme}, {Busso}, {Carry},
  {Cellino}, {Clementini}, {Cowell}, {Creevey}, {Cuypers}, {Davidson}, {De
  Ridder}, {de Torres}, {Delchambre}, {Dell'Oro}, {Ducourant}, {Fr{\'e}mat},
  {Garc{\'\i}a-Torres}, {Gosset}, {Halbwachs}, {Hambly}, {Harrison}, {Hauser},
  {Hestroffer}, {Hodgkin}, {Huckle}, {Hutton}, {Jasniewicz}, {Jordan},
  {Kontizas}, {Korn}, {Lanzafame}, {Manteiga}, {Moitinho}, {Muinonen},
  {Osinde}, {Pancino}, {Pauwels}, {Petit}, {Recio-Blanco}, {Robin}, {Sarro},
  {Siopis}, {Smith}, {Smith}, {Sozzetti}, {Thuillot}, {van Reeven}, {Viala},
  {Abbas}, {Abreu Aramburu}, {Accart}, {Aguado}, {Allan}, {Allasia},
  {Altavilla}, {{\'A}lvarez}, {Alves}, {Anderson}, {Andrei}, {Anglada Varela},
  {Antiche}, {Antoja}, {Ant{\'o}n}, {Arcay}, {Atzei}, {Ayache}, {Bach},
  {Baker}, {Balaguer-N{\'u}{\~n}ez}, {Barache}, {Barata}, {Barbier}, {Barblan},
  {Baroni}, {Barrado y Navascu{\'e}s}, {Barros}, {Barstow}, {Becciani},
  {Bellazzini}, {Bellei}, {Bello Garc{\'\i}a}, {Belokurov}, {Bendjoya},
  {Berihuete}, {Bianchi}, {Bienaym{\'e}}, {Billebaud}, {Blagorodnova},
  {Blanco-Cuaresma}, {Boch}, {Bombrun}, {Borrachero}, {Bouquillon}, {Bourda},
  {Bouy}, {Bragaglia}, {Breddels}, {Brouillet}, {Br{\"u}semeister},
  {Bucciarelli}, {Budnik}, {Burgess}, {Burgon}, {Burlacu}, {Busonero}, {Buzzi},
  {Caffau}, {Cambras}, {Campbell}, {Cancelliere}, {Cantat-Gaudin}, {Carlucci},
  {Carrasco}, {Castellani}, {Charlot}, {Charnas}, {Charvet}, {Chassat},
  {Chiavassa}, {Clotet}, {Cocozza}, {Collins}, {Collins}, {Costigan}, {Crifo},
  {Cross}, {Crosta}, {Crowley}, {Dafonte}, {Damerdji}, {Dapergolas}, {David},
  {David}, {De Cat}, {de Felice}, {de Laverny}, {De Luise}, {De March}, {de
  Martino}, {de Souza}, {Debosscher}, {del Pozo}, {Delbo}, {Delgado},
  {Delgado}, {di Marco}, {Di Matteo}, {Diakite}, {Distefano}, {Dolding}, {Dos
  Anjos}, {Drazinos}, {Dur{\'a}n}, {Dzigan}, {Ecale}, {Edvardsson}, {Enke},
  {Erdmann}, {Escolar}, {Espina}, {Evans}, {Eynard Bontemps}, {Fabre},
  {Fabrizio}, {Faigler}, {Falc{\~a}o}, {Farr{\`a}s Casas}, {Faye}, {Federici},
  {Fedorets}, {Fern{\'a}ndez-Hern{\'a}ndez}, {Fernique}, {Fienga}, {Figueras},
  {Filippi}, {Findeisen}, {Fonti}, {Fouesneau}, {Fraile}, {Fraser}, {Fuchs},
  {Furnell}, {Gai}, {Galleti}, {Galluccio}, {Garabato}, {Garc{\'\i}a-Sedano},
  {Gar{\'e}}, {Garofalo}, {Garralda}, {Gavras}, {Gerssen}, {Geyer}, {Gilmore},
  {Girona}, {Giuffrida}, {Gomes}, {Gonz{\'a}lez-Marcos},
  {Gonz{\'a}lez-N{\'u}{\~n}ez}, {Gonz{\'a}lez-Vidal}, {Granvik}, {Guerrier},
  {Guillout}, {Guiraud}, {G{\'u}rpide}, {Guti{\'e}rrez-S{\'a}nchez}, {Guy},
  {Haigron}, {Hatzidimitriou}, {Haywood}, {Heiter}, {Helmi}, {Hobbs},
  {Hofmann}, {Holl}, {Holland}, {Hunt}, {Hypki}, {Icardi}, {Irwin}, {Jevardat
  de Fombelle}, {Jofr{\'e}}, {Jonker}, {Jorissen}, {Julbe}, {Karampelas},
  {Kochoska}, {Kohley}, {Kolenberg}, {Kontizas}, {Koposov}, {Kordopatis},
  {Koubsky}, {Kowalczyk}, {Krone-Martins}, {Kudryashova}, {Kull}, {Bachchan},
  {Lacoste-Seris}, {Lanza}, {Lavigne}, {Le Poncin-Lafitte}, {Lebreton},
  {Lebzelter}, {Leccia}, {Leclerc}, {Lecoeur-Taibi}, {Lemaitre}, {Lenhardt},
  {Leroux}, {Liao}, {Licata}, {Lindstr{\o}m}, {Lister}, {Livanou}, {Lobel},
  {L{\"o}ffler}, {L{\'o}pez}, {Lopez-Lozano}, {Lorenz}, {Loureiro},
  {MacDonald}, {Magalh{\~a}es Fernandes}, {Managau}, {Mann}, {Mantelet},
  {Marchal}, {Marchant}, {Marconi}, {Marie}, {Marinoni}, {Marrese},
  {Marschalk{\'o}}, {Marshall}, {Mart{\'\i}n-Fleitas}, {Martino}, {Mary},
  {Matijevi{\v{c}}}, {Mazeh}, {McMillan}, {Messina}, {Mestre}, {Michalik},
  {Millar}, {Miranda}, {Molina}, {Molinaro}, {Molinaro}, {Moln{\'a}r},
  {Moniez}, {Montegriffo}, {Monteiro}, {Mor}, {Mora}, {Morbidelli}, {Morel},
  {Morgenthaler}, {Morley}, {Morris}, {Mulone}, {Muraveva}, {Musella},
  {Narbonne}, {Nelemans}, {Nicastro}, {Noval}, {Ord{\'e}novic},
  {Ordieres-Mer{\'e}}, {Osborne}, {Pagani}, {Pagano}, {Pailler}, {Palacin},
  {Palaversa}, {Parsons}, {Paulsen}, {Pecoraro}, {Pedrosa}, {Pentik{\"a}inen},
  {Pereira}, {Pichon}, {Piersimoni}, {Pineau}, {Plachy}, {Plum}, {Poujoulet},
  {Pr{\v{s}}a}, {Pulone}, {Ragaini}, {Rago}, {Rambaux}, {Ramos-Lerate},
  {Ranalli}, {Rauw}, {Read}, {Regibo}, {Renk}, {Reyl{\'e}}, {Ribeiro},
  {Rimoldini}, {Ripepi}, {Riva}, {Rixon}, {Roelens}, {Romero-G{\'o}mez},
  {Rowell}, {Royer}, {Rudolph}, {Ruiz-Dern}, {Sadowski}, {Sagrist{\`a}
  Sell{\'e}s}, {Sahlmann}, {Salgado}, {Salguero}, {Sarasso}, {Savietto},
  {Schnorhk}, {Schultheis}, {Sciacca}, {Segol}, {Segovia}, {Segransan},
  {Serpell}, {Shih}, {Smareglia}, {Smart}, {Smith}, {Solano}, {Solitro},
  {Sordo}, {Soria Nieto}, {Souchay}, {Spagna}, {Spoto}, {Stampa}, {Steele},
  {Steidelm{\"u}ller}, {Stephenson}, {Stoev}, {Suess}, {S{\"u}veges}, {Surdej},
  {Szabados}, {Szegedi-Elek}, {Tapiador}, {Taris}, {Tauran}, {Taylor},
  {Teixeira}, {Terrett}, {Tingley}, {Trager}, {Turon}, {Ulla}, {Utrilla},
  {Valentini}, {van Elteren}, {Van Hemelryck}, {van Leeuwen}, {Varadi},
  {Vecchiato}, {Veljanoski}, {Via}, {Vicente}, {Vogt}, {Voss}, {Votruba},
  {Voutsinas}, {Walmsley}, {Weiler}, {Weingrill}, {Werner}, {Wevers},
  {Whitehead}, {Wyrzykowski}, {Yoldas}, {{\v{Z}}erjal}, {Zucker}, {Zurbach},
  {Zwitter}, {Alecu}, {Allen}, {Allende Prieto}, {Amorim},
  {Anglada-Escud{\'e}}, {Arsenijevic}, {Azaz}, {Balm}, {Beck}, {Bernstein},
  {Bigot}, {Bijaoui}, {Blasco}, {Bonfigli}, {Bono}, {Boudreault}, {Bressan},
  {Brown}, {Brunet}, {Bunclark}, {Buonanno}, {Butkevich}, {Carret}, {Carrion},
  {Chemin}, {Ch{\'e}reau}, {Corcione}, {Darmigny}, {de Boer}, {de Teodoro}, {de
  Zeeuw}, {Delle Luche}, {Domingues}, {Dubath}, {Fodor}, {Fr{\'e}zouls},
  {Fries}, {Fustes}, {Fyfe}, {Gallardo}, {Gallegos}, {Gardiol}, {Gebran},
  {Gomboc}, {G{\'o}mez}, {Grux}, {Gueguen}, {Heyrovsky}, {Hoar}, {Iannicola},
  {Isasi Parache}, {Janotto}, {Joliet}, {Jonckheere}, {Keil}, {Kim},
  {Klagyivik}, {Klar}, {Knude}, {Kochukhov}, {Kolka}, {Kos}, {Kutka}, {Lainey},
  {LeBouquin}, {Liu}, {Loreggia}, {Makarov}, {Marseille}, {Martayan},
  {Martinez-Rubi}, {Massart}, {Meynadier}, {Mignot}, {Munari}, {Nguyen},
  {Nordlander}, {Ocvirk}, {O'Flaherty}, {Olias Sanz}, {Ortiz}, {Osorio},
  {Oszkiewicz}, {Ouzounis}, {Palmer}, {Park}, {Pasquato}, {Peltzer}, {Peralta},
  {P{\'e}turaud}, {Pieniluoma}, {Pigozzi}, {Poels}, {Prat}, {Prod'homme},
  {Raison}, {Rebordao}, {Risquez}, {Rocca-Volmerange}, {Rosen}, {Ruiz-Fuertes},
  {Russo}, {Sembay}, {Serraller Vizcaino}, {Short}, {Siebert}, {Silva},
  {Sinachopoulos}, {Slezak}, {Soffel}, {Sosnowska}, {Strai{\v{z}}ys}, {ter
  Linden}, {Terrell}, {Theil}, {Tiede}, {Troisi}, {Tsalmantza}, {Tur},
  {Vaccari}, {Vachier}, {Valles}, {Van Hamme}, {Veltz}, {Virtanen}, {Wallut},
  {Wichmann}, {Wilkinson}, {Ziaeepour}, \&
  {Zschocke}}]{gaia_collaboration_mission_2016}
{Gaia Collaboration}, {Prusti}, T., {de Bruijne}, J.~H.~J., {et~al.} 2016,
  \aap, 595, A1

\bibitem[{{Gamen} {et~al.}(2015){Gamen}, {Barb{\`a}}, {Walborn}, {Morrell},
  {Arias}, {Ma{\'\i}z Apell{\'a}niz}, {Sota}, \&
  {Alfaro}}]{2015A&A...583L...4G}
{Gamen}, R., {Barb{\`a}}, R.~H., {Walborn}, N.~R., {et~al.} 2015, \aap, 583, L4

\bibitem[{García {et~al.}(2021)García, Simaz~Bunzel, Chaty, Porter, \&
  Chassande-Mottin}]{garcia_progenitors_2021}
García, F., Simaz~Bunzel, A., Chaty, S., Porter, E., \& Chassande-Mottin, E.
  2021, A\&A, 649, A114

\bibitem[{{Gies} \& {Bolton}(1986)}]{1986ApJS...61..419G}
{Gies}, D.~R. \& {Bolton}, C.~T. 1986, \apjs, 61, 419

\bibitem[{{Gregory}(2002)}]{2002ApJ...575..427G}
{Gregory}, P.~C. 2002, \apj, 575, 427

\bibitem[{Grimm {et~al.}(2002)Grimm, Gilfanov, \& Sunyaev}]{grimm_milky_2002}
Grimm, H.-J., Gilfanov, M., \& Sunyaev, R. 2002, \aap, 391, 923

\bibitem[{{Grundstrom} {et~al.}(2007){Grundstrom}, {Boyajian}, {Finch}, {Gies},
  {Huang}, {McSwain}, {O'Brien}, {Riddle}, {Trippe}, {Williams}, {Wingert}, \&
  {Zaballa}}]{2007ApJ...660.1398G}
{Grundstrom}, E.~D., {Boyajian}, T.~S., {Finch}, C., {et~al.} 2007, \apj, 660,
  1398

\bibitem[{{Grunhut} {et~al.}(2014){Grunhut}, {Bolton}, \&
  {McSwain}}]{2014A&A...563A...1G}
{Grunhut}, J.~H., {Bolton}, C.~T., \& {McSwain}, M.~V. 2014, \aap, 563, A1

\bibitem[{Hansen \& Phinney(1997)}]{hansen_pulsar_1997}
Hansen, B. M.~S. \& Phinney, E.~S. 1997, MNRAS, 291, 569

\bibitem[{{Hobbs} {et~al.}(2005){Hobbs}, {Lorimer}, {Lyne}, \&
  {Kramer}}]{2005MNRAS.360..974H}
{Hobbs}, G., {Lorimer}, D.~R., {Lyne}, A.~G., \& {Kramer}, M. 2005, \mnras,
  360, 974

\bibitem[{{Houk}(1978)}]{1978mcts.book.....H}
{Houk}, N. 1978, {Michigan catalogue of two-dimensional spectral types for the
  HD stars}

\bibitem[{{Hu} {et~al.}(2017){Hu}, {Chou}, {Ng}, {Lin}, \&
  {Yen}}]{2017ApJ...844...16H}
{Hu}, C.-P., {Chou}, Y., {Ng}, C.~Y., {Lin}, L. C.-C., \& {Yen}, D. C.-C. 2017,
  \apj, 844, 16

\bibitem[{{Hunter}(2007)}]{2007CSE.....9...90H}
{Hunter}, J.~D. 2007, Computing in Science and Engineering, 9, 90

\bibitem[{{Hutchings}(1984)}]{1984PASP...96..312H}
{Hutchings}, J.~B. 1984, \pasp, 96, 312

\bibitem[{{Hutchings} {et~al.}(1981){Hutchings}, {Cowley}, {Crampton}, \&
  {Williams}}]{1981PASP...93..741H}
{Hutchings}, J.~B., {Cowley}, A.~P., {Crampton}, D., \& {Williams}, G. 1981,
  \pasp, 93, 741

\bibitem[{{Hutchings} {et~al.}(1987){Hutchings}, {Crampton}, {Cowley}, \&
  {Thompson}}]{1987PASP...99..420H}
{Hutchings}, J.~B., {Crampton}, D., {Cowley}, A.~P., \& {Thompson}, I.~B. 1987,
  \pasp, 99, 420

\bibitem[{{Hutchings} {et~al.}(1982){Hutchings}, {Crampton}, {Cowley},
  {Cowley}, \& {Bord}}]{1982PASP...94..541H}
{Hutchings}, J.~B., {Crampton}, D., {Cowley}, D., {Cowley}, A.~P., \& {Bord},
  D.~J. 1982, \pasp, 94, 541

\bibitem[{{Hynes} {et~al.}(2002){Hynes}, {Clark}, {Barsukova}, {Callanan},
  {Charles}, {Collier Cameron}, {Fabrika}, {Garcia}, {Haswell}, {Horne},
  {Miroshnichenko}, {Negueruela}, {Reig}, {Welsh}, \&
  {Witherick}}]{2002A&A...392..991H}
{Hynes}, R.~I., {Clark}, J.~S., {Barsukova}, E.~A., {et~al.} 2002, \aap, 392,
  991

\bibitem[{Igoshev(2020)}]{igoshev_observed_2020}
Igoshev, A.~P. 2020, MNRAS, 494, 3663

\bibitem[{{in't Zand} {et~al.}(2000){in't Zand}, {Halpern}, {Eracleous},
  {McCollough}, {Augusteijn}, {Remillard}, \& {Heise}}]{2000A&A...361...85I}
{in't Zand}, J.~J.~M., {Halpern}, J., {Eracleous}, M., {et~al.} 2000, \aap,
  361, 85

\bibitem[{{in't Zand} {et~al.}(2007){in't Zand}, {Kuiper}, {den Hartog},
  {Hermsen}, \& {Corbet}}]{2007A&A...469.1063I}
{in't Zand}, J.~J.~M., {Kuiper}, L., {den Hartog}, P.~R., {Hermsen}, W., \&
  {Corbet}, R.~H.~D. 2007, \aap, 469, 1063

\bibitem[{{Islam} \& {Paul}(2016)}]{2016MNRAS.461..816I}
{Islam}, N. \& {Paul}, B. 2016, \mnras, 461, 816

\bibitem[{Ivanova {et~al.}(2013)Ivanova, Justham, Chen, De~Marco, Fryer,
  Gaburov, Ge, Glebbeek, Han, Li, Lu, Marsh, Podsiadlowski, Potter, Soker,
  Taam, Tauris, van~den Heuvel, \& Webbink}]{ivanova_common_2013}
Ivanova, N., Justham, S., Chen, X., {et~al.} 2013, A\&AR, 21, 59

\bibitem[{{Janka}(2013)}]{2013MNRAS.434.1355J}
{Janka}, H.-T. 2013, \mnras, 434, 1355

\bibitem[{{Janot-Pacheco} {et~al.}(1981){Janot-Pacheco}, {Ilovaisky}, \&
  {Chevalier}}]{1981A&A....99..274J}
{Janot-Pacheco}, E., {Ilovaisky}, S.~A., \& {Chevalier}, C. 1981, \aap, 99, 274

\bibitem[{{Jaschek} \& {Egret}(1982)}]{1982IAUS...98..261J}
{Jaschek}, M. \& {Egret}, D. 1982, in Be Stars, ed. M.~{Jaschek} \& H.~G.
  {Groth}, Vol.~98, 261

\bibitem[{{Johnston} {et~al.}(1994){Johnston}, {Manchester}, {Lyne},
  {Nicastro}, \& {Spyromilio}}]{1994MNRAS.268..430J}
{Johnston}, S., {Manchester}, R.~N., {Lyne}, A.~G., {Nicastro}, L., \&
  {Spyromilio}, J. 1994, \mnras, 268, 430

\bibitem[{Jones {et~al.}(2001)Jones, Oliphant, Peterson, {et~al.}}]{scipy}
Jones, E., Oliphant, T., Peterson, P., {et~al.} 2001, {SciPy}: Open source
  scientific tools for {Python}

\bibitem[{{Kaaret} {et~al.}(2000){Kaaret}, {Cusumano}, \&
  {Sacco}}]{2000ApJ...542L..41K}
{Kaaret}, P., {Cusumano}, G., \& {Sacco}, B. 2000, \apjl, 542, L41

\bibitem[{{Kaaret} {et~al.}(1999){Kaaret}, {Piraino}, {Halpern}, \&
  {Eracleous}}]{1999ApJ...523..197K}
{Kaaret}, P., {Piraino}, S., {Halpern}, J., \& {Eracleous}, M. 1999, \apj, 523,
  197

\bibitem[{Kabiraj \& Paul(2020)}]{kabiraj_broad-band_2020}
Kabiraj, S. \& Paul, B. 2020, MNRAS, 497, 1059

\bibitem[{{Kalogera}(1996)}]{1996ApJ...471..352K}
{Kalogera}, V. 1996, \apj, 471, 352

\bibitem[{{Kaper} {et~al.}(2006){Kaper}, {van der Meer}, \&
  {Najarro}}]{2006A&A...457..595K}
{Kaper}, L., {van der Meer}, A., \& {Najarro}, F. 2006, \aap, 457, 595

\bibitem[{{Kaur} {et~al.}(2008){Kaur}, {Paul}, {Kumar}, \&
  {Sagar}}]{2008MNRAS.386.2253K}
{Kaur}, R., {Paul}, B., {Kumar}, B., \& {Sagar}, R. 2008, \mnras, 386, 2253

\bibitem[{{Kawata} {et~al.}(2019){Kawata}, {Bovy}, {Matsunaga}, \&
  {Baba}}]{2019MNRAS.482...40K}
{Kawata}, D., {Bovy}, J., {Matsunaga}, N., \& {Baba}, J. 2019, \mnras, 482, 40

\bibitem[{Kelley(1986)}]{1986ASIC..167...75K}
Kelley, R.~L. 1986, 167, 75, the Evolution of Galactic X-Ray Binaries

\bibitem[{Kiziltan {et~al.}(2013)Kiziltan, Kottas, De~Yoreo, \&
  Thorsett}]{kiziltan_neutron_2013}
Kiziltan, B., Kottas, A., De~Yoreo, M., \& Thorsett, S.~E. 2013, ApJ, 778, 66

\bibitem[{{Koenigsberger} {et~al.}(2003){Koenigsberger}, {Canalizo}, {Arrieta},
  {Richer}, \& {Georgiev}}]{2003RMxAA..39...17K}
{Koenigsberger}, G., {Canalizo}, G., {Arrieta}, A., {Richer}, M.~G., \&
  {Georgiev}, L. 2003, \rmxaa, 39, 17

\bibitem[{{Koenigsberger} {et~al.}(2006){Koenigsberger}, {Georgiev}, {Moreno},
  {Richer}, {Toledano}, {Canalizo}, \& {Arrieta}}]{2006A&A...458..513K}
{Koenigsberger}, G., {Georgiev}, L., {Moreno}, E., {et~al.} 2006, \aap, 458,
  513

\bibitem[{{Krti{\v{c}}ka} {et~al.}(2015){Krti{\v{c}}ka}, {Kub{\'a}t}, \&
  {Krti{\v{c}}kov{\'a}}}]{2015A&A...579A.111K}
{Krti{\v{c}}ka}, J., {Kub{\'a}t}, J., \& {Krti{\v{c}}kov{\'a}}, I. 2015, \aap,
  579, A111

\bibitem[{La~Palombara \& Mereghetti(2017)}]{la_palombara_swift_2017}
La~Palombara, N. \& Mereghetti, S. 2017, A\&A, 602, A114

\bibitem[{Lai {et~al.}(2001)Lai, Chernoff, \& Cordes}]{lai_pulsar_2001}
Lai, D., Chernoff, D.~F., \& Cordes, J.~M. 2001, ApJ, 549, 1111

\bibitem[{Landi {et~al.}(2017)Landi, Bassani, Bazzano, Bird, Fiocchi, Malizia,
  Panessa, Sguera, \& Ubertini}]{landi_investigating_2017}
Landi, R., Bassani, L., Bazzano, A., {et~al.} 2017, MNRAS, 470, 1107

\bibitem[{Levine {et~al.}(1993)Levine, Rappaport, Deeter, Boynton, \&
  Nagase}]{1993ApJ...410..328L}
Levine, A., Rappaport, S., Deeter, J.~E., Boynton, P.~E., \& Nagase, F. 1993,
  ApJ, 410, 328

\bibitem[{Levine {et~al.}(2000)Levine, Rappaport, \&
  Zojcheski}]{2000ApJ...541..194L}
Levine, A.~M., Rappaport, S.~A., \& Zojcheski, G. 2000, ApJ, 541, 194

\bibitem[{Lindegren {et~al.}(2021)Lindegren, Bastian, Biermann, Bombrun,
  de~Torres, Gerlach, Geyer, Hernández, Hilger, Hobbs, Klioner, Lammers,
  McMillan, Ramos-Lerate, Steidelmüller, Stephenson, \& van
  Leeuwen}]{Lindegren_gaia_2021}
Lindegren, L., Bastian, U., Biermann, M., {et~al.} 2021, A\&A, 649, 31

\bibitem[{Liu {et~al.}(2006)Liu, van Paradijs, \& van~den
  Heuvel}]{liu_catalogue_2006}
Liu, Q.~Z., van Paradijs, J., \& van~den Heuvel, E. P.~J. 2006, \aap, 455, 1165

\bibitem[{Liu {et~al.}(2015)Liu, Tauris, Röpke, Moriya, Kruckow, Stancliffe,
  \& Izzard}]{liu_interaction_2015}
Liu, Z.-W., Tauris, T.~M., Röpke, F.~K., {et~al.} 2015, A\&A, 584, A11

\bibitem[{{Lopes de Oliveira} {et~al.}(2006){Lopes de Oliveira}, {Motch},
  {Haberl}, {Negueruela}, \& {Janot-Pacheco}}]{2006A&A...454..265L}
{Lopes de Oliveira}, R., {Motch}, C., {Haberl}, F., {Negueruela}, I., \&
  {Janot-Pacheco}, E. 2006, \aap, 454, 265

\bibitem[{{Lyne} \& {Lorimer}(1994)}]{1994Natur.369..127L}
{Lyne}, A.~G. \& {Lorimer}, D.~R. 1994, \nat, 369, 127

\bibitem[{{Lyubimkov} {et~al.}(1997){Lyubimkov}, {Rostopchin}, {Roche}, \&
  {Tarasov}}]{1997MNRAS.286..549L}
{Lyubimkov}, L.~S., {Rostopchin}, S.~I., {Roche}, P., \& {Tarasov}, A.~E. 1997,
  \mnras, 286, 549

\bibitem[{Marchant {et~al.}(2016)Marchant, Langer, Podsiadlowski, Tauris, \&
  Moriya}]{marchant_new_2016}
Marchant, P., Langer, N., Podsiadlowski, P., Tauris, T.~M., \& Moriya, T.~J.
  2016, A\&A, 588, A50

\bibitem[{Marrese {et~al.}(2019)Marrese, Marinoni, Fabrizio, \&
  Altavilla}]{marrese_gaia_2019}
Marrese, P.~M., Marinoni, S., Fabrizio, M., \& Altavilla, G. 2019, \aap, 621,
  A144

\bibitem[{{Martins} {et~al.}(2005){Martins}, {Schaerer}, \&
  {Hillier}}]{2005A&A...436.1049M}
{Martins}, F., {Schaerer}, D., \& {Hillier}, D.~J. 2005, \aap, 436, 1049

\bibitem[{{Masetti} {et~al.}(2008){Masetti}, {Mason}, {Morelli}, {Cellone},
  {McBride}, {Palazzi}, {Bassani}, {Bazzano}, {Bird}, {Charles}, {Dean},
  {Galaz}, {Gehrels}, {Landi}, {Malizia}, {Minniti}, {Panessa}, {Romero},
  {Stephen}, {Ubertini}, \& {Walter}}]{2008A&A...482..113M}
{Masetti}, N., {Mason}, E., {Morelli}, L., {et~al.} 2008, \aap, 482, 113

\bibitem[{{Masetti} {et~al.}(2010){Masetti}, {Parisi}, {Palazzi},
  {Jim{\'e}nez-Bail{\'o}n}, {Chavushyan}, {Bassani}, {Bazzano}, {Bird}, {Dean},
  {Charles}, {Galaz}, {Landi}, {Malizia}, {Mason}, {McBride}, {Minniti},
  {Morelli}, {Schiavone}, {Stephen}, \& {Ubertini}}]{2010A&A...519A..96M}
{Masetti}, N., {Parisi}, P., {Palazzi}, E., {et~al.} 2010, \aap, 519, A96

\bibitem[{{Masetti} {et~al.}(2009){Masetti}, {Parisi}, {Palazzi},
  {Jim{\'e}nez-Bail{\'o}n}, {Morelli}, {Chavushyan}, {Mason}, {McBride},
  {Bassani}, {Bazzano}, {Bird}, {Dean}, {Galaz}, {Gehrels}, {Landi}, {Malizia},
  {Minniti}, {Schiavone}, {Stephen}, \& {Ubertini}}]{2009A&A...495..121M}
{Masetti}, N., {Parisi}, P., {Palazzi}, E., {et~al.} 2009, \aap, 495, 121

\bibitem[{Masetti {et~al.}(2013)Masetti, Parisi, Palazzi, Jiménez-Bailón,
  Chavushyan, McBride, Rojas, Steward, Bassani, Bazzano, Bird, Charles, Galaz,
  Landi, Malizia, Mason, Minniti, Morelli, Schiavone, Stephen, \&
  Ubertini}]{masetti_unveiling_2013}
Masetti, N., Parisi, P., Palazzi, E., {et~al.} 2013, \aap, 556, A120

\bibitem[{{McBride} {et~al.}(2007){McBride}, {Wilms}, {Kreykenbohm}, {Coe},
  {Rothschild}, {Kretschmar}, {Pottschmidt}, {Fisher}, \&
  {Hamson}}]{2007A&A...470.1065M}
{McBride}, V.~A., {Wilms}, J., {Kreykenbohm}, I., {et~al.} 2007, \aap, 470,
  1065

\bibitem[{{Miller-Jones} {et~al.}(2018){Miller-Jones}, {Deller}, {Shannon},
  {Dodson}, {Mold{\'o}n}, {Rib{\'o}}, {Dubus}, {Johnston}, {Paredes}, {Ransom},
  \& {Tomsick}}]{2018MNRAS.479.4849M}
{Miller-Jones}, J.~C.~A., {Deller}, A.~T., {Shannon}, R.~M., {et~al.} 2018,
  \mnras, 479, 4849

\bibitem[{Mirabel {et~al.}(2002)Mirabel, Mignani, Rodrigues, Combi, Rodríguez,
  \& Guglielmetti}]{mirabel_runaway_2002}
Mirabel, I.~F., Mignani, R., Rodrigues, I., {et~al.} 2002, 395, 595

\bibitem[{{Moritani} {et~al.}(2018){Moritani}, {Kawano}, {Chimasu}, {Kawachi},
  {Takahashi}, {Takata}, \& {Carciofi}}]{2018PASJ...70...61M}
{Moritani}, Y., {Kawano}, T., {Chimasu}, S., {et~al.} 2018, \pasj, 70, 61

\bibitem[{{Negueruela} {et~al.}(2008){Negueruela}, {Casares}, {Verrecchia},
  {Blay}, {Israel}, \& {Covino}}]{2008ATel.1876....1N}
{Negueruela}, I., {Casares}, J., {Verrecchia}, F., {et~al.} 2008, ATel, 1876, 1

\bibitem[{{Negueruela} \& {Okazaki}(2001)}]{2001A&A...369..108N}
{Negueruela}, I. \& {Okazaki}, A.~T. 2001, \aap, 369, 108

\bibitem[{{Negueruela} {et~al.}(2011){Negueruela}, {Rib{\'o}}, {Herrero},
  {Lorenzo}, {Khangulyan}, \& {Aharonian}}]{2011ApJ...732L..11N}
{Negueruela}, I., {Rib{\'o}}, M., {Herrero}, A., {et~al.} 2011, \apjl, 732, L11

\bibitem[{{Negueruela} {et~al.}(1999){Negueruela}, {Roche}, {Fabregat}, \&
  {Coe}}]{1999MNRAS.307..695N}
{Negueruela}, I., {Roche}, P., {Fabregat}, J., \& {Coe}, M.~J. 1999, \mnras,
  307, 695

\bibitem[{{Nespoli} {et~al.}(2008){Nespoli}, {Fabregat}, \&
  {Mennickent}}]{2008A&A...486..911N}
{Nespoli}, E., {Fabregat}, J., \& {Mennickent}, R.~E. 2008, \aap, 486, 911

\bibitem[{Ng \& Romani(2007)}]{ng_birth_2007}
Ng, C.~Y. \& Romani, R.~W. 2007, ApJ, 660, 1357

\bibitem[{{Nikolaeva} {et~al.}(2013){Nikolaeva}, {Bikmaev}, {Melnikov},
  {Galeev}, {Zhuchkov}, \& {Irtuganov}}]{2013BCrAO.109...27N}
{Nikolaeva}, E.~A., {Bikmaev}, I.~F., {Melnikov}, S.~S., {et~al.} 2013,
  Bulletin Crimean Astrophysical Observatory, 109, 27

\bibitem[{Noutsos {et~al.}(2013)Noutsos, Schnitzeler, Keane, Kramer, \&
  Johnston}]{noutsos_pulsar_2013}
Noutsos, A., Schnitzeler, D. H. F.~M., Keane, E.~F., Kramer, M., \& Johnston,
  S. 2013, MNRAS, 430, 2281

\bibitem[{{Okazaki} \& {Negueruela}(2001)}]{2001A&A...377..161O}
{Okazaki}, A.~T. \& {Negueruela}, I. 2001, \aap, 377, 161

\bibitem[{{Pacheco} {et~al.}(1982){Pacheco}, {Chevalier}, \&
  {Ilovaisky}}]{1982IAUS...98..151P}
{Pacheco}, E.~J., {Chevalier}, C., \& {Ilovaisky}, S.~A. 1982, in Be Stars, ed.
  M.~{Jaschek} \& H.~G. {Groth}, Vol.~98, 151--154

\bibitem[{{Parkes} {et~al.}(1978){Parkes}, {Murdin}, \&
  {Mason}}]{1978MNRAS.184P..73P}
{Parkes}, G.~E., {Murdin}, P.~G., \& {Mason}, K.~O. 1978, \mnras, 184, 73P

\bibitem[{{Parkes} {et~al.}(1980){Parkes}, {Murdin}, \&
  {Mason}}]{1980MNRAS.190..537P}
{Parkes}, G.~E., {Murdin}, P.~G., \& {Mason}, K.~O. 1980, \mnras, 190, 537

\bibitem[{{Pellizza} {et~al.}(2006){Pellizza}, {Chaty}, \&
  {Negueruela}}]{2006A&A...455..653P}
{Pellizza}, L.~J., {Chaty}, S., \& {Negueruela}, I. 2006, \aap, 455, 653

\bibitem[{{Perez} \& {Granger}(2007)}]{2007CSE.....9c..21P}
{Perez}, F. \& {Granger}, B.~E. 2007, Computing in Science and Engineering, 9,
  21

\bibitem[{{Podsiadlowski} {et~al.}(2004){Podsiadlowski}, {Langer},
  {Poelarends}, {Rappaport}, {Heger}, \& {Pfahl}}]{2004ApJ...612.1044P}
{Podsiadlowski}, P., {Langer}, N., {Poelarends}, A.~J.~T., {et~al.} 2004, \apj,
  612, 1044

\bibitem[{{Podsiadlowski} {et~al.}(2005){Podsiadlowski}, {Pfahl}, \&
  {Rappaport}}]{2005ASPC..328..327P}
{Podsiadlowski}, P., {Pfahl}, E., \& {Rappaport}, S. 2005, in ASPC Series, Vol.
  328, Binary Radio Pulsars, ed. F.~A. {Rasio} \& I.~H. {Stairs}, 327

\bibitem[{{Polcaro} {et~al.}(1990){Polcaro}, {Rossi}, {Giovannelli},
  {Ferrari-Toniolo}, {La Padula}, {Persi}, {Manchanda}, {Golinskaya}, {Kurt},
  {Misykima}, {Shafer}, {Shamolin}, {Smirnov}, {Sheffer}, {Boyarchuck}, \&
  {Gershberg}}]{1990A&A...231..354P}
{Polcaro}, V.~F., {Rossi}, C., {Giovannelli}, F., {et~al.} 1990, \aap, 231, 354

\bibitem[{{Portegies Zwart} \& {Verbunt}(1996)}]{1996A&A...309..179P}
{Portegies Zwart}, S.~F. \& {Verbunt}, F. 1996, \aap, 309, 179

\bibitem[{{Porter}(1996)}]{1996MNRAS.280L..31P}
{Porter}, J.~M. 1996, \mnras, 280, L31

\bibitem[{Pradhan {et~al.}(2013)Pradhan, Maitra, Paul, \&
  Paul}]{pradhan_revisiting_2013}
Pradhan, P., Maitra, C., Paul, B., \& Paul, B.~C. 2013, MNRAS, 436, 945

\bibitem[{{Raichur} \& {Paul}(2010)}]{2010MNRAS.406.2663R}
{Raichur}, H. \& {Paul}, B. 2010, \mnras, 406, 2663

\bibitem[{{Ray} \& {Chakrabarty}(2002)}]{2002ApJ...581.1293R}
{Ray}, P.~S. \& {Chakrabarty}, D. 2002, \apj, 581, 1293

\bibitem[{{Reid} \& {Brunthaler}(2004)}]{2004ApJ...616..872R}
{Reid}, M.~J. \& {Brunthaler}, A. 2004, \apj, 616, 872

\bibitem[{{Reig} {et~al.}(2017){Reig}, {Blay}, \&
  {Blinov}}]{2017A&A...598A..16R}
{Reig}, P., {Blay}, P., \& {Blinov}, D. 2017, \aap, 598, A16

\bibitem[{{Reig} {et~al.}(2004){Reig}, {Negueruela}, {Fabregat}, {Chato},
  {Blay}, \& {Mavromatakis}}]{2004A&A...421..673R}
{Reig}, P., {Negueruela}, I., {Fabregat}, J., {et~al.} 2004, \aap, 421, 673

\bibitem[{{Reig} {et~al.}(2005{\natexlab{a}}){Reig}, {Negueruela}, {Fabregat},
  {Chato}, \& {Coe}}]{2005A&A...440.1079R}
{Reig}, P., {Negueruela}, I., {Fabregat}, J., {Chato}, R., \& {Coe}, M.~J.
  2005{\natexlab{a}}, \aap, 440, 1079

\bibitem[{{Reig} {et~al.}(2005{\natexlab{b}}){Reig}, {Negueruela},
  {Papamastorakis}, {Manousakis}, \& {Kougentakis}}]{2005A&A...440..637R}
{Reig}, P., {Negueruela}, I., {Papamastorakis}, G., {Manousakis}, A., \&
  {Kougentakis}, T. 2005{\natexlab{b}}, \aap, 440, 637

\bibitem[{Repetto {et~al.}(2017)Repetto, Igoshev, \&
  Nelemans}]{repetto_galactic_2017}
Repetto, S., Igoshev, A.~P., \& Nelemans, G. 2017, 467, 298

\bibitem[{Rivinius {et~al.}(2013)Rivinius, Carciofi, \&
  Martayan}]{rivinius_classical_2013}
Rivinius, T., Carciofi, A.~C., \& Martayan, C. 2013, A\&AR, 21, 69

\bibitem[{{Robin} {et~al.}(2017){Robin}, {Bienaym{\'e}},
  {Fern{\'a}ndez-Trincado}, \& {Reyl{\'e}}}]{2017A&A...605A...1R}
{Robin}, A.~C., {Bienaym{\'e}}, O., {Fern{\'a}ndez-Trincado}, J.~G., \&
  {Reyl{\'e}}, C. 2017, \aap, 605, A1

\bibitem[{Robin {et~al.}(2003)Robin, Reylé, Derrière, \&
  Picaud}]{robin_synthetic_2003}
Robin, A.~C., Reylé, C., Derrière, S., \& Picaud, S. 2003, A\&A, v.409,
  p.523-540 (2003), 409, 523

\bibitem[{Rohatgi(2021)}]{Rohatgi2020}
Rohatgi, A. 2021, Webplotdigitizer: Version 4.5

\bibitem[{{Romero-G{\'o}mez} {et~al.}(2019){Romero-G{\'o}mez}, {Mateu},
  {Aguilar}, {Figueras}, \& {Castro-Ginard}}]{2019A&A...627A.150R}
{Romero-G{\'o}mez}, M., {Mateu}, C., {Aguilar}, L., {Figueras}, F., \&
  {Castro-Ginard}, A. 2019, \aap, 627, A150

\bibitem[{Safi~Harb {et~al.}(1996)Safi~Harb, Ogelman, \&
  Dennerl}]{1996ApJ...456L..37S}
Safi~Harb, S., Ogelman, H., \& Dennerl, K. 1996, ApJ, 456, L37

\bibitem[{Sanjurjo-Ferrrín {et~al.}(2017)Sanjurjo-Ferrrín, Torrejón,
  Postnov, Oskinova, Rodes-Roca, \&
  Bernabeu}]{sanjurjo-ferrrin_xmm-newton_2017}
Sanjurjo-Ferrrín, G., Torrejón, J.~M., Postnov, K., {et~al.} 2017, A\&A, 606,
  A145

\bibitem[{Schneider {et~al.}(2015)Schneider, Izzard, Langer, \&
  de~Mink}]{2015ApJ...805...20S}
Schneider, F. R.~N., Izzard, R.~G., Langer, N., \& de~Mink, S.~E. 2015, ApJ,
  805, 20

\bibitem[{Schwope {et~al.}(2020)Schwope, Worpel, Webb, Koliopanos, \&
  Guillot}]{schwope_identification_2020}
Schwope, A.~D., Worpel, H., Webb, N.~A., Koliopanos, F., \& Guillot, S. 2020,
  \aap, 637, A35

\bibitem[{{Sidoli} \& {Paizis}(2018)}]{2018MNRAS.481.2779S}
{Sidoli}, L. \& {Paizis}, A. 2018, \mnras, 481, 2779

\bibitem[{Skrutskie {et~al.}(2006)Skrutskie, Cutri, Stiening, Weinberg,
  Schneider, Carpenter, Beichman, Capps, Chester, Elias, Huchra, Liebert,
  Lonsdale, Monet, Price, Seitzer, Jarrett, Kirkpatrick, Gizis, Howard, Evans,
  Fowler, Fullmer, Hurt, Light, Kopan, Marsh, McCallon, Tam, Van~Dyk, \&
  Wheelock}]{skrutskie_two_2006}
Skrutskie, M.~F., Cutri, R.~M., Stiening, R., {et~al.} 2006, AS, 131, 1163

\bibitem[{{Slettebak}(1982)}]{1982ApJS...50...55S}
{Slettebak}, A. 1982, \apjs, 50, 55

\bibitem[{{Sota} {et~al.}(2014){Sota}, {Ma{\'\i}z Apell{\'a}niz}, {Morrell},
  {Barb{\'a}}, {Walborn}, {Gamen}, {Arias}, \& {Alfaro}}]{2014ApJS..211...10S}
{Sota}, A., {Ma{\'\i}z Apell{\'a}niz}, J., {Morrell}, N.~I., {et~al.} 2014,
  \apjs, 211, 10

\bibitem[{Staubert {et~al.}(2019)Staubert, Trümper, Kendziorra, Klochkov,
  Postnov, Kretschmar, Pottschmidt, Haberl, Rothschild, Santangelo, Wilms,
  Kreykenbohm, \& Fürst}]{staubert_cyclotron_2019}
Staubert, R., Trümper, J., Kendziorra, E., {et~al.} 2019, A\&A, 622, A61

\bibitem[{{Stickland} {et~al.}(1997){Stickland}, {Lloyd}, \&
  {Radziun-Woodham}}]{1997MNRAS.286L..21S}
{Stickland}, D., {Lloyd}, C., \& {Radziun-Woodham}, A. 1997, \mnras, 286, L21

\bibitem[{{Stoyanov} {et~al.}(2014){Stoyanov}, {Zamanov}, {Latev}, {Abedin}, \&
  {Tomov}}]{2014AN....335.1060S}
{Stoyanov}, K.~A., {Zamanov}, R.~K., {Latev}, G.~Y., {Abedin}, A.~Y., \&
  {Tomov}, N.~A. 2014, Astronomische Nachrichten, 335, 1060

\bibitem[{{Strader} {et~al.}(2015){Strader}, {Chomiuk}, {Cheung}, {Salinas}, \&
  {Peacock}}]{2015ApJ...813L..26S}
{Strader}, J., {Chomiuk}, L., {Cheung}, C.~C., {Salinas}, R., \& {Peacock}, M.
  2015, \apjl, 813, L26

\bibitem[{Tauris {et~al.}(1999)Tauris, Fender, van~den Heuvel, Johnston, \&
  Wu}]{tauris_circinus_1999}
Tauris, T.~M., Fender, R.~P., van~den Heuvel, E. P.~J., Johnston, H.~M., \& Wu,
  K. 1999, 310, 1165

\bibitem[{Tauris \& Takens(1998)}]{tauris_runaway_1998}
Tauris, T.~M. \& Takens, R.~J. 1998, A\&A, v.330, p.1047-1059 (1998), 330, 1047

\bibitem[{Taylor(2005)}]{taylor_topcat_2005}
Taylor, M.~B. 2005, Astronomical Data Analysis Software and Systems XIV, 347,
  29

\bibitem[{{Townsend} {et~al.}(2011){Townsend}, {Coe}, {Corbet}, \&
  {Hill}}]{2011MNRAS.416.1556T}
{Townsend}, L.~J., {Coe}, M.~J., {Corbet}, R.~H.~D., \& {Hill}, A.~B. 2011,
  \mnras, 416, 1556

\bibitem[{Uchida {et~al.}(2021)Uchida, Takahashi, Fukazawa, \&
  Makishima}]{uchida_study_2021}
Uchida, N., Takahashi, H., Fukazawa, Y., \& Makishima, K. 2021, PASJ, 73, 1389

\bibitem[{van~den Eijnden {et~al.}(2021)van~den Eijnden, Degenaar, Russell,
  Wijnands, Bahramian, Miller-Jones, Hernández~Santisteban, Gallo, Atri,
  Plotkin, Maccarone, Sivakoff, Miller, Reynolds, Russell, Maitra, Heinke,
  Armas~Padilla, \& Shaw}]{van_den_eijnden_new_2021}
van~den Eijnden, J., Degenaar, N., Russell, T.~D., {et~al.} 2021, MNRAS, 507,
  3899

\bibitem[{van~den Heuvel {et~al.}(2000)van~den Heuvel, Portegies~Zwart,
  Bhattacharya, \& Kaper}]{van_den_heuvel_origin_2000}
van~den Heuvel, E. P.~J., Portegies~Zwart, S.~F., Bhattacharya, D., \& Kaper,
  L. 2000, \aap, 364, 563

\bibitem[{{van der Walt} {et~al.}(2011){van der Walt}, {Colbert}, \&
  {Varoquaux}}]{2011CSE....13b..22V}
{van der Walt}, S., {Colbert}, S.~C., \& {Varoquaux}, G. 2011, Computing in
  Science and Engineering, 13, 22

\bibitem[{{Vieira} {et~al.}(2003){Vieira}, {Corradi}, {Alencar}, {Mendes},
  {Torres}, {Quast}, {Guimar{\~a}es}, \& {da Silva}}]{2003AJ....126.2971V}
{Vieira}, S.~L.~A., {Corradi}, W.~J.~B., {Alencar}, S.~H.~P., {et~al.} 2003,
  \aj, 126, 2971

\bibitem[{{Waisberg} \& {Romani}(2015)}]{2015ApJ...805...18W}
{Waisberg}, I.~R. \& {Romani}, R.~W. 2015, \apj, 805, 18

\bibitem[{Walter {et~al.}(2015)Walter, Lutovinov, Bozzo, \&
  Tsygankov}]{walter_high-mass_2015}
Walter, R., Lutovinov, A.~A., Bozzo, E., \& Tsygankov, S.~S. 2015, A\&AR,
  Volume 23, article id.2, 99 pp., 23, 2

\bibitem[{{Walter} {et~al.}(2015){Walter}, {Lutovinov}, {Bozzo}, \&
  {Tsygankov}}]{2015A&ARv..23....2W}
{Walter}, R., {Lutovinov}, A.~A., {Bozzo}, E., \& {Tsygankov}, S.~S. 2015,
  \aapr, 23, 2

\bibitem[{{Wang} \& {Gies}(1998)}]{1998PASP..110.1310W}
{Wang}, Z.~X. \& {Gies}, D.~R. 1998, \pasp, 110, 1310

\bibitem[{Weng {et~al.}(2022)Weng, Qian, Wang, Torres, Papitto, Jiang, Xu, Li,
  Yan, Liu, Ge, \& Yuan}]{weng_radio_2022}
Weng, S.-S., Qian, L., Wang, B.-J., {et~al.} 2022, Nature Astronomy,
  2022NatAs.tmp...71W

\bibitem[{Wenger {et~al.}(2000)Wenger, Ochsenbein, Egret, Dubois, Bonnarel,
  Borde, Genova, Jasniewicz, Laloë, Lesteven, \& Monier}]{wenger_simbad_2000}
Wenger, M., Ochsenbein, F., Egret, D., {et~al.} 2000, A\&AS, 143, 9

\bibitem[{Willcox {et~al.}(2021)Willcox, Mandel, Thrane, Deller, Stevenson, \&
  Vigna-Gómez}]{willcox_constraints_2021}
Willcox, R., Mandel, I., Thrane, E., {et~al.} 2021, ApJ, 920, L37

\bibitem[{{Wilson} {et~al.}(1998){Wilson}, {Finger}, {Harmon}, {Chakrabarty},
  \& {Strohmayer}}]{1998ApJ...499..820W}
{Wilson}, C.~A., {Finger}, M.~H., {Harmon}, B.~A., {Chakrabarty}, D., \&
  {Strohmayer}, T. 1998, \apj, 499, 820

\bibitem[{{Zorec} {et~al.}(2005){Zorec}, {Fr{\'e}mat}, \&
  {Cidale}}]{2005A&A...441..235Z}
{Zorec}, J., {Fr{\'e}mat}, Y., \& {Cidale}, L. 2005, \aap, 441, 235

\end{thebibliography}

\clearpage

\begin{appendix}

\section{Fitting NS HMXB kick distribution with Maxwellian functions}\label{sect:appendix1}

In Section \ref{sect:discuss:subsect:kick} we characterised the observed NS kick magnitudes in HMXBs using Gamma statistics. Here, we present the same results, only using Maxwellian statistics.

The initial motivation to use Maxwellian to reproduce a distribution of kick velocities comes from historical studies focused on isolated pulsars. For instance, \cite{hansen_pulsar_1997} show that the natal kick derived on a set of 86 isolated pulsars could be reproduced by a Maxwellian that would imply a characteristic velocity of 250 to 300\,km\,s$^{-1}$. The authors also note they find no evidence of a low-velocity contribution. However, in our case, we deal with systems that survived through the SN event.

As such, for the sake of comparison, we attempted to fit the probability density function (PDF) of observed NS HMXB kicks with Maxwellian statistics, which we parameterized by the Maxwellian scale ($scale$) and localisation ($loc$). For that we used the Python \textsc{Scipy.stats} package \citep{scipy}, that provides a Maxwellian PDF defined by Equation \eqref{eq:maxwell}.

\begin{equation}
\label{eq:maxwell}
\begin{split}
PDF(y) &= \frac{1}{scale} \sqrt{\frac{2}{\pi}} \, y^2 \, e^{-y^2/2} ,
\\
y &= \frac{x - loc}{scale}
\end{split}
\end{equation}

\begin{figure}
\includegraphics[width=0.95\columnwidth]{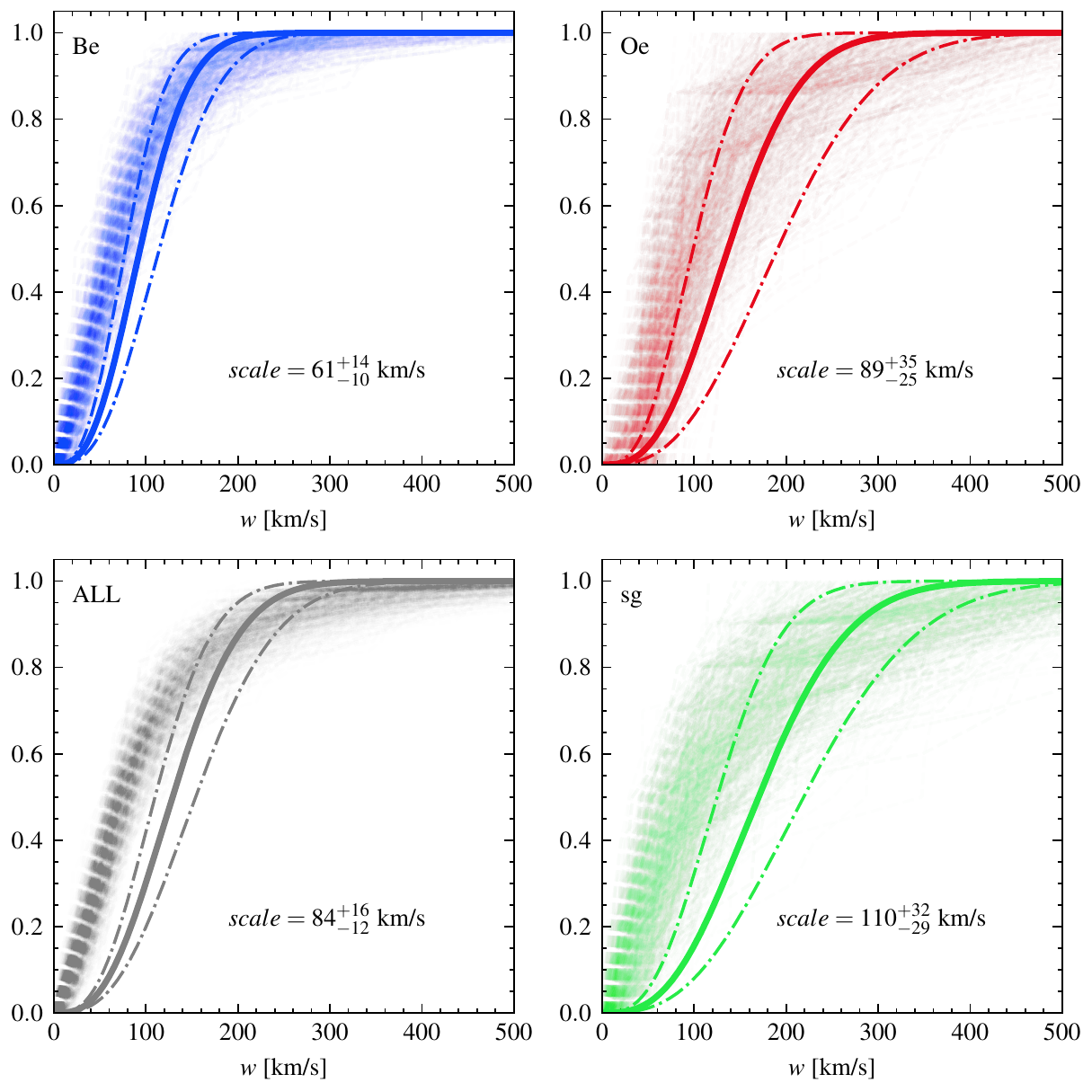}
\caption{Cumulative distributions of kicks obtained with the bootstrapping over each population (transparent) against the median results of the fit by a standardized Maxwell function (opaque lines, 1-$\sigma$ errors are dotted).}\label{fig:KickCDFmaxwell1}

\includegraphics[width=0.95\columnwidth]{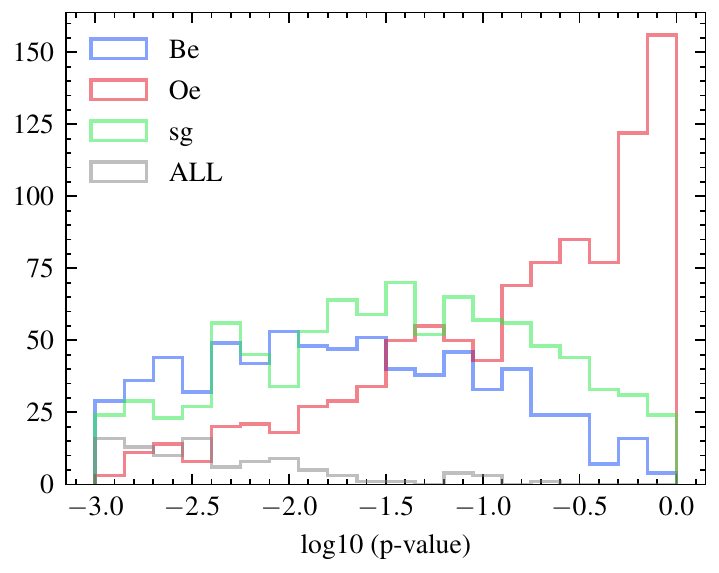}
\caption{Distribution of the p-values associated to KS-tests performed after each iteration of Maxwellian fit (1 free parameter).}\label{fig:pvalue_maxwell_1param}
\end{figure}

In the first attempt, we fix the $loc$ parameter to 0, which produces a standardised Maxwell distribution. As shown in Figure \ref{fig:KickCDFmaxwell1}, while the high velocity regime is somewhat correctly reproduced, it fails at taking into account the low velocity tail. This results in a overall very poor fit as demonstrated by the distribution of $p$-values from individual KS-tests performed at each fit iteration (see Figure \ref{fig:pvalue_maxwell_1param}).

\begin{figure}
\includegraphics[width=0.95\columnwidth]{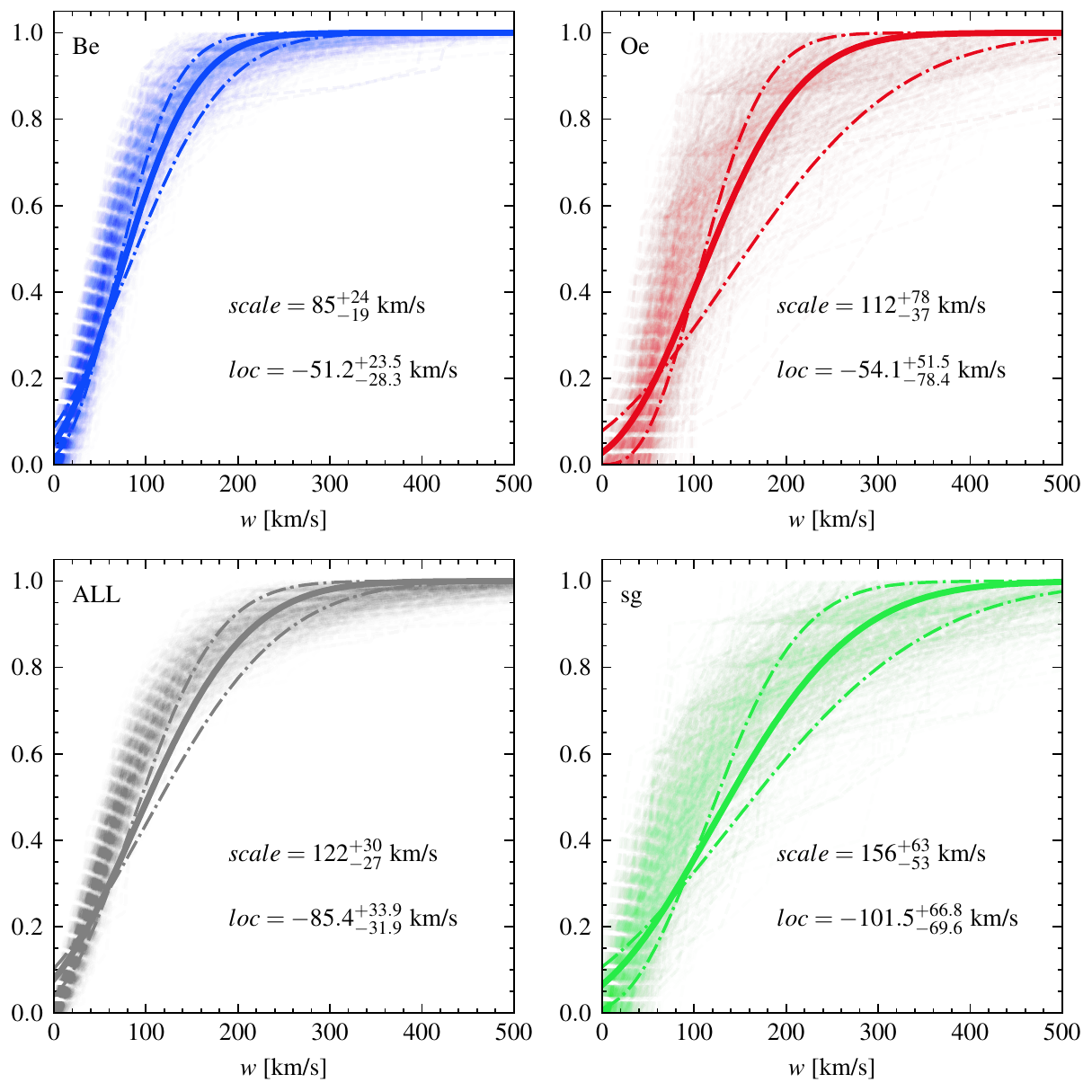}
\caption{Cumulative distributions of kicks obtained with the bootstrapping over each population (transparent) against the median results of the fit by a shifted Maxwell function (opaque lines, 1-$\sigma$ errors are dotted).}\label{fig:KickCDFmaxwell2}
\end{figure}

\begin{figure}
\includegraphics[width=0.95\columnwidth]{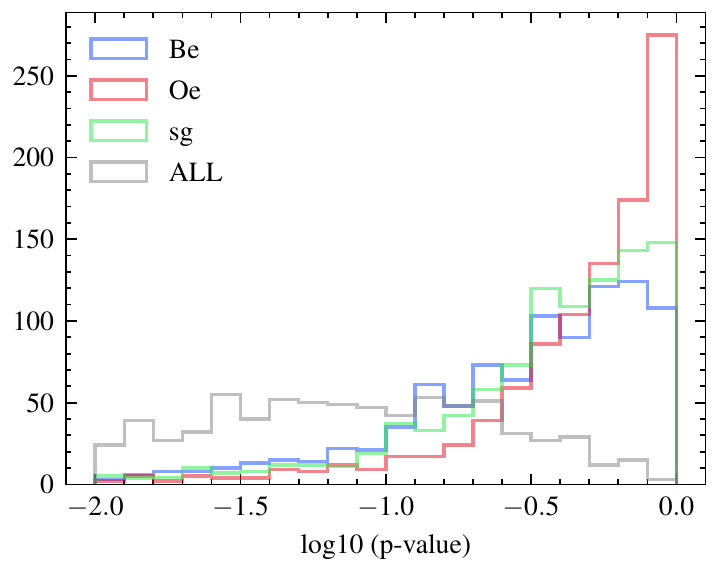}
\caption{Distribution of the p-values associated to KS-tests performed after each iteration of Maxwellian fit (2 free parameters).}\label{fig:pvalue_maxwell_2param}
\end{figure}

We then free the $loc$ parameter to fit shifted Maxwell distributions instead. The results in Figure \ref{fig:KickCDFmaxwell2} show an improvement over the single-parameter Maxwellian; this is also reflected in the distribution of the associated $p$-values (see Figure \ref{fig:pvalue_maxwell_2param}). While the kick distributions are better reproduced than with a standard Maxwellian, the goodness of fit is at best unconvincing for OeHMXBs, rather poor for BeHMXBs and sgHMXBs, and extremely bad when considering the whole sample of HMXBs. Moreover, giving a physical interpretation to a negative offset ($loc < 0$~km~s$^{-1}$) in the Maxwellian distribution of kick velocities is challenging.

Thus, in the scope of this paper, we reckon that the observed population of the Galactic NS HMXBs do not produce natal kick distributions that can be represented by Maxwellian statistics.

\end{appendix}

\end{document}